\documentclass[final,12pt]{elsarticle}
\usepackage{moreverb}
\usepackage{amssymb,bm,amsbsy,graphicx,color,amsmath,amsfonts,framed,scalerel,empheq,colortbl}
\usepackage{mathrsfs}
\usepackage{latexsym}
\usepackage{pseudocode}
\usepackage{epsfig,epsf,rotating}
\usepackage{subfig}
\usepackage{multirow}
\usepackage{bbm}
\usepackage{booktabs}	
\usepackage{tikz}
\usetikzlibrary{shapes,arrows,calc}
\usepackage{setspace}
\usepackage{cancel}
\usepackage{stmaryrd}

\usepackage{float}
\usepackage{epstopdf}
\usepackage[longend,linesnumbered]{algorithm2e}
\usepackage{algpseudocode}

\usepackage{tabularx,tcolorbox}
\tcbuselibrary{skins,breakable}

\usepackage{accents}
\usepackage{tabularx}
\makeatletter
\newcommand*\bigcdot{\mathpalette\bigcdot@{.9}}
\newcommand*\bigcdot@[2]{\mathbin{\vcenter{\hbox{\scalebox{#2}{$\m@th#1\bullet$}}}}}
\makeatother

\newcommand*\widefbox[1]{\fbox{\hspace{2em}#1\hspace{2em}}}
\newcommand{\dbar}[1]{\bar{\bar{#1}}}

\newcommand{\Cross}{\mathbin{\tikz [x=1.4ex,y=1.4ex,line width=.25ex] \draw (0.1,0.1) -- (0.9,0.9) (0.1,0.9) -- (0.9,0.1);}}
\newcommand{\vect}[1]{\boldsymbol{#1}}
\renewcommand{\vec}[1]{\boldsymbol{#1}}

\DeclareMathOperator*{\argmax}{arg\,max}

\makeatletter
\def\mathcolor#1#{\@mathcolor{#1}}
\def\@mathcolor#1#2#3{%
	\protect\leavevmode
	\begingroup
	\color#1{#2}#3%
	\endgroup
}
\makeatother

\usepackage[colorlinks,bookmarksopen,bookmarksnumbered,citecolor=red,urlcolor=red]{hyperref}

\usepackage{geometry}
\newtheorem{corollary}{Corollary}

\newtheorem{remark}{Remark}

\definecolor{gr_red}{RGB}{206, 90, 87}
\definecolor{gr_green}{RGB}{0, 168, 119}
\definecolor{gr_blue}{RGB}{0, 127, 255}

\begin{document}

\begin{frontmatter}

\title{Generalised tangent stabilised nonlinear elasticity: An automated framework for controlling material and geometric instabilities}

\author{%
Roman Poya\,$^{a}$\footnote{Corresponding author: roman.poya@siemens.com},
Rogelio Ortigosa$^b$\footnote{Corresponding author: rogelio.ortigosa@upct.es}, Antonio J. Gil\,$^c$, Theodore Kim\,$^d$, Javier Bonet$^e$%
}
\address{%
$^a$Meshing Framework, Simulation \& Test Solutions, Siemens Digital Industries Software, Cambridge, United Kingdom\\
$^b$Multiphysics Simulation and Optimization Lab, Technical University of Cartagena, Cartagena (Murcia), Spain\\
$^c$ Zienkiewicz Centre for Modelling, Data and AI, Faculty of Science and Engineering, Swansea University, Bay Campus, United Kingdom,\\
$^d$ Computer Graphics Group, Yale School of Engineering $\&$ Applied Science, Yale University, New Haven, Connecticut, United States,\\
$^e$ Centre Internacional de M\'{e}todes Num\'{e}rics en Enginyeria (CIMNE), Barcelona, Spain
}


\begin{abstract}
Tangent stabilised large strain isotropic elasticity was recently proposed by \citet{Poya2023b} wherein by working directly with principal stretches the entire eigenstructure of constitutive and geometric/initial stiffness terms were found in \emph{closed-form}, giving fresh insights into exact convexity conditions of highly non-convex functions in discrete settings. Consequently, owing to these newly found tangent eigenvalues an analytic tangent stabilisation was proposed (for common non-convex strain energies that exhibit material and/or geometric instabilities) bypassing incumbent numerical approaches routinely used in nonlinear finite element analysis. This formulation appears to be extremely robust for quasi-static simulation of complex deformations even with no load increments and time stepping while still capturing instabilities (similar to dynamic analysis) automatically in ways that are infeasible for path-following techniques in practice. In this work, we generalise the notion of tangent stabilised elasticity to virtually all known invariant formulations of nonlinear elasticity. We show that, closed-form eigen-decomposition of tangents is easily available irrespective of invariant formulation or integrity basis. In particular, we work out closed-form tangent eigensystems for isotropic Total Lagrangian deformation gradient ($\vec{F}$)-based and right Cauchy-Green ($\vec{C}$)-based as well as Updated Lagrangian left Cauchy-Green ($\vec{b}$)-based formulations and present their exact convexity conditions postulated in terms of their corresponding tangent and initial stiffness eigenvalues. In addition, we introduce the notion of geometrically stabilised polyconvex large strain elasticity for models that are materially stable but exhibit geometric instabilities for whom we construct their initial stiffness in a spectrally-decomposed form analytically. We further extend this framework to the case of transverse isotropy where once again, closed-form tangent eigensystems are found for common transversely isotropic invariants. In this context, we augment the recent work on mixed variational formulations in principal stretches for deformable and rigid bodies, by presenting a mixed variational formulation for models with arbitrarily directed inextensible fibres. Since, tangent stabilisation unleashes an unparalleled capability for extreme deformations new numerical techniques are required to guarantee element-inversion-safe analysis. To this end, we propose a discretisation-aware load-stepping together with a line search scheme for a robust industry-grade implementation of tangent stabilised elasticity over general polyhedral meshes. Extensive comparisons with path-following techniques provide conclusive evidence that utilising tangent stabilised elasticity can offer both faster and automated results.
\end{abstract}


\end{frontmatter}

\section{Introduction}\label{sec:intro}
Simulating large deformation material and structural response robustly in the context of second order solvers remains a challenging problem. In recent years, nonlinear elastic models have been extensively used for newer applications where extreme deformations are encountered to a degree seldom seen in conventional engineering structures. Non-exhaustive examples of this include, designing meta-materials governed by their geometry rather than composition with multi-stable configurations \cite{PhysRevB.101.064101}, structural topology optimisation wherein the underlying strain energy function is forced to behave in compliance with shapes dictated by void regions and hard physical constraints \cite{Lahuerta2013,WANG2014453,ORTIGOSA2019SIMP}, and physics-based rendering and animation wherein volumetric objects are crashed to a plane, line or even a point. Often, for these applications, obtaining the final deformation profile is more important than the yield or limit points. There is substantial evidence that current numerical approaches in the engineering simulation community fall short of modelling such phenomena \cite{Lahuerta2013,WANG2014453,Farrell2015a}. 

Polyconvex large strain elastic models \cite{SchroderPolyconvex11,BonetPolyconvex15,BonetPolyconvex15b,Kraus2019,GilPolyconvex22} are the most recent incarnations in robust simulation of nonlinear deformations and facilitate a systematic way for describing materially stable models (i.e.\ models with Positive Semi-Definite (PSD) constitutive tangent operator) while also circumventing numerical phenomena such as shear and volumetric lockings due to their enhanced albeit substantially expensive numerical treatment. However, these models by nature leave out or do not preclude geometric instabilities. Here, it serves the purposes to illustrate what is implied by these two forms of instabilities. Mathematically speaking, given a composite strain energy function $e(I(f))$ where $f$ is in general a deformation measure and $I$ the integrity basis \cite{golub1996, nocedal06}, we can obtain the first and second derivatives of $e$ with respect to $f$ through the chain rule as
\begin{align*}
\frac{\partial e}{\partial f} &= \frac{\partial e}{\partial I}\frac{\partial I}{\partial f}, \quad
\frac{\partial^2 e}{\partial f \partial f} =  \mathcolor{gr_green}{ \frac{\partial I}{\partial f} \left( \frac{\partial^2 e}{\partial I\partial I} \right) \frac{\partial I}{\partial f} } + \mathcolor{gr_red}{  \frac{\partial e}{\partial I}  \frac{\partial^2 I}{\partial f \partial f} }.
\end{align*}

In the above expression the first (green) term comprises of second derivative of $e$ with respect to $I$ and hence, represents the constitutive tangent contribution encoding sources of \mathcolor{gr_green}{material instability}, if any, while the second (red) term only has the first derivative of $e$ with respect to $I$ which is usually called the geometric or initial component responsible for \mathcolor{gr_red}{geometric instability}, if any. Virtually, all formulations of nonlinear elasticity (in fact, all metric-based optimisation formulations) possess these two components. Physically, the names material and geometric instabilities come from materials that exhibit strain softening such as brittle failure in geomechanics and snap-through and snap-back behaviour of thin-walled and engineering structures in structural analysis. As mentioned, for polyconvex energy functions the constitutive component of the tangent is always PSD hence, they are called ``materially-stable" models. For convex energy functions both constitutive and geometric (if any) components are PSD however, such models are often too limiting \cite{Poya2023b}. In the general 3-dimensional setting, eliminating such sources of instability from strain energy functions requires an a posteriori \emph{numerical} eigen-decomposition of both constitutive tangent and initial stiffness operators and subsequently clamping the negative eigenvalues to zero (or taking their absolute value as in \cite{HChen24a} for a bigger stabilisation). This is often called positive semi-definite projection of tangents \cite{nocedal06,Teran05,stoma12}. Interestingly, aside from specialised applications, this approach has seldom or rather never been followed in nonlinear analysis of solids in the engineering community as it amounts to ``approximation" of the two derivatives which in turn deteriorates convergence of second order solvers such as Newton-Raphson (NR). In engineering simulations, Newton-Raphson with continuation (i.e. applying the load in increments) remains the most well-established nonlinear solution procedure.

\begin{figure}
%
\centering
\includegraphics[scale=1.15]{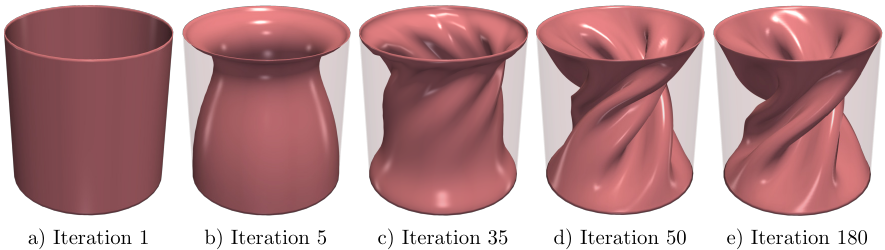}
\caption{Quasi-static simulation with tangent stabilised elasticity with a single step (1 load increment) utilising Newton-Raphson (i.e. Newton Raphson with tangents projected to positive semi-definite cone - P-NR) showing evolution of coarse wrinkles on a cylindrical twist}
\label{fig:cyltwist}%
\end{figure} %

Recently, in \citet{Poya2023b} the authors introduced a new way of looking at large strain elasticity by using the principal stretch information obtained via the Singular-Value-Decomposition (SVD) of the deformation gradient. Using this information alone the complete eigensystem of the tangent operators (constitutive and geometric) was worked out explicitly and in \emph{closed-form}. Consequently, a new tangent stabilisation technique was proposed to circumvent indefiniteness of the global tangent operator and in turn non-convergence of Newton-Raphson when faced with material and/or geometric instabilities. This technique involves fast per quadrature-point projection of the local tangents to positive semi-definite cone using analytically constructed tangent eigenvalues. The resulting second order minimisation approach is the so-called analytically Projected Newton-Raphson (PN or P-NR) \cite{Poya2023b, BSmith19}. In the following sections, you may notice that the terms P-NR and tangent stabilisation are used interchangeably. P-NR even in quasi-static regimes, behaves like dynamic analysis without adding any inertia effect. While the quadratic convergence of NR is compromised this technique enables handling large deformations even with a single pass of Newton i.e.\ a single load increment with no manual tuning or parameter adjustment required. \autoref{fig:cyltwist} shows such an example of Projected Newton with a single load increment and the evolution of wrinkles at various iterations. To verify the robustness of this technique, a large-scale distortion optimisation benchmark with extreme deformations over 10000 computational meshes was carried in \citet{Poya2022a} where the proposed framework passed the entire benchmark with 100\% success rate with no exception. To best of the our knowledge, no incumbent approach in structural analysis has ever undergone such a comprehensive and successful benchmarking. Moreover, blending this technique with continuation approaches specially for displacement-driven problems is straight-forward but often not required. Under some pathological conditions applying the entire imposed displacements in one increment can result in element inversion or self-intersection for which continuation is often helpful. In this work, we present a continuation technique that achieves exactly this goal.

Incumbent structural mechanics approaches for simulations beyond the onset of material and geometric instabilities is based on path-following techniques such as the popular arc-length method for quasi-static analysis \cite{RIKS1979529,CRISFIELD198155,CRISFIELD1983a,FENG199543,FENG1996479,DESOUZANETO199981,CARRERA1994217} or alternatively resorting to dynamic analysis \cite{crisfield_book, Belyt_Book, Bonet-Book21}. The arc length method in particular, is a curve parametrisation strategy that restricts the converged solution to follow a certain path. For quasi-static analysis, arc length is currently the only powerful method used for overcoming non-convergence of NR. However, despite receiving considerable attention since the early 1980s, the method is still far from being either robust and/or automatic on problems of industrial complexity \cite{PRETTI2022106674,Farrell2015a}. Back-tracing an already traced path due to lack of a globally-accepted appropriate load factor choice, the dependence of the method on increment size, and no globally well-established strategy for Dirichlet-driven problems (or when external forces are absent) are amongst outstanding issues of the method not fully resolved yet \cite{crisfield_book}. However, the more pressing issue entails the non-convergence of the method under extreme deformations in emerging applications as repeatedly documented \cite{Lahuerta2013,WANG2014453,ORTIGOSA2019SIMP,ORTIGOSA2020112924}, forcing researchers to adopt simpler consistently linearised elasticity formulations instead \cite{WANG2014453,ORTIGOSA2020112924,Poya2016,POYA201875,POYA2019613}. In this work, we make the case that, tangent stabilised elasticity can be a powerful alternative for such applications with excellent automation capacity requiring no tuning and parameter input.

It is worth noting that, in the context of nonlinear finite elements, many stabilisation techniques have been proposed throughout the years specially for mixed finite elements \cite{BREZZI199027,SchroderMixed17a,AURICCHIO20051075,AURICCHIO2010314,Wall2000a,SIMO1993359,Armero2000a,CERVERA20102559} which are specific to the treatment of instabilities arising from certain families of mixed methods. Some of these instabilities are numerical such as artificial hourglassing nevertheless, they still emanate from indefiniteness of tangent operators \cite{Pfefferkorn2019a,Bieber2023a}. These approaches have more of a case-by-case treatment depending on the mixed finite element spaces used. Our work is distinct from these works and does not deal with such artificial instabilities. Instead, we tackle the general problem of physical instabilities in hyperelastic solids and their subsequent tangent stabilisation. However, the proposed stabilisation can be used in the context of mixed methods (as shown in \cite{Poya2023b} for the novel class of mixed finite elements in principal stretches) eliminating sources of artificial instability too. In \S\ref{sec:6}, we describe a mixed variational principle for transversely isotropic solids with inextensible fibres wherein we perform a similar stabilisation. We adhere to the most prudent way of eliminating tangent indefiniteness favouring automation and robustness while compromising quadratic convergence. 

In this work, we lay out the groundwork for generalised tangent stabilised elasticity for virtually all known formulations of large strain elasticity by expanding the recently proposed framework of \citet{Poya2023b}. Intuitively, this entails finding analytical eigensystems of tangent elasticity operators emanating from all of these formulations. In particular, the key contributions of this work can be summarised as follows:
\begin{enumerate}
\item Closed-form tangent eigensystems for Total Lagrangian deformation gradient ($\vec{F}$)-based and right Cauchy-Green ($\vec{C}$)-based as well as Updated Lagrangian left Cauchy-Green ($\vec{b}$)-based formulations of nonlinear elasticity.
\item Exact convexity conditions for all of the above formulations postulated in terms of their corresponding tangent and initial stiffness eigenvalues.
\item Introducing geometrically stabilised polyconvex large strain elasticity, for polyconvex formulations and presenting compact expressions for their initial stiffness eigensystem.
\item Extension to the case of transverse isotropy and once again, working out closed-form tangent eigensystems for common transversely isotropic invariants.
\item Discretisation-aware load stepping scheme for inversion-safe simulation with tangent stabilised elasticity, and finally
\item Mixed Hu-Washizu variational formulation and finite element implementation for large strain elastic models with inextensible fibres augmenting the mixed finite element framework of \citet{Poya2023b} for deformable and rigid models.
\end{enumerate}

Complete implementation of all formulations is available for prototyping via the open-source Florence finite element package \url{https://github.com/romeric/florence} while a performance-oriented and more thorough implementation is part of \href{https://plm.sw.siemens.com/en-US/simcenter/mechanical-simulation/simcenter-3d/}{Simcenter3D Pre/Post (Meshing)} suite.

The structure of the paper is as follows. In \S\ref{sec:2}, the mapping and kinematics of motion is introduced, followed by the ingredients that make up our formulation. In \S\ref{sec:3}, we present large strain elasticity, the general convexity conditions and transverse isotropy restrictions imposed on the strain energy densities. In \S\ref{sec:4}, we briefly review the work of \cite{Poya2023b} and explicitly work out eigensystems for various isotropic invariant formulations of large strain elasticity. In \S\ref{sec:5}, we present the same for transversely isotropic invariants. In \ref{sec:8}, we introduce a new continuation technique with Newton-Raphson procedure and a line search scheme that guarantees element-inversion free simulations through safe load stepping. Various numerical examples are presented in \S\ref{sec:9}. Finally \S\ref{sec:10} concludes the paper. In \ref{sec:6}-\ref{sec:7}, variational principles including a mixed Hu-Washizu variational formulation for transversely isotropic large strain elasticity with inextensible fibres is presented and its finite element discretisation is discussed. 

\section{Continuum mechanics preliminaries}\label{sec:2}
\begin{figure}
\centering
\includegraphics[scale=0.99]{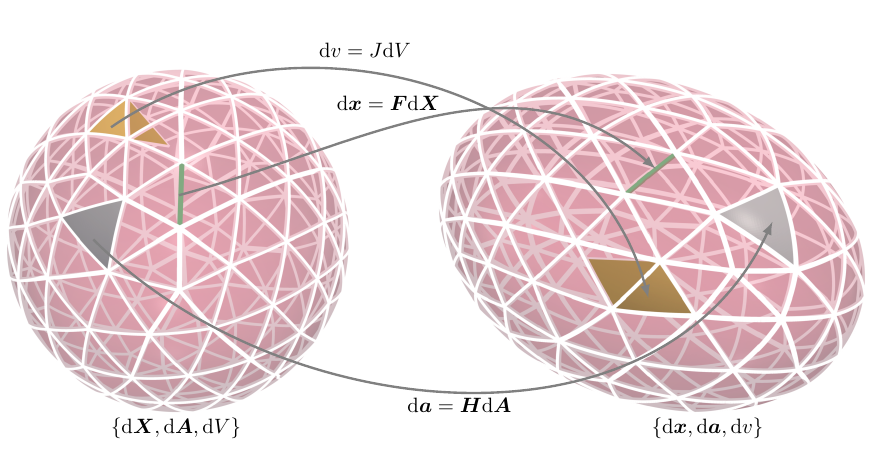}
\caption{Mapping original and deformed configurations and fundamental kinematic measures of motion in $\mathbb{R}^3$ Euclidean space. $\{\vec{F},\vec{H},J\}$ map edges, areas and volumes from original $\{\text{d}\vec{X},\text{d}\vec{A},\text{d}V\}$ to deformed $\{\text{d}\vec{x},\text{d}\vec{a},\text{d}v\}$ configuration, respectively.}
\label{fig:potatomap}
\end{figure}

Consider the 3-dimensional deformation of an elastic medium from its initial configuration occupying a volume $V$, of boundary $\partial V$, into a final configuration at volume $v$, of boundary $\partial v$ (see Fig.~\ref{fig:potatomap}). Following Bonet et. al. \cite{BonetPolyconvex15, BonetPolyconvex15b} the standard notation and definitions for fundamental kinematic measures namely, the deformation gradient $\vec{F}$, its cofactor $\vec{H}$ and its determinant $J$ are used \footnote{Unless stated otherwise, throughout this work, lower case bold letters ($\vec{a}$) denote vectors, capital bold letters ($A$) second order tensors and capital blackboard-bold letters ($\mathbb{A}$) fourth order tensors. Calligraphic letters ($\mathcal{A}$) are used to represent tangent elasticity operators, $\lambda_i$s are used to denote the singular-values of the deformation gradient tensor (i.e.\ principal stretches) and $\bar{\lambda}_i$s are used to denote the eigenvalues of Hessian and tangent elasticity operators. Principal and non-principal directions are represented with $\vec{N}$ and $\vec{M}$, respectively. Square brackets around bold letters with subscripts such as ($[\vec{N}]_i$) indicate distinct vectors/tensors spanning in $d$-dimensional space as opposed to indices of the stated vector/tensor ($N_i$).}
 \begin{align}
\vec{F} = \frac{\partial \vec{x}}{\partial \vec{X}} = \vec{\nabla}_0 \vec{x},  \quad \vec{H} = J \vec{F}^{-T}= \frac{1}{2}\vec{F} \Cross \vec{F},  \quad J = \text{det}\vec{F}.
\end{align}
where $\vec{x}$ represents the current position of a particle originally at $\vec{X}$ and $\vec{\nabla}_0$ denotes the gradient with respect to material coordinates and $\Cross$ is the tensor cross product operator $[\vec{A} \Cross \vec{B}]_{ij} = \mathcal{E}_{ijk} \mathcal{E}_{IJK} A_{jJ} B_{kK}$ where $\mathcal{E}$ is the third order Levi-Civita tensor \cite{deBoer, BonetPolyconvex15b}. Virtual and incremental variations of $\vec{x}$ will be denoted by $\delta \vec{u}$ and $\Delta \vec{u}$, respectively.  It is assumed that $\vec{x},\delta \vec{u}$ and $\Delta \vec{u}$ satisfy appropriate displacement based boundary conditions in $\partial_u V$. Additionally, the body is under the action of certain body forces per unit undeformed volume $\vec{f}_0$ and traction per unit undeformed area $\vec{t}_0$ in $\partial_t V$, where $\partial_t V \cup \partial_u V = \partial V$ and $\partial_t V \cap \partial_u V = \emptyset$.
Crucial to our development is the left polar decomposition of the deformation gradient tensor 
\begin{align}
\vec{F} = \vec{R}\vec{U},
\end{align}
into a rotation tensor $\vec{R}$ and a stretch tensor $\vec{U}$. This decomposition is often obtained from the Singular-Value-Decomposition (SVD) of $\vec{F}$ as follows
\begin{align}
\vec{F} = \hat{\vec{U}}\vec{\Lambda}\hat{\vec{V}}^T, \quad \vec{R} = \hat{\vec{U}}\hat{\vec{V}}^T,  \quad \vec{U} = \hat{\vec{V}}\vec{\Lambda}\hat{\vec{V}}^T,
\label{eq:svd}
\end{align}
where $\hat{\vec{U}}$ and $\hat{\vec{V}}$ are orthogonal tensors namely left and right singular-matrices, respectively and the tensor $\vec{\Lambda}$ encodes the singular-values of $\vec{F}$ i.e.\ the principal stretches namely $\lambda_1 \geq \lambda_2 \geq \lambda_3 \geq 0$ such that $\lambda_i = \Lambda_{ii}$ (we have used the symbol $\hat{\vec{U}}$ for left singular matrix (and accordingly $\hat{\vec{V}}$ for right singular matrix) to distinguish with the stretch tensor $\vec{U}$). Note that, in mechanics literature, the more standard definitions for the stretch tensor $\vec{U}$ and the deformation gradient tensor $\vec{F}$ in principal direction appear as
\begin{align}
\vec{U} = \sum_{i=1}^d \lambda_i [\vec{N}]_i \otimes [\vec{N}]_i, \quad \vec{F} = \sum_{i=1}^d \lambda_i [\vec{n}]_i \otimes [\vec{N}]_i,
\end{align}
where $[\vec{N}]_i$ are the material principal directions in fact corresponding to the columns of $\hat{\vec{V}}$ which is obtained from the SVD and $d$ denotes the Euclidean space dimension (i.e.\ 1,2,3). Similarly, vectors $[\vec{n}]_i$ represent the spatial principal directions $[\vec{n}]_i = \vec{R} [\vec{N}]_i$.

\section{Large strain elasticity}\label{sec:3}
There is often a set of requirements that the internal strain energy density $e$ has to fulfil. In this section we briefly review two such requirements most pertinent to our work.

The first requirement is that of convexity. Convexity implies that for energy function $e(\vec{F})$ the following condition should be met  \cite{Hartmann2003a,DacorognaBooka,GilPolyconvex22} 
%
\begin{align}\label{eqn:convexity}
e(\theta\vec{F}_1 + (1-\theta)\vec{F}_2) \leq \theta e(\vec{F}_1) +  (1-\theta) e(\vec{F}_2);  \quad \forall \vec{F}_1,\vec{F}_2 \in\mathbb{R}^{d\times d}; \quad \theta \in [0,1].
\end{align}
which for differentiable functions can be expressed alternatively as 
\begin{align}\label{eqn:convexity in first derivatives}
	\Bigg(\frac{\partial e(\vec{F})}{\partial\vec{F}}\bigg|_{\vec{F}_1} - \frac{\partial e(\vec{F})}{\partial\vec{F}}\bigg|_{\vec{F}_2}\Bigg): \big(\vec{F}_1-\vec{F}_2\big)\geq 0;\,\qquad \forall \vec{F}_1,\vec{F}_2\in\mathbb{R}^{d\times d},
\end{align}
and for twice differentiable functions can be expressed alternatively as
\begin{align}
D^2 e(\vec{F})[\delta \vec{F}; \delta \vec{F}] = \delta \vec{F} \bigcdot \vec{\mathcal{C}} \bigcdot \delta \vec{F} \geq 0; \quad \forall \vec{F}, \delta \vec{F} \quad \in \mathbb{R}^{d\times d},
\label{eq:convexcond1}
\end{align}
where $ \bigcdot$ is the dual product. Equation \eqref{eq:convexcond1} essentially mandates positive semi-definiteness of the elasticity tensor $\vec{\mathcal{C}}$ in \eqref{eq:lambCorig}.

The second requirement is that of objectivity or frame invariance which implies independence of the energy density with respect to arbitrary rotations $\vec{Q}$ of the spatial configuration, which can be formulated as
\begin{align}
e(\vec{Q} \vec{F} )  = e(\vec{F}); \quad \forall  \vec{Q} \in SO(3)
\end{align}
with $SO(3)$ the special (proper) orthogonal group \cite{Gurtin-Book, Ogden-Book}. This requirement dictates that the strain energy $e$ should be expressed in terms of symmetric strain measures such as the right Cauchy-Green strain tensor $\vec{C}=\vec{F}^T\vec{F}$ as $e(\vec{F}) = \breve{e}(\vec{C})$. For the anisotropic energies further restrictions, reflecting the consequences of the principle of material symmetry, have to be taken into account. In particular, the energy $e$ must be independent of any rotations/reflections $\vec{Q} \in \mathcal{G}_m \subset$ O(3) of the material configuration, where $\mathcal{G}_m$ denotes the corresponding symmetry group and O(3) the full orthogonal group \cite{Zheng1994,Schroder20083486,GilPolyconvex22}. The anisotropic restriction is formally formulated as
\begin{align}
\breve{e}(\vec{C}) = \breve{e}(\vec{QCQ}^T); \quad \forall \vec{Q} \in \mathrm{O(3)}
\end{align}
Similar to \cite{GilPolyconvex22}, in this work, we restrict ourselves to the symmetry $D_{\infty h}$, which corresponds to the usual definition of transverse isotropy and the symmetry $C_\infty$, the so-called rotational symmetry. Moreover, these two symmetry groups can be characterised by appropriate \emph{structural tensors} which encapsulate the symmetry attributes of their corresponding groups, which in are given by the second-order tensor $\vec{M} \otimes \vec{M}$ (for $D_{\infty h}$) and the first-order tensor (vector) $\vec{M}$ (for $C_\infty$). It is now straightforward to obtain an irreducible list of isotropic invariants (the so-called \emph{integrity basis}) for the characterisation of energy densities.
\begin{subequations}
\begin{align}
\mathrm{isotropic\; elasticity}:& \qquad \mathrm{tr}(\vec{C}), \mathrm{tr}(\vec{C}^2), \mathrm{tr}(\vec{C}^3), \\
\mathrm{transversely-isotropic\; elasticity}: & \qquad  \mathrm{tr}(\vec{C}(\vec{M} \otimes \vec{M})), \mathrm{tr}(\vec{C}^2(\vec{M} \otimes \vec{M})),
\end{align}%
\end{subequations}
where tr represents the trace operator. We denote the above 5 invariants as
\begin{align}
J_1 = \mathrm{tr}(\vec{C}), \quad
J_2 = \mathrm{tr}(\vec{C}^2), \quad
J_3 = \mathrm{tr}(\vec{C}^3), \quad
J_4 =  \mathrm{tr}(\vec{C}(\vec{M} \otimes \vec{M})), \quad
J_5 = \mathrm{tr}(\vec{C}^2(\vec{M} \otimes \vec{M})),
\end{align}%
and introduce the internal energy $\breve{e}$ as
\begin{align}
e( \vec{\nabla}_0\vec{x} ) = \breve{e}(J_1, J_2, J_3, J_4, J_5)
\end{align}%
Many particularisations of the above integrity bases are possible and have surfaced in the literature. In the next two sections, we review virtually all commonly used formulations of isotropic and transversely isotropic large strain elasticity. We proceed to obtain closed-form eigen-decompositions for their constitutive tangent and initial stiffness operators. This enables postulating convexity in terms of these newly found eigenvalues which is used to devise the corresponding tangent stabilisations for each formulation.

\section{Stabilised large strain isotropic elasticity} \label{sec:4}
\subsection{Large strain isotropic elasticity in principle stretches: The new way}
We start with large strain isotropic elasticity formulation of \citet{Poya2023b} which will serve as the basis for our subsequent formulations. Following \cite{Poya2023b}, we express the internal strain energy density $e$ dependent on the state of deformation $\vec{\nabla}_0 \vec{x}$ and encapsulating the constitutive information of an isotropic hyperelastic material necessary to close the system of governing equations, in terms of principal stretches of the deformation gradient as follows 
\begin{align}
e(\vec{\nabla}_0\vec{x}) = W(\lambda_1, \lambda_2, \lambda_3) = W(\mathcal{A_F}),
\label{eq:energydef1}
\end{align}
where for brevity we have defined the set of principal stretches as $\mathcal{A_F}=\{\lambda_1,\cdots, \lambda_d\}$. Furthermore, a set of work conjugate stresses $\mathcal{S_F}=\{\Sigma_{\lambda_1},\cdots, \Sigma_{\lambda_d}\}$ can be introduced corresponding to each principal stretch such that
\begin{align}
\Sigma_{\lambda_1} = \frac{\partial W}{\partial  \lambda_1}, \quad
\Sigma_{\lambda_2} = \frac{\partial W}{\partial  \lambda_2}, \quad
\Sigma_{\lambda_3} = \frac{\partial W}{\partial  \lambda_3},
\end{align}
which can be used to obtain the more standard first Piola-Kirchhoff stress tensor $\vec{P}$ by first recalling that
\begin{align}
De[\delta \vec{u}] = \vec{P} : \vec{\nabla}_0 \delta \vec{u}; \qquad \vec{P} = \frac{\partial e}{\partial \vec{F}} \bigg|_{\vec{F}=\vec{\nabla}_0\vec{x}}.
\end{align}
which leads to the evaluation of $\vec{P}$ as \cite{Poya2023b}
\begin{align}
\vec{P} =  \sum_{i=1}^d  \Sigma_{\lambda_i} \frac{\partial \lambda_i}{\partial \vec{F}} = \sum_{i=1}^d \Lambda_{P_{ii}} [\vec{n}]_i \otimes [\vec{N}]_i = \hat{\vec{U}} \vec{\Lambda_P} \hat{\vec{V}}^T,
\label{eq:lambpk1}%
\end{align}%
where $\Lambda_{P_{ii}} = \Sigma_{\lambda_i}$ is the diagonal principal stress tensor revealing that, the first Piola Kirchhoff stress tensor $\vec{P}$ is coaxial with $\vec{F}$. Consequently, we term this formulation as ``$\vec{F}$-based formulation in principal stretches" in sharp contrast with the classical right Cauchy-Green or ``$\vec{C}$-based formulation in principal stretches" that uses the squares of the singular-values. The total tangent elasticity operator $\vec{\mathcal{C}}$ needed in the context of second order minimisation techniques such as Newton-Raphson can be obtain in a similar fashion by first recalling that 
\begin{align}
D^2e[\delta \vec{u}; \Delta \vec{u}] = \vec{\nabla}_0 \delta \vec{u} : D\vec{P}[\Delta \vec{u}] =  \vec{\nabla}_0 \delta \vec{u} :  \vec{\mathcal{C}}  :  \vec{\nabla}_0 \Delta \vec{u}; \qquad \vec{\mathcal{C}} = \frac{\partial \vec{P}}{\partial \vec{F}} \bigg|_{\vec{F}=\vec{\nabla}_0\vec{x}} = \frac{\partial^2 e}{\partial \vec{F}\partial \vec{F}} \bigg|_{\vec{F}=\vec{\nabla}_0\vec{x}},
\label{eq:lambCorig}
\end{align}
which leads to the evaluation of $\vec{\mathcal{C}}$ as an algebraic summation of the constitutive $\vec{\mathcal{C}}_k$ and the geometric stiffness $\vec{\mathcal{C}}_p$ components as \cite{Poya2023b}
\begin{align}
\vec{\mathcal{C}} = \vec{\mathcal{C}}_k + \vec{\mathcal{C}}_p  &=
\begin{bmatrix}
\dfrac{\partial \lambda_1}{\partial \vec{F}}, &
\cdots, &
\dfrac{\partial \lambda_d}{\partial \vec{F}}
\end{bmatrix}
[\vec{H}_W]
\begin{bmatrix}
\dfrac{\partial \lambda_1}{\partial \vec{F}} \\
\vdots \\
\dfrac{\partial \lambda_d}{\partial \vec{F}} 
\end{bmatrix}%
+ \sum_{i=1}^d \Sigma_{\lambda_i} \frac{\partial^2 \lambda_i}{\partial \vec{F} \partial \vec{F}} \nonumber \\
&= 
\underbrace{%
\sum_{i=1}^d  \sum_{j=1}^d \frac{\partial^2 W}{\partial \lambda_i \lambda_j}  \frac{\partial \lambda_i}{\partial \vec{F}} \otimes  \frac{\partial \lambda_j}{\partial \vec{F}}%
}_{ \vec{\mathcal{C}}_k }
+
\underbrace{%
 \sum_{i=1}^d \Sigma_{\lambda_i} \frac{\partial^2 \lambda_i}{\partial \vec{F} \partial \vec{F}},%
 }_{ \vec{\mathcal{C}}_p }
\label{eq:lambC2}%
\end{align} 
where $\vec{H}_W$ denotes the symmetric but not necessarily PSD $d \times d$ Hessian operator containing the second derivatives of $W$ with respect to principal stretches
\begin{align}
\vec{H}_W = 
\begin{bmatrix}
\partial^2_{ \lambda_1 \lambda_1}W & \cdots & \partial^2_{ \lambda_1 \lambda_d}W \\
&  \ddots & \vdots \\
sym &  & \partial^2_{ \lambda_d \lambda_d}W
\end{bmatrix},
\label{eq:invarHw}
\end{align}
where $\partial^2_{\lambda_i \lambda_j}\hat{W} = \frac{\partial^2\,\hat{W}}{\partial \lambda_j \partial \lambda_j}$
and the last term in \eqref{eq:lambC2} is the ``initial stress'' or ``geometric stiffness'' term. As described in \cite{Poya2023b}, convexity mandates positive semi-definiteness of \eqref{eq:lambC2} and hence, of both constitutive $\vec{\mathcal{C}}_k$ (via the Hessian $\vec{H}_W$) and initial $\vec{\mathcal{C}}_p$ contributions. The derivatives of principal stretches with respect to the deformation gradient $\frac{\partial \lambda_i}{\partial \vec{F}}$ needed for consistent linearisation can be obtained by taking directional derivatives which in 2-dimensions yields
%
\begin{subequations}
\begin{align}
D\lambda_1[\delta\vec{u}] = \frac{\partial \lambda_1}{\partial \vec{F}} : \vec{\nabla}_0 \delta \vec{u}, \qquad 
[\vec{D}]_1 = \frac{\partial \lambda_1}{\partial \vec{F}} &= [\vec{n}]_1 \otimes [\vec{N}]_1 = \hat{\vec{U}} 
\begin{bmatrix}
1 & 0 \\
0 & 0
\end{bmatrix} 
\hat{\vec{V}}^T, \\
D\lambda_2[\delta\vec{u}] = \frac{\partial \lambda_2}{\partial \vec{F}} : \vec{\nabla}_0 \delta \vec{u}, \qquad 
[\vec{D}]_2 = \frac{\partial \lambda_2}{\partial \vec{F}} &= [\vec{n}]_2 \otimes [\vec{N}]_2 =  \hat{\vec{U}}
\begin{bmatrix}
0 & 0 \\
0 & 1
\end{bmatrix}
\hat{\vec{V}}^T, 
\end{align}%
\label{eq:lambdder1a}%
\end{subequations}%
and similarly in 3-dimensions
\begin{subequations}
\begin{align}
D\lambda_1[\delta\vec{u}] = \frac{\partial \lambda_1}{\partial \vec{F}} : \vec{\nabla}_0 \delta \vec{u}, \qquad 
[\vec{D}]_1 = \frac{\partial \lambda_1}{\partial \vec{F}} &= [\vec{n}]_1 \otimes [\vec{N}]_1 = \hat{\vec{U}} 
\begin{bmatrix}
1 & 0 & 0 \\
0 & 0 & 0 \\
0 & 0 & 0 
\end{bmatrix} 
\hat{\vec{V}}^T, \\
D\lambda_2[\delta\vec{u}] = \frac{\partial \lambda_2}{\partial \vec{F}} : \vec{\nabla}_0 \delta \vec{u}, \qquad 
[\vec{D}]_2 = \frac{\partial \lambda_2}{\partial \vec{F}} &= [\vec{n}]_2 \otimes [\vec{N}]_2 = \hat{\vec{U}} 
\begin{bmatrix}
0 & 0 & 0 \\
0 & 1 & 0 \\
0 & 0 & 0 
\end{bmatrix} 
\hat{\vec{V}}^T, \\
D\lambda_3[\delta\vec{u}] = \frac{\partial \lambda_3}{\partial \vec{F}} : \vec{\nabla}_0 \delta \vec{u}, \qquad 
[\vec{D}]_3 = \frac{\partial \lambda_3}{\partial \vec{F}} &= [\vec{n}]_3 \otimes [\vec{N}]_3 = \hat{\vec{U}} 
\begin{bmatrix}
0 & 0 & 0 \\
0 & 0 & 0 \\
0 & 0 & 1 
\end{bmatrix} 
\hat{\vec{V}}^T, 
\end{align}%
\label{eq:lambdder1b}%
\end{subequations}%
where we have denoted $[\vec{D}]_i = \frac{\partial \lambda_i}{\partial \vec{F}}$ for the the \emph{scaling} rank-one tensors to ease further developments down the line.
The elegance of this formulation is that both the constitutive tangent $\vec{\mathcal{C}}_k$ and initial stiffness $\vec{\mathcal{C}}_p$ can be represented in spectrally-decomposed form. The Hessian operator $\vec{H}_W$ is a small $d\times d$ matrix and simple to find its analytical eigenvalues both in 2 and 3 dimensions; refer to \cite{Poya2023b} for analytical formulae. The initial stiffness on the other hand, has a null space of 2 in 2-dimensions (only two eigenvalues) and a null space of 3 in 3-dimensions (only 6 eigenvalues) and can be written as an algebraic summation of eigenvalues and eigenmatrices as \cite{Poya2023b} 
\begin{align}
\vec{\mathcal{C}_p} = \sum_{i=1}^k \bar{\lambda}_i^{\vec{\mathcal{C}_p}} [\vec{L}]_i \otimes [\vec{L}]_i + \sum_{i=1}^{k} \bar{\lambda}_{i+k}^{\vec{\mathcal{C}_p}}  [\vec{T}]_i \otimes [\vec{T}]_i, \quad \text{where} \quad k=1\; \text{in\;2D;\;}\; k=3\; \text{in\;3D},
\label{eq:initstiff1}
\end{align}
where $\bar{\lambda}_i^{\vec{\mathcal{C}_p}}$ are the eigenvalues of the initial stiffness term which in 2-dimensions are found concisely as
\begin{align}
\bar{\lambda}_1^{\vec{\mathcal{C}_p}} = \frac{\Sigma_{\lambda_1} - \Sigma_{\lambda_2}}{\lambda_1 - \lambda_2}, \qquad
\bar{\lambda}_2^{\vec{\mathcal{C}_p}} = \frac{\Sigma_{\lambda_1} + \Sigma_{\lambda_2}}{\lambda_1 + \lambda_2},
\label{eq:isPSD2d}
\end{align}
and similarly in 3-dimensions the 6 eigenvalues of the initial stiffness are found concisely as
\begin{align}
\bar{\lambda}_1^{\vec{\mathcal{C}_p}} &= \frac{\Sigma_{\lambda_2} - \Sigma_{\lambda_3}}{\lambda_2 - \lambda_3}, \qquad
\bar{\lambda}_2^{\vec{\mathcal{C}_p}} = \frac{\Sigma_{\lambda_1} - \Sigma_{\lambda_3}}{\lambda_1 - \lambda_3},
\qquad
\bar{\lambda}_3^{\vec{\mathcal{C}_p}} = \frac{\Sigma_{\lambda_1} - \Sigma_{\lambda_2}}{\lambda_1 - \lambda_2}, \nonumber \\
\bar{\lambda}_4^{\vec{\mathcal{C}_p}} &= \frac{\Sigma_{\lambda_2} + \Sigma_{\lambda_3}}{\lambda_2 + \lambda_3}, \qquad
\bar{\lambda}_5^{\vec{\mathcal{C}_p}}  = \frac{\Sigma_{\lambda_1} + \Sigma_{\lambda_3}}{\lambda_1 + \lambda_3},
\qquad
\bar{\lambda}_6^{\vec{\mathcal{C}_p}}  = \frac{\Sigma_{\lambda_1} + \Sigma_{\lambda_2}}{\lambda_1 + \lambda_2},
\label{eq:isPSD3d}
\end{align}
where $[\vec{L}]_i$s, $[\vec{T}]_i$s are the eigenmatrices or the so-called the \emph{flip} and \emph{twist} tensors (in the sense of \citet{BSmith19}) which in 2-dimensions are given by
\begin{align}
[\vec{L}]_1 = \frac{1}{\sqrt{2}}\hat{\vec{U}}
\begin{bmatrix}
0 & 1 \\
1 & 0
\end{bmatrix}
\hat{\vec{V}}^{T},
\qquad
[\vec{T}]_1 = \frac{1}{\sqrt{2}}\hat{\vec{U}}
\begin{bmatrix}
0 & -1 \\
1 & 0
\end{bmatrix}
\hat{\vec{V}}^{T},
\label{eq:fliptwisttensors2d}%
\end{align}
and similarly in 3-dimensions
\begin{subequations}
\begin{align}
[\vec{L}]_1 &= \frac{1}{\sqrt{2}}\hat{\vec{U}}
\begin{bmatrix}
0 & 0 &  0 \\
0 & 0 &  1 \\
0 & 1 &  0
\end{bmatrix}
\hat{\vec{V}}^{T}, \quad
[\vec{T}]_1 = \frac{1}{\sqrt{2}}\hat{\vec{U}}
\begin{bmatrix}
0 & 0 &  0 \\
0 & 0 & -1 \\
0 & 1 &  0
\end{bmatrix}
\hat{\vec{V}}^{T}, \\
[\vec{L}]_2 &= \frac{1}{\sqrt{2}}\hat{\vec{U}}
\begin{bmatrix}
0 & 0 &  1 \\
0 & 0 &  0 \\
1 & 0 &  0
\end{bmatrix}
\hat{\vec{V}}^{T},  \quad
[\vec{T}]_2 = \frac{1}{\sqrt{2}}\hat{\vec{U}}
\begin{bmatrix}
0 & 0 & -1 \\
0 & 0 &  0 \\
1 & 0 &  0
\end{bmatrix}
\hat{\vec{V}}^{T},  \\
[\vec{L}]_3 &= \frac{1}{\sqrt{2}}\hat{\vec{U}}
\begin{bmatrix}
0 &  1 &  0 \\
1 &  0 &  0 \\
0 &  0 &  0
\end{bmatrix}
\hat{\vec{V}}^{T}, \quad
[\vec{T}]_3 = \frac{1}{\sqrt{2}}\hat{\vec{U}}
\begin{bmatrix}
0 & -1 &  0 \\
1 &  0 &  0 \\
0 &  0 &  0
\end{bmatrix}
\hat{\vec{V}}^{T}.%
\end{align}%
\label{eq:fliptwisttensors3d}%
\end{subequations}%
It is important to note that, equations \eqref{eq:isPSD2d}-\eqref{eq:fliptwisttensors2d} and \eqref{eq:isPSD3d}-\eqref{eq:fliptwisttensors3d} describe the eigen-decomposition of the initial stiffness operator for \emph{any} isotropic hyperelastic energy. Given that, \eqref{eq:isPSD2d} and \eqref{eq:isPSD3d} are eigenvalues hence, their positiveness coupled with positive semi-definiteness of \eqref{eq:invarHw} describes the convexity condition of the strain energy. In other words, convexity can be postulated in terms of these newly found tangent eigenvalues concisely as \cite{Poya2023b}. \\ \ 

\begin{corollary}
A sufficiently smooth and twice differentiable two-dimensional energy function $e(\vec{F})$ is convex in $\vec{F}$ if and only if there exists a function $W(\lambda_1, \lambda_2)=e(\vec{F})$ that satisfies
\begin{align}
\dfrac{\Sigma_{\lambda_1} + \Sigma_{\lambda_2}}{\lambda_1 + \lambda_2} \geq 0, \qquad
\dfrac{\Sigma_{\lambda_1} - \Sigma_{\lambda_2}}{\lambda_1 - \lambda_2} \geq 0, \qquad
\partial^2_{ \lambda_1 \lambda_1}W \; \partial^2_{ \lambda_2 \lambda_2}W \geq \left(  \partial^2_{ \lambda_1 \lambda_2}W  \right)^2.
\label{eq:convcondps2d}
\end{align}
\end{corollary}
\begin{corollary}
A sufficiently smooth and twice differentiable three-dimensional energy function $e(\vec{F})$ is convex in $\vec{F}$ if and only if there exists a function $W(\lambda_1, \lambda_2, \lambda_3)=e(\vec{F})$ that satisfies
\begin{align}
\dfrac{\Sigma_{\lambda_1} + \Sigma_{\lambda_2}}{\lambda_1 + \lambda_2} \geq 0, \qquad
\dfrac{\Sigma_{\lambda_1} + \Sigma_{\lambda_3}}{\lambda_1 + \lambda_3} \geq 0, \qquad
\dfrac{\Sigma_{\lambda_2} + \Sigma_{\lambda_3}}{\lambda_2 + \lambda_3} \geq 0, \nonumber \\
\dfrac{\Sigma_{\lambda_1} - \Sigma_{\lambda_2}}{\lambda_1 - \lambda_2} \geq 0, \qquad
\dfrac{\Sigma_{\lambda_1} - \Sigma_{\lambda_3}}{\lambda_1 - \lambda_3} \geq 0, \qquad
\dfrac{\Sigma_{\lambda_2} - \Sigma_{\lambda_3}}{\lambda_2 - \lambda_3} \geq 0, \nonumber \\
\partial^2_{ \lambda_1 \lambda_1}W \; \partial^2_{ \lambda_2 \lambda_2}W \; \partial^2_{ \lambda_3 \lambda_3}W 
+ 2 \partial^2_{ \lambda_1 \lambda_2}W \; \partial^2_{ \lambda_1 \lambda_3}W \; \partial^2_{ \lambda_2 \lambda_3}W \geq \nonumber \\
\partial^2_{ \lambda_1 \lambda_1}W  \; \left(  \partial^2_{ \lambda_2 \lambda_3}W \right)^2 +
\partial^2_{ \lambda_2 \lambda_2}W  \; \left(  \partial^2_{ \lambda_1 \lambda_3}W \right)^2 +
\partial^2_{ \lambda_3 \lambda_3}W  \; \left(  \partial^2_{ \lambda_1 \lambda_2}W \right)^2.
\label{eq:convcondps3d}
\end{align}
\end{corollary}
The proof of above corollaries can be found in \cite{Poya2023b}. The above convexity conditions are often too restrictive to impose on an energy function. The elegance of this formulation however is that we have obtained the tangent eigenvalues \emph{and} their linkage to convexity essentially cost-free. Hence, alternatively we can proceed to work with the common (typically non-convex) strain energy functions and instead stabilise the resulting tangents. In other words, to guarantee positive semi-definiteness of $\vec{H}_W$, a tangent stabilisation procedure can be devised via performing an $L^2$ projection to PSD cone which essentially entails pruning tangent and initial stiffness eigenvalues which are not positive. This has traditionally been achieved numerically by performing eigen-decomposition on the much larger tangent elasticity tensor $\vec{\mathcal{C}}_k$ and clamping the negative eigenvalues to zeros \cite{Teran05,stoma12}. For the Hessian operator $\vec{H}_W$ this can be achieved as
\begin{align}
\vec{H}^{\texttt{PSD}}_W = \sum_{i=1}^d \text{max}(\bar{\lambda}^{\vec{H}_W}_i, 0) \; [\vec{e}]_i \otimes [\vec{e}]_i,
\label{eq:invarHwPSD}
\end{align}
where $\bar{\lambda}^{\vec{H}_W}_i$ are the eigenvalues and $[\vec{e}]_i$ are the eigenvectors of the Hessian $\vec{H}_W$. Fortunately, the eigenvalues of the constitutive part of the elasticity tensor $\vec{\mathcal{C}}_k$ can be found in closed-form solely through analytic eigen-decomposition of $\vec{H}_{W}$ since $\vec{\mathcal{C}}_k$ has a null space of 2 in 2-dimensions and 6 in 3-dimensions. Hence, once $\vec{H}^{\texttt{PSD}}_W$ has been obtained the positive semi-definite constitutive tangent elasticity tensor $\vec{\mathcal{C}}^{\texttt{PSD}}_k$ can be built in $d$-dimensions simply as
\begin{align}
\vec{\mathcal{C}}^{\texttt{PSD}}_k &=
\begin{bmatrix}
\dfrac{\partial \lambda_1}{\partial \vec{F}}, &
\cdots, &
\dfrac{\partial \lambda_d}{\partial \vec{F}}
\end{bmatrix}
[\vec{H}^{\texttt{PSD}}_W]
\begin{bmatrix}
\dfrac{\partial \lambda_1}{\partial \vec{F}} \\
\vdots \\
\dfrac{\partial \lambda_d}{\partial \vec{F}} 
\end{bmatrix}. 
\label{eq:ckPSD}
\end{align}
From the numerical implementation point of view, it is more convenient to build $\vec{\mathcal{C}}_k$ and $\vec{\mathcal{C}}^{\texttt{PSD}}_k$ in a spectrally decomposed form as follows
\begin{align}
\vec{\mathcal{C}}_k = \sum_{i=1}^d\sum_{j=1}^d \bar{\lambda}^{\vec{H}_W}_i [\vec{e}]_i^j [\vec{D}]_j \otimes  [\vec{e}]_i^j [\vec{D}]_j, \qquad
\vec{\mathcal{C}}^{\texttt{PSD}}_k = \sum_{i=1}^d\sum_{j=1}^d  \text{max}(\bar{\lambda}^{\vec{H}_W}_i, 0) [\vec{e}]_i^j [\vec{D}]_j \otimes  [\vec{e}]_i^j [\vec{D}]_j,
\label{eq:ckbuilding}
\end{align}
where $ [\vec{e}]_i^j$ implies the $j$th component of the eigenvector $ [\vec{e}]_i$. A similar PSD projection can be performed on the initial stiffness term. This is an easier case as the eigensystem of initial stiffness operator is already obtained in a decomposed form as presented in \eqref{eq:initstiff1}. Consequently, PSD projection is obtained via
\begin{align}
\vec{\mathcal{C}}_p^{\texttt{PSD}} = \sum_{i=1}^k \text{max}(\bar{\lambda}_i^{\vec{\mathcal{C}_p}}, 0) [\vec{L}]_i \otimes [\vec{L}]_i + \sum_{i=1}^{k} \text{max}(\bar{\lambda}_{i+k}^{\vec{\mathcal{C}_p}}, 0)  [\vec{T}]_i \otimes [\vec{T}]_i.
\label{eq:cpPSD}
\end{align}

\subsection{Large strain isotropic elasticity using Cauchy-Green strain}
In practice, perhaps the most widely used and commonly implemented formulation of large strain elasticity is based on Cauchy-Green strain \cite{mar83, Ogden-Book, Bonet-Book16, Wriggers-Book, peric_book}. In this section, we demonstrate that the technique presented in \citet{Poya2023b} for spectrally-decomposed tangent operator construction is generic and can be applied to virtually any family of invariants in particular, to strain energies expressed in terms of Cauchy-Green and its invariants. To this end, as customary in nonlinear mechanics literature, we first express the strain energy $e$ as
\begin{align}
e(\vec{\nabla}_0\vec{x}) = \breve{W}(\vec{C}) = \bar{W}(\lambda_1^2,\lambda_2^2,\lambda_3^2) =\bar{W}(\mathcal{A_C}),
\end{align}
where once again, as customary, the strain energy $e$ has been further expressed in terms of the set of squares of principal stretches $\mathcal{A_C}$. The Cauchy-Green strain tensor $\vec{C}$ can now be described in terms of principal information as
\begin{align}
\vec{C} = \sum_{i=1}^d  \lambda_i^2 [\vec{N}]_i \otimes [\vec{N}]_i = \hat{\vec{V}} \vec{\Lambda}_{\vec{C}} \hat{\vec{V}}^T,
\end{align}
where $\Lambda_{\vec{C}_{ii}} = \lambda_i^2$. For brevity and without going into the details, the first and second directional derivatives of the energy $\bar{W}$ needed in the context of second order minimisation methods such as Newton-Raphson can be computed as; c.f.\ \cite{Bonet-Book16, Wriggers-Book}
\begin{subequations}
\begin{align}
D\breve{W}[\delta \vec{u}] &= 2 \frac{\partial \breve{W}}{\partial \vec{C}} : \frac{1}{2} D\vec{C}[\delta \vec{u}] \\
D^2\breve{W}[\delta \vec{u}; \Delta \vec{u}] &= \frac{1}{2} D\vec{C}[\delta \vec{u}] : 4 \frac{\partial^2 \breve{W}}{\partial \vec{C}\partial \vec{C}} : \frac{1}{2} D\vec{C}[\Delta \vec{u}] + 2 \frac{\partial \breve{W}}{\partial \vec{C}} : \frac{1}{2} D^2\vec{C}[\delta \vec{u}; \Delta \vec{u}], 
\end{align}%
\label{eq:dderWbar}%
\end{subequations}%
and noting that, the directional derivatives of $\vec{C}$ are given by (see for instance, \cite{Bonet-Book16} page 118 or \cite{Wriggers-Book}, page 77), \eqref{eq:dderWbar} simplifies to
\begin{subequations}
\begin{align}
\label{eq:dderWbar2a}%
D\breve{W}[\delta \vec{u}] &= \vec{S} : \frac{1}{2} D\vec{C}[\delta \vec{u}] \\
\label{eq:dderWbar2b}%
D^2\breve{W}[\delta \vec{u}; \Delta \vec{u}] &= \frac{1}{2} D\vec{C}[\delta \vec{u}] : \breve{\vec{\mathcal{C}}}_k : \frac{1}{2} D\vec{C}[\Delta \vec{u}] + \vec{\nabla}_0 \delta \vec{u} : \breve{\vec{\mathcal{C}}}_p : \vec{\nabla}_0 \Delta \vec{u}  
\end{align}%
\label{eq:dderWbar2}%
\end{subequations}
leading to the evaluation of the second Piola-Kirchhoff stress tensor $\vec{S}$, the constitutive tangent $ \breve{\vec{\mathcal{C}}}_k$ and the initial stiffness $ \breve{\vec{\mathcal{C}}}_p$ operators
\begin{align}
\vec{S} = 2 \frac{\partial \breve{W}}{\partial \vec{C}}, \qquad 
\breve{\vec{\mathcal{C}}}_k = 4 \frac{\partial^2 \breve{W}}{\partial \vec{C}\partial \vec{C}}, \qquad
\breve{\vec{\mathcal{C}}}_p = \vec{I} \otimes \vec{S},
\label{eq:pk2}
\end{align}
Similar to \eqref{eq:lambC2}, a sufficient condition for convexity is then the positive semi-definiteness of \eqref{eq:dderWbar2b} via the constitutive $\breve{\vec{\mathcal{C}}}_k$ and initial $\breve{\vec{\mathcal{C}}}_p$ contributions.

Our goal now is to find the closed-form eigensystem of both $ \breve{\vec{\mathcal{C}}}_k$ and $ \breve{\vec{\mathcal{C}}}_p$. The recipe here is to resort once again to principal stretch-based approach (which we term as the $\vec{C}$-based formulation in principal stretches) and similar to before introduce a set of work conjugate stresses $\mathcal{S_C}=\{\bar{\Sigma}_{\lambda_1} ,\cdots,\bar{\Sigma}_{\lambda_d}\}$ first but this time corresponding to the squares of principal stretches such that
\begin{align}
\bar{\Sigma}_{\lambda_1} = 2\frac{\partial \bar{W}}{\partial  \lambda_1^2}, \quad
\bar{\Sigma}_{\lambda_2} = 2\frac{\partial \bar{W}}{\partial  \lambda_2^2}, \quad
\bar{\Sigma}_{\lambda_3} = 2\frac{\partial \bar{W}}{\partial  \lambda_3^2},
\end{align}
which can be used to define the second Piola-Kirchhoff stress tensor $\vec{S}$ similar to \eqref{eq:lambpk1}
\begin{align}
\vec{S} =  \sum_{i=1}^d  \bar{\Sigma}_{\lambda_i} \frac{\partial \lambda_i^2}{\partial \vec{C}} =  \sum_{i=1}^d  \bar{\Sigma}_{\lambda_i} [\vec{N}]_i \otimes [\vec{N}]_i = \hat{\vec{V}} \vec{\Lambda_S} \hat{\vec{V}}^T.
\label{eq:lambpk2}%
\end{align}%
Similar to \eqref{eq:lambdder1a}-\eqref{eq:lambdder1b}, the derivatives of squares of principal stretches with respect to $\vec{C}$ needed for consistent linearisation can be calculated by taking directional derivatives which in 2-dimensions leads to
\begin{subequations}
\begin{align}
[\bar{\vec{D}}]_1 = \frac{\partial \lambda_1^2}{\partial \vec{C}} &= [\vec{N}]_1 \otimes [\vec{N}]_1 = \hat{\vec{V}} 
\begin{bmatrix}
1 & 0 \\
0 & 0
\end{bmatrix} 
\hat{\vec{V}}^T, \\
[\bar{\vec{D}}]_2 = \frac{\partial \lambda_2^2}{\partial \vec{C}} &= [\vec{N}]_2 \otimes [\vec{N}]_2 =  \hat{\vec{V}}
\begin{bmatrix}
0 & 0 \\
0 & 1
\end{bmatrix}
\hat{\vec{V}}^T, 
\end{align}%
\label{eq:lamb2dder1aC}%
\end{subequations}%
and similarly in 3-dimensions
\begin{subequations}
\begin{align}
[\bar{\vec{D}}]_1 = \frac{\partial \lambda_1^2}{\partial \vec{C}} &= [\vec{N}]_1 \otimes [\vec{N}]_1 = \hat{\vec{V}} 
\begin{bmatrix}
1 & 0 & 0 \\
0 & 0 & 0 \\
0 & 0 & 0 
\end{bmatrix} 
\hat{\vec{V}}^T, \\
[\bar{\vec{D}}]_2 = \frac{\partial \lambda_2^2}{\partial \vec{C}} &= [\vec{N}]_2 \otimes [\vec{N}]_2 = \hat{\vec{V}} 
\begin{bmatrix}
0 & 0 & 0 \\
0 & 1 & 0 \\
0 & 0 & 0 
\end{bmatrix} 
\hat{\vec{V}}^T, \\
[\bar{\vec{D}}]_3 = \frac{\partial \lambda_3^2}{\partial \vec{C}} &= [\vec{N}]_3 \otimes [\vec{N}]_3 = \hat{\vec{V}} 
\begin{bmatrix}
0 & 0 & 0 \\
0 & 0 & 0 \\
0 & 0 & 1 
\end{bmatrix} 
\hat{\vec{V}}^T, 
\end{align}%
\label{eq:lamb2dder1bC}%
\end{subequations}%
where we have denoted $[\bar{\vec{D}}]_i = \frac{\partial \lambda_i^2}{\partial \vec{C}}$ for the the \emph{scaling} rank-one tensors to highlight their similarities with $[{\vec{D}}]_i$ found earlier. Hence, following a procedure similar to \citet{Poya2023b} for finding the eigensystem of $\frac{\partial^2 W}{\partial \vec{F}\partial \vec{F}}$, the constitutive tangent $\vec{\breve{\mathcal{C}}}_k$ operator emerges as an algebraic summation of two further terms 
\begin{align}
\breve{\vec{\mathcal{C}}}_k = \frac{\partial^2 \bar{W}}{\partial \vec{C} \partial \vec{C}} &= 4
\begin{bmatrix}
\dfrac{\partial \lambda_1^2}{\partial \vec{C}}, &
\cdots, &
\dfrac{\partial \lambda_d^2}{\partial \vec{C}}
\end{bmatrix}
[\vec{H}_{\bar{W}}]
\begin{bmatrix}
\dfrac{\partial \lambda_1^2}{\partial \vec{C}} \\
\vdots \\
\dfrac{\partial \lambda_d^2}{\partial \vec{C}} 
\end{bmatrix}%
+ \sum_{i=1}^d \bar{\Sigma}_{\lambda_i} \frac{\partial^2 \lambda_i^2}{\partial \vec{C} \partial \vec{C}} \nonumber \\
&= 
\sum_{i=1}^d  \sum_{j=1}^d 4\frac{\partial^2 \bar{W}}{\partial \lambda_i^2 \lambda_j^2}  \frac{\partial \lambda_i^2}{\partial \vec{C}} \otimes  \frac{\partial \lambda_j^2}{\partial \vec{C}}%
+
 \sum_{i=1}^d \bar{\Sigma}_{\lambda_i} \frac{\partial^2 \lambda_i^2}{\partial \vec{C} \partial \vec{C}}.%
\label{eq:lambCC2}%
\end{align} 
where $\vec{H}_{\bar{W}}$ denotes the symmetric but not necessarily PSD $d \times d$ Hessian operator containing the second derivatives of $\bar{W}$ with respect to squares of principal stretches
\begin{align}
\vec{H}_{\bar{W}} =
\begin{bmatrix}
\partial^2_{ \lambda_1^2 \lambda_1^2}\bar{W}  & \cdots & \partial^2_{ \lambda_1^2 \lambda_d^2}\bar{W}  \\
&  \ddots & \vdots \\
sym &  & \partial^2_{ \lambda_d^2 \lambda_d^2}\bar{W}
\end{bmatrix}.
\label{eq:invarHwC}
\end{align}
The similarity between \eqref{eq:lambC2} and \eqref{eq:lambCC2} and \eqref{eq:invarHw} and \eqref{eq:invarHwC} is apparent. In fact, the structure of \eqref{eq:lambCC2} reveals a further decomposition of $\vec{\breve{\mathcal{C}}}_k$ into a constitutive and an initial/geometric-type part when expressed in terms of principal stretches. Unlike the $\vec{F}$-based formulation of \eqref{eq:lambC2}, this $\vec{C}$-based formulation mixes up first and second order derivatives. This can be demonstrated by looking at the explicit structure of $\vec{H}_{\bar{W}}$ in \eqref{eq:invarHwC} which can be expanded in terms of derivatives of principal stretches directly (rather than their squares)
\begin{align}
\vec{H}_{\bar{W}} =
\begin{bmatrix}
\dfrac{1}{\lambda_1^2} \partial^2_{ \lambda_1 \lambda_1}\bar{W} - \dfrac{1}{2\lambda_1^3} \bar{\Sigma}_{\lambda_1} & \cdots & \dfrac{1}{\lambda_1 \lambda_d} \partial^2_{ \lambda_1 \lambda_d}\bar{W} \\
&  \ddots & \vdots \\
sym &  & \dfrac{1}{\lambda_d^2} \partial^2_{ \lambda_d \lambda_d}\bar{W} - \dfrac{1}{2\lambda_d^3} \bar{\Sigma}_{\lambda_d}
\end{bmatrix}.
\label{eq:invarHwC2}
\end{align}
Following the same procedure as \citet{Poya2023b} after some tedious but straight-forward algebra we can show that the analytic eigensytem of $\vec{C}$-based formulations has a strikingly similar structure to $\vec{F}$-based formulation with minor differences. The first $d$ eigenvalues and eigenmatrices are found from the matrix $\vec{H}_{\bar{W}}$ in \eqref{eq:invarHwC} (or the first term in \eqref{eq:lambCC2}). The next few constitutive tangent eigenvalues come from the second term in \eqref{eq:lambCC2}. This term has a null space of 3 in dimensions (only 1 eigenvalue) and a null space of 6 in 3-dimensions (only 3 eigenvalues). Its only eigenvalue in 2-dimensions can be written concisely as
\begin{align}
\bar{\lambda}_3^{\vec{\breve{\mathcal{C}}}_k} = \frac{\bar{\Sigma}_{\lambda_1} - \bar{\Sigma}_{\lambda_2}}{\lambda_1^2 - \lambda_2^2},
\label{eq:isPSD2dC}
\end{align}
and similarly in 3-dimensions the 3 eigenvalues of the initial stiffness are found concisely as
\begin{align}
\bar{\lambda}_4^{\vec{\breve{\mathcal{C}}}_k} &= \frac{\bar{\Sigma}_{\lambda_2} - \bar{\Sigma}_{\lambda_3}}{\lambda_2^2 - \lambda_3^2}, \qquad
\bar{\lambda}_5^{\vec{\breve{\mathcal{C}}}_k} = \frac{\bar{\Sigma}_{\lambda_1} - \bar{\Sigma}_{\lambda_3}}{\lambda_1^2 - \lambda_3^2},
\qquad
\bar{\lambda}_6^{\vec{\breve{\mathcal{C}}}_k} = \frac{\bar{\Sigma}_{\lambda_1} - \bar{\Sigma}_{\lambda_2}}{\lambda_1^2 - \lambda_2^2}.
\label{eq:isPSD3dC}
\end{align}
The only eigenmatrix in 2-dimensions take the form
\begin{subequations}
\begin{align}
[\bar{\vec{L}}]_1 &= \hat{\vec{V}}
\begin{bmatrix}
0 & 1 \\
1 & 0
\end{bmatrix}
\hat{\vec{V}}^{T},
\end{align}%
\label{eq:fliptwisttensors2dC}%
\end{subequations}
and similarly in 3-dimensions the 3 eigenmatrices take the form
\begin{subequations}
\begin{align}
[\vec{\bar{L}}]_1 = \hat{\vec{V}}
\begin{bmatrix}
0 & 0 &  0 \\
0 & 0 &  1 \\
0 & 1 &  0
\end{bmatrix}
\hat{\vec{V}}^{T}, \quad
[\vec{\bar{L}}]_2 = \hat{\vec{V}}
\begin{bmatrix}
0 & 0 &  1 \\
0 & 0 &  0 \\
1 & 0 &  0
\end{bmatrix}
\hat{\vec{V}}^{T},  \quad
[\vec{\bar{L}}]_3 = \hat{\vec{V}}
\begin{bmatrix}
0 &  1 &  0 \\
1 &  0 &  0 \\
0 &  0 &  0
\end{bmatrix}
\hat{\vec{V}}^{T}.
\end{align}%
\label{eq:fliptwisttensors3dC}%
\end{subequations}%
which are essentially the $\vec{C}$-based \emph{flip} tensors similar to \eqref{eq:isPSD3d}. Interestingly due to material frame invariance of $\vec{C}$-based formulation the corresponding $\vec{C}$-based \emph{twist} tensors do not exist. Finally, the last 3 eigenvalues come from the initial stiffness operator $\vec{\breve{\mathcal{C}}}_p$. Noting its structure in \eqref{eq:pk2} (that is the appearance of the identity tensor $I$), its eigensystem is trivial and in fact already available. The initial stiffness operator $\vec{\breve{\mathcal{C}}}_p$ has 3 eigenvalues with a multiplicity of 3 and they are indeed the principal conjugate stresses $\bar{\Sigma}_{\lambda_i}$s i.e.
\begin{align}
\bar{\lambda}_i^{\vec{\breve{\mathcal{C}}}_p} &= \bar{\Sigma}_{\lambda_i} \quad i=1, \cdots, d
\label{eq:isPSD3dC2}
\end{align}
and the corresponding eigenmatrices are simply $\hat{\vec{V}}$. Consequently, the constitutive tangent elasticity tensor $\vec{\breve{\mathcal{C}}}_k$ can be constructed in a spectrally decomposed form as
\begin{align}
\breve{\vec{\mathcal{C}}}_k = \frac{\partial^2 \bar{W}}{\partial \vec{C} \partial \vec{C}} &= 4
\begin{bmatrix}
\dfrac{\partial \lambda_1^2}{\partial \vec{C}}, &
\cdots, &
\dfrac{\partial \lambda_d^2}{\partial \vec{C}}
\end{bmatrix}
[\vec{H}_{\bar{W}}]
\begin{bmatrix}
\dfrac{\partial \lambda_1^2}{\partial \vec{C}} \\
\vdots \\
\dfrac{\partial \lambda_d^2}{\partial \vec{C}}
\end{bmatrix}%
+
 \sum_{i=1}^k \bar{\lambda}_{i+d}^{\vec{\breve{\mathcal{C}}}_k} [\bar{\vec{L}}]_i \otimes [\bar{\vec{L}}]_i.
 \label{eq:cpC}
\end{align}
To guarantee positive semi-definiteness of $\vec{H}_{\bar{W}}$ a similar tangent stabilisation procedure described before can be performed via an $L^2$ projection to PSD cone. For the Hessian operator this can be achieved as
\begin{align}
\vec{H}^{\texttt{PSD}}_{\bar{W}} = \sum_{i=1}^d \text{max}(\bar{\lambda}^{\vec{H}_{\bar{W}}}_i, 0) \; [\vec{\bar{e}}]_i \otimes [\vec{\bar{e}}]_i,
\label{eq:invarHwCPSD}
\end{align}
where $\bar{\lambda}^{\vec{H}_{\bar{W}}}_i$ are the eigenvalues and $[\vec{\bar{e}}]_i$ are the eigenvectors of the Hessian $\vec{H}_{\bar{W}}$. Similar to $\vec{F}$-based formulation, the eigenvalues of the constitutive part of the elasticity tensor $\breve{\vec{\mathcal{C}}}_k$ can be found in closed-form solely through analytic eigen-decomposition of $\vec{H}_{\bar{W}}$ can be built in $d$-dimensions simply as
\begin{align}
\breve{\vec{\mathcal{C}}}_k^{\texttt{PSD}} = 4\left( \frac{\partial^2 \bar{W}}{\partial \vec{C} \partial \vec{C}} \right)^{\texttt{PSD}} &= 4
\begin{bmatrix}
\dfrac{\partial \lambda_1^2}{\partial \vec{C}}, &
\cdots, &
\dfrac{\partial \lambda_d^2}{\partial \vec{C}}
\end{bmatrix}
[\vec{H}_{\bar{W}}^{\texttt{PSD}}]
\begin{bmatrix}
\dfrac{\partial \lambda_1^2}{\partial \vec{C}} \\
\vdots \\
\dfrac{\partial \lambda_d^2}{\partial \vec{C}}
\end{bmatrix}%
+
 \sum_{i=1}^k \text{max}(\bar{\lambda}_{i+d}^{\vec{\bar{\mathcal{C}}}_k}, 0) [\bar{\vec{L}}]_i \otimes [\bar{\vec{L}}]_i.
\label{eq:ckCPSD}
\end{align}
Alternatively, for numerical implementation a more convenient format of building $\breve{\vec{\mathcal{C}}}_k^{\texttt{PSD}} $ is
\begin{align}
\breve{\vec{\mathcal{C}}}_k^{\texttt{PSD}}  = 4 \sum_{i=1}^d\sum_{j=1}^d  \text{max}(\bar{\lambda}^{\vec{H}_{\bar{W}} }_i, 0) [\bar{\vec{e}}]_i^j [\bar{\vec{D}}]_j \otimes  [\bar{\vec{e}}]_i^j [\bar{\vec{D}}]_j
+ \sum_{i=1}^k \text{max}(\bar{\lambda}_{i+d}^{\vec{\bar{\mathcal{C}}}_k}, 0) [\bar{\vec{L}}]_i \otimes [\bar{\vec{L}}]_i.
\label{eq:ckbuilding}
\end{align}
A similar PSD projection can be performed on the initial stiffness term. This is a straight-forward case. Consequently, PSD projection is obtained via
\begin{align}
\breve{\vec{\mathcal{C}}}_p^{\texttt{PSD}} = \vec{I} \otimes \vec{S}^{\texttt{PSD}} = \vec{I} \otimes ( \hat{\vec{V}} \vec{\Lambda_S}^{\texttt{PSD}}  \hat{\vec{V}}^T).
\label{eq:cpCPSD}
\end{align}
From \eqref{eq:ckCPSD}-\eqref{eq:cpCPSD} it is evident that the $\vec{C}$-based formulation in total comprises of 5 eigenvalues in 2-dimensions and 9 eigenvalues in 3-dimensions (as opposed to 4 and 9 for $\vec{F}$-based formulation).
Given that, \eqref{eq:isPSD2dC} and \eqref{eq:isPSD3dC} are eigenvalues hence their positiveness coupled with positive semi-definiteness of \eqref{eq:invarHw} and \eqref{eq:cpCPSD} describes the convexity condition of the strain energy. In other words, convexity can be postulated in terms of these newly found eigenvalues as \cite{Poya2023b}. \\ \

\begin{corollary}
For a smooth and twice differentiable two-dimensional energy function $\breve{W}(\vec{C})$ to be convex in $\vec{C}$, it is sufficient that there exists a function $\bar{W}(\lambda_1^2, \lambda_2^2)=\breve{W}(\vec{C})$ that satisfies
\begin{align}
\bar{\Sigma}_{\lambda_1} \geq 0, \qquad
\bar{\Sigma}_{\lambda_2} \geq 0, \qquad
\dfrac{\bar{\Sigma}_{\lambda_1} - \bar{\Sigma}_{\lambda_2}}{\lambda_1^2 - \lambda_2^2} \geq 0, \qquad
\partial^2_{ \lambda_1^2 \lambda_1^2}\bar{W} \; \partial^2_{ \lambda_2^2 \lambda_2^2}{\bar{W}} \geq \left(  \partial^2_{ \lambda_1^2 \lambda_2^2}{\bar{W}}  \right)^2.
\label{eq:convcondps2dC}
\end{align}
\end{corollary}
\begin{corollary}
For a smooth and twice differentiable three-dimensional energy function $\breve{W}(\vec{C})$ to be convex in $\vec{C}$, it is sufficient that there exists a function $\bar{W}(\lambda_1^2, \lambda_2^2, \lambda_3^2)=\breve{W}(\vec{C})$ that satisfies
\begin{align}
\bar{\Sigma}_{\lambda_1} \geq 0, \qquad
\bar{\Sigma}_{\lambda_2} \geq 0, \qquad
\bar{\Sigma}_{\lambda_3} \geq 0, \nonumber \\
\dfrac{\bar{\Sigma}_{\lambda_1} - \bar{\Sigma}_{\lambda_2}}{\lambda_1^2 - \lambda_2^2} \geq 0, \qquad
\dfrac{\bar{\Sigma}_{\lambda_1} - \bar{\Sigma}_{\lambda_3}}{\lambda_1^2 - \lambda_3^2} \geq 0, \qquad
\dfrac{\bar{\Sigma}_{\lambda_2} - \bar{\Sigma}_{\lambda_3}}{\lambda_2^2 - \lambda_3^2} \geq 0, \nonumber \\
\partial^2_{ \lambda_1^2 \lambda_1^2}\bar{W} \; \partial^2_{ \lambda_2^2 \lambda_2^2}\bar{W} \; \partial^2_{ \lambda_3^2 \lambda_3^2}\bar{W} + 2 \partial^2_{ \lambda_1^2 \lambda_2^2}\bar{W} \; \partial^2_{ \lambda_1^2 \lambda_3^2}\bar{W} \; \partial^2_{ \lambda_2^2 \lambda_3^2}\bar{W} \geq \nonumber \\
\partial^2_{ \lambda_1^2 \lambda_1^2}\bar{W}  \; \left(  \partial^2_{ \lambda_2^2 \lambda_3^2}\bar{W} \right)^2 +
\partial^2_{ \lambda_2^2 \lambda_2^2}\bar{W}  \; \left(  \partial^2_{ \lambda_1^2 \lambda_3^2}\bar{W} \right)^2 +
\partial^2_{ \lambda_3^2 \lambda_3^2}\bar{W}  \; \left(  \partial^2_{ \lambda_1^2 \lambda_2^2}\bar{W} \right)^2.
\label{eq:convcondps3dC}
\end{align}
\end{corollary}
\begin{remark}\label{rem:cinvar}
It is often convenient and rather customary to express the strain energy in terms of invariants of Cauchy-Green tensor $\vec{C}$ such as
\begin{align}
e(\vec{\nabla}_0 \vec{x}) = \psi(I_C, II_C, III_C)
\end{align}
where the set of invariants is defined as
\begin{align}
I_C = \mathrm{tr}(\vec{C}) = \sum_{i=1}^d \lambda_i^2, \quad II_C = \vec{C}:\vec{C} = \sum_{i=1}^d \lambda_i^4, \quad III_C = C = \mathrm{det}(\vec{C}) = \prod_{i=1}^d \lambda_i^2,
\end{align}
where $C$ is determinant of $\vec{C}$. Furthermore, we can introduce $\vec{G} = C\vec{C}^{-1} = \frac{1}{2} \vec{C} \Cross \vec{C}$ to represent the cofactor of $\vec{C}$. Finding analytical eigensystems emanating from this integrity basis is now straight-forward using the algebra developed so far. Specifically noting that, consistent linearisation of $\psi$ yields
\begin{subequations}
\begin{align}
\frac{\partial \psi}{\partial \vec{C}} &= \frac{\partial \psi}{\partial I_C}\vec{I} + 2 \frac{\partial \psi}{\partial II_C}\vec{C} + \frac{\partial \psi}{\partial III_C} \vec{G}, \\
\frac{\partial^2 \psi}{\partial \vec{C} \partial \vec{C}} &=
\underbrace{%
\begin{bmatrix}
\vec{I}, & 2\vec{C}, & \vec{G}
\end{bmatrix}
[\vec{H}_{W}^{\psi}]
\begin{bmatrix}
\vec{I} \\ 2\vec{C} \\ \vec{G}
\end{bmatrix}%
}_{\vec{\mathcal{C}}_k^{\psi}}
+ \underbrace{%
\left( 2 \frac{\partial \psi}{\partial II_C} \mathbb{I} +  \frac{\partial \psi}{\partial III_C} \left( \mathbb{I} \Cross \vec{C} \right)  \right)
}_{\vec{\mathcal{C}}_p^{\psi}},
\label{eq:lambCC3}
\end{align}%
\end{subequations}
where $\vec{H}_{W}^{\psi}$ represents the symmetric (but not necessarily positive definite) $d\times d$ Hessian operator
\begin{align}
\vec{H}_{W}^{\psi} =
\begin{bmatrix}
\dfrac{\partial^2 \psi}{\partial I_C\partial I_C} & \dfrac{\partial^2 \psi}{\partial I_C\partial II_C} & \dfrac{\partial^2 \psi}{\partial I_C\partial III_C} \\
 & \dfrac{\partial^2 \psi}{\partial II_C\partial II_C} & \dfrac{\partial^2 \psi}{\partial II_C\partial III_C} \\
sym  & & \dfrac{\partial^2 \psi}{\partial III_C\partial III_C}
\end{bmatrix}.
\end{align}%
Similar to \eqref{eq:lambC2}, a sufficient condition for convexity is then positive semi-definiteness of \eqref{eq:lambCC3} via the constitutive ${\vec{\mathcal{C}}}_k^{\psi}$ and initial ${\vec{\mathcal{C}}}_p^{\psi}$ contributions. The constitutive part once again, has a $d\times d$ Hessian whose eigen-decomposition can be found analytically similar to \eqref{eq:invarHw} and \eqref{eq:invarHwC} using the formula presented in \cite{Poya2023b}. The eigensystem of the initial stiffness part can be found by the algebra developed earlier for $\vec{C}$-based formulation in particular, the term $\mathbb{I} \Cross \vec{C}$ represents the Hessian of $C=J^2$ which can be easily obtained using our earlier arrangements. Later in \S  \ref{sec:polyconvex}, we undertake determining the eigensystem of the initial stiffness term emerging from polyconvex energies which has the exact same structure. Ultimately, this formulation creates another level of ``indirection" but its eigensystem can always be worked out using principal information alone. This in general holds true for all isotropic invariant formulations.
\end{remark}

\subsection{Updated Lagrangian formulation for large strain isotropic elasticity}
In the engineering simulation community often stress and strain measures are used that are physically more meaningful such as the Cauchy ($\vec{\sigma}$) or the Kirchhoff ($\vec{\tau}$) stress and engineering ($\vec{\varepsilon}$) or alternatively the corresponding work-conjugate Almansi or left Cauchy-Green ($\vec{b}=\vec{FF}^T$) strain. We will henceforth call this formulation the $\vec{b}$-based formulation and introduce $b=\text{det}(\vec{b})$ and $\vec{g} = b\vec{b}^{-T} = \frac{1}{2}\vec{b}\Cross \vec{b}$. The above mentioned stresses can be obtained using the standard push-forward relationships \cite{Bonet-Book16, Ogden-Book, Wriggers-Book, peric_book}
\begin{align}
\vec{\tau} = J \vec{\sigma} = \vec{FSF}^T = \vec{PF}^T.
\end{align}
Note that, the structure of $\vec{FSF}^T$ reveals a crucial identity. That is, if the second Piola-Kirchhoff stress tensor $\vec{S}$ is PSD then from elementary linear algebra \cite{golub1996} it follows that both $\vec{\tau}$, and $\vec{\sigma}$ are PSD (subject to positiveness of $J$ which is always assumed). This is important as the eigenvalues of the initial stiffness operator $\breve{\vec{\mathcal{C}}}_p$ in the $\vec{C}$-based formulation presented in the previous section in \eqref{eq:isPSD3dC2} were precisely the singular-values of $\vec{S}$. This implies that, we only need to perform PSD projection in the material or Lagrangian configuration and hence, reveals that push-forward operations involving $\vec{C}$-based formulation retain convexity as well as any stabilisation applied. 
In particular, noting the standard push-forward operation between material $ \breve{\vec{\mathcal{C}}}_k$ and spatial $ \dbar{\vec{\mathcal{C}}}_k$ elasticity tensors \cite{Bonet-Book16, Wriggers-Book}
\begin{align}
\bar{\bar{\vec{\mathcal{C}}}}_k = [\vec{F}]_{iI} [\vec{F}]_{jJ} [\vec{F}]_{pP} [\vec{F}]_{qQ} \breve{\vec{\mathcal{C}}}_k,
\end{align}
the updated Lagrangian versions of the \emph{scaling} \eqref{eq:lamb2dder1aC}-\eqref{eq:lamb2dder1bC} and \emph{flip} \eqref{eq:fliptwisttensors2dC}-\eqref{eq:fliptwisttensors3dC} eigenmatrices are obtained as follows in 2-dimensions
\begin{subequations}
\begin{align}
[\dbar{\vec{D}}]_1 &= \hat{\vec{U}}
\begin{bmatrix}
1 & 0 \\
0 & 0
\end{bmatrix}
\hat{\vec{U}}^{T}, \quad
[\dbar{\vec{D}}]_2 = \hat{\vec{U}}
\begin{bmatrix}
0 & 0 \\
0 & 1
\end{bmatrix}
\hat{\vec{U}}^{T}, \quad
[\dbar{\vec{L}}]_1 = \hat{\vec{U}}
\begin{bmatrix}
0 & 1 \\
1 & 0
\end{bmatrix}
\hat{\vec{U}}^{T},
\end{align}%
\label{eq:fliptwisttensors2dsigma}%
\end{subequations}
and similarly in 3-dimensions the 3 eigenmatrices take the form
\begin{subequations}
\begin{align}
[\vec{\dbar{D}}]_1 &= \hat{\vec{U}}
\begin{bmatrix}
1 & 0 &  0 \\
0 & 0 &  0 \\
0 & 0 &  0
\end{bmatrix}
\hat{\vec{U}}^{T}, \quad
[\vec{\dbar{D}}]_2 = \hat{\vec{U}}
\begin{bmatrix}
0 & 0 &  0 \\
0 & 1 &  0 \\
0 & 0 &  0
\end{bmatrix}
\hat{\vec{U}}^{T}, \quad
[\vec{\dbar{D}}]_3 = \hat{\vec{U}}
\begin{bmatrix}
0 & 0 &  0 \\
0 & 0 &  0 \\
0 & 0 &  1
\end{bmatrix}
\hat{\vec{U}}^{T}, \\
[\vec{\dbar{L}}]_1 &= \hat{\vec{U}}
\begin{bmatrix}
0 & 0 &  0 \\
0 & 0 &  1 \\
0 & 1 &  0
\end{bmatrix}
\hat{\vec{U}}^{T}, \quad
[\vec{\dbar{L}}]_2 = \hat{\vec{U}}
\begin{bmatrix}
0 & 0 &  1 \\
0 & 0 &  0 \\
1 & 0 &  0
\end{bmatrix}
\hat{\vec{U}}^{T}, \quad
[\vec{\dbar{L}}]_3 = \hat{\vec{U}}
\begin{bmatrix}
0 &  1 &  0 \\
1 &  0 &  0 \\
0 &  0 &  0
\end{bmatrix}
\hat{\vec{U}}^{T}.
\end{align}%
\label{eq:fliptwisttensors3dsigma}%
\end{subequations}%
Furthermore, the spatial Hessian $\vec{H}_{\dbar{W}}$ is related to the material Hessian $\vec{H}_{\bar{W}}$ in \eqref{eq:invarHwC} via the following push-forward operation
\begin{align}
\vec{H}_{\bar{\bar{W}}} = \vec{\Lambda}^2 \vec{H}_{\bar{W}} \vec{\Lambda}^2,
\label{eq:pushfd_Hw}
\end{align}
where $ \vec{\Lambda}^2$ is the diagonal tensor containing the squares of principal stretches $\Lambda^2_{ii} = \lambda_i^2$. The eigenvalues of the spatial constitutive operator corresponding to \emph{flip} modes are related to their material counterparts in \eqref{eq:ckCPSD} via
\begin{subequations}
\begin{align}
\bar{\lambda}_3^{\vec{\dbar{\mathcal{C}}}_k} = (\lambda_1\lambda_2)^2 \bar{\lambda}_3^{\vec{\breve{\mathcal{C}}}_k}, \quad \text{in\; 2D} \\
\bar{\lambda}_4^{\vec{\dbar{\mathcal{C}}}_k} = (\lambda_2\lambda_3)^2 \bar{\lambda}_4^{\vec{\breve{\mathcal{C}}}_k}, \quad
\bar{\lambda}_5^{\vec{\dbar{\mathcal{C}}}_k} = (\lambda_1\lambda_3)^2 \bar{\lambda}_5^{\vec{\breve{\mathcal{C}}}_k}, \quad
\bar{\lambda}_6^{\vec{\dbar{\mathcal{C}}}_k} = (\lambda_1\lambda_2)^2 \bar{\lambda}_6^{\vec{\breve{\mathcal{C}}}_k}, \quad \text{in\; 3D}
\end{align}%
\label{eq:pushfd_flips}%
\end{subequations}
Similarly, the corresponding eigenvalues for spatial initial $ \dbar{\vec{\mathcal{C}}}_p$ stiffness operator is related to the eigenvaules of its material counterpart $ \breve{\vec{\mathcal{C}}}_p$ in \eqref{eq:pk2} in 2-dimensions via
\begin{subequations}
\begin{align}
\bar{\lambda}_4^{\vec{\dbar{\mathcal{C}}}_p} =  \lambda_1^2 \bar{\lambda}_4^{\vec{\breve{\mathcal{C}}}_p}, \quad
\bar{\lambda}_5^{\vec{\dbar{\mathcal{C}}}_p} =  \lambda_2^2 \bar{\lambda}_5^{\vec{\breve{\mathcal{C}}}_p},  \quad \text{in\; 2D} \\
\bar{\lambda}_7^{\vec{\dbar{\mathcal{C}}}_p} =  \lambda_1^2 \bar{\lambda}_7^{\vec{\breve{\mathcal{C}}}_p}, \quad
\bar{\lambda}_8^{\vec{\dbar{\mathcal{C}}}_p} =  \lambda_2^2 \bar{\lambda}_8^{\vec{\breve{\mathcal{C}}}_p}, \quad
\bar{\lambda}_9^{\vec{\dbar{\mathcal{C}}}_p} =  \lambda_3^2 \bar{\lambda}_9^{\vec{\breve{\mathcal{C}}}_p},  \quad \text{in\; 3D}.
\end{align}%
\label{eq:pushfd_IS}%
\end{subequations}
Finally, the convexity of updated Lagrangian formulation for $\vec{b}$-based invariants can be postulated in terms of the above eigenvalues exactly in a similar fashion to \eqref{eq:convcondps2dC}-\eqref{eq:convcondps3dC} in that, as mentioned before, in this setting, convexity does not change with push-forward operations.
\begin{remark}
One difference of the current $\vec{C}$-based (and the updated Lagrangian) formulation from classic formulations of elasticity is that, the resulting tangent operators are $d^2\times d^2$ (assuming that, we flatten fourth order tensors column-wise) instead of their usual Voigt notations ($3\times3$ in 2-dimensions and $6\times6$ in 3-dimensions).\ However, Voigt matrices can easily be retrieved once PSD projection has been performed using the standard Voigt transformations; c.f. \citet{Bonet-Book16} (page 245).\ From computer implementation standpoint, many performance-oriented optimisations are possible in order to directly create stabilised Voigt matrices; see \ref{app:voigt}.
\end{remark}

\subsection{Polyconvex large strain elasticity} \label{sec:polyconvex}
An alternative formulation for large strain elasticity that has surfaced in recent years is the polyconvex formulations which is based on the approximation of the minors of the deformation gradient or the fundamental kinematic measures namely $\{\vec{F}, \vec{H}, J\}$, mainly attributed to the work of \citet{SchroderPolyconvex11} and \citet{BonetPolyconvex15}. These formulations have a compelling property in that, they are by-design (\emph{ab-initio}) materially stable i.e. their corresponding constitutive tangent operators are guaranteed to be PSD. Moreover, polyconvexity ensures the well-posedness of the boundary-value problem in the large strain regime \cite{BallPolyconvex1976, BallPolyconvex1976b, BonetPolyconvex15, BonetPolyconvex15b, SchroderPolyconvex11, Schroder20083486, BonetPolyconvex15c}. However, polyconvex models do not preclude geometric instabilities implying that the initial stiffness operators obtained using these formulations can still be indefinite. In this section, we extend the tangent stabilisation of \cite{Poya2023b} to the case of polyconvex formulations wherein once again closed-form eigensystems can be obtained for the initial stiffness term emerging from these formulations. Hence, giving rise to the notion of geometrically stable large strain polyconvex elasticity.

Following \citet{BonetPolyconvex15}, the internal energy density $e$ can be written in terms of the fundamental kinematics as
\begin{align}\label{eqn:polyconvexity_FHJ}
e(\vec{\nabla}_0\vec{x}) = \tilde{W}(\vec{F},\vec{H},J),
\end{align}
where $\tilde{W}$ represents a polyconvex energy function in terms of the extended set of arguments $\mathcal{A_{FHJ}} =\{\vec{F},\vec{H},J\}$.  Similar to our earlier development, it is once again, convenient to introduce a set of work-conjugate stresses  $\mathcal{S_{FHJ}} =\{\vec{\Sigma_F}, \vec{\Sigma_H}, \Sigma_J \}$
\begin{align}
\vec{\Sigma_F} = \frac{\partial \tilde{W}}{\partial  \vec{F}}, \quad
\vec{\Sigma_H} = \frac{\partial \tilde{W}}{\partial  \vec{H}}, \quad
\Sigma_{J} = \frac{\partial \tilde{W}}{\partial  J},
\end{align}
which can now be used to obtain the more standard first Piola-Kirchhoff stress tensor $\vec{P}$ \cite{BonetPolyconvex15}
\begin{align}\label{eqn:FHJ_energy_lin_1_1}
\vec{P} = \vec{\Sigma_F}+\vec{\Sigma_H}\Cross \vec{F}+ \Sigma_J\vec{H}.
\end{align}
In the context of second order optimisation methods, such as Newton-Raphson-like schemes, the tangent elasticity operator $\vec{\hat{\mathcal{C}}}$ is also often required which can be obtained through similar directional derivatives as an algebraic summation of constitutive $\vec{\hat{\mathcal{C}}}_k$ and initial stiffness $\vec{\hat{\mathcal{C}}}_p$ components \cite{BonetPolyconvex15}
\begin{align}
\vec{\tilde{\mathcal{C}}} = \vec{\tilde{\mathcal{C}}}_k + \vec{\tilde{\mathcal{C}}}_p  &=
	\underbrace{\begin{bmatrix}
			\mathbb{I}, &
			\mathbb{I}\Cross\vec{F}, &
			\vec{H}
		\end{bmatrix}
		[\mathbb{H}_{\tilde{W}}]
		\begin{bmatrix}
			\mathbb{I}\\
			\mathbb{I}\Cross\vec{F}\\
		\vec{H}
	\end{bmatrix} }_{\hat{\vec{\mathcal{C}}}_k}
	+ \underbrace{\mathbb{I}\Cross\Big( \vec{\Sigma_H}  + \Sigma_J \vec{F}\Big)}_{\hat{\vec{\mathcal{C}}}_p}  \nonumber \\
&= \tilde{W}_{\vect{FF}}+\vect{F}\Cross\left(\tilde{W}_{\vect{HH}}\Cross\vect{F}\right)+
\tilde{W}_{JJ}\vect{H}\otimes\vect{H} + 2(\tilde{W}_{\vect{FH}}\Cross\vect{F})^{\text{sym}} \nonumber \\
&+2(\tilde{W}_{\vect{F}J}\otimes\vect{H})^{\text{sym}} + 2\left(\left(\vect{F}\Cross \tilde{W}_{\vect{H}J}\right)\otimes\vect{H}\right)^{\text{sym}} +   \vec{\tilde{\mathcal{C}}}_p, %
\label{eq:lambCFHJ}
\end{align}
where $\vec{\mathbb{H}}_{\hat{W}}$ represents the symmetric positive definite Hessian operator given by
\begin{align}
[\vec{\mathbb{H}}_{\tilde{W}}] &=
\begin{bmatrix}
\partial^2_{\vec{FF}}\tilde{W} & \partial^2_{\vec{FH}}\tilde{W} &\partial^2_{\vec{F}J}\tilde{W} \\
\partial^2_{\vec{HF}}\tilde{W} & \partial^2_{\vec{HH}}\tilde{W} &\partial^2_{\vec{H}J}\tilde{W} \\
\partial^2_{J\vec{F}}\tilde{W} & \partial^2_{J\vec{H}}\tilde{W} &\partial^2_{JJ}\tilde{W} \\
\end{bmatrix},
\label{eq:invarHwFHJ}
\end{align}
where $\partial^2_{\vec{AB}}\tilde{W} = \frac{\partial^2\,\tilde{W}}{\partial\vec{A}\partial\vec{B}}$, where $\vec{A}$ and $\vec{B}$ can represent any two elements from the set $\mathcal{A_{FHJ}}$. Similar to \eqref{eq:lambC2} and \eqref{eq:lambCC2}, a sufficient condition for convexity is then positive semi-definiteness of \eqref{eq:lambCFHJ} via the constitutive $\tilde{\vec{\mathcal{C}}}_k$ (via the Hessian $\mathbb{H}_{\tilde{W}}$) and initial $\tilde{\vec{\mathcal{C}}}_p$ contributions.

It can be proven that for polyconvex models the constitutive tangent operator $ \vec{\tilde{\mathcal{C}}}_k$ emanating from $\mathbb{H}_{\tilde{W}}$ is always PSD; see \cite{BonetPolyconvex15, Poya2023b}.  As a result, polyconvex formulation remains the only formulation that truly separates material instabilities encapsulated in the constitutive term $ \vec{\tilde{\mathcal{C}}}_k$ from geometric instabilities included in the initial stress term $ \vec{\tilde{\mathcal{C}}}_p$. It is however, well-known that polyconvexity does not entail convexity with respect to $\vec{F}$ \cite{Poya2023b} that is, the term $ \vec{\tilde{\mathcal{C}}}_p$ is not necessarily PSD.

\subsection{Stabilisation of initial stiffness in polyconvex large strain elasticity}
Our goal now is to come up with a stabilisation strategy for the initial stiffness term $\vec{\tilde{\mathcal{C}}}_p$. To do so, it is necessary to recall that, in polyconvex formulations the kinematics measures $\vec{F}$, $\vec{H}$ and $J$ are often treated as separate variables whose information typically cannot be retrieved solely from $\vec{F}$. In particular, in the context of variational formulations two separate approaches can be followed. The mixed Hu-Washizu variational formulation  \cite{BonetPolyconvex15,SchroderPolyconvex11} and the standard displacement-based formulation \cite{BonetPolyconvex15,POYA201875}. Although we are yet to discuss variational formulations, it is nevertheless important to keep this in mind. We consider the mixed Hu-Washizu variational formulation first where the kinematic measures are treated as separate variables. The stabilisation then follows closely the algebra developed so far by choosing to split the initial stiffness $\vec{\tilde{\mathcal{C}}}_{p}$ term in \eqref{eq:lambCFHJ} into two as
\begin{align}
\vec{\tilde{\mathcal{C}}}_{p} = \vec{\tilde{\mathcal{C}}}_{p_1} + \vec{\tilde{\mathcal{C}}}_{p_2}  &= \mathbb{I} \Cross \vec{\Sigma}_{\vec{H}} + \mathbb{I} \Cross \Sigma_J \vec{F}.
\end{align}
This split is mathematically accurate as discussed shortly. Considering the second term first
\begin{align}
\vec{\tilde{\mathcal{C}}}_{p_2}  &= \mathbb{I} \Cross \Sigma_J \vec{F},
\end{align}
which is in fact the Hessian of $J$ i.e. $\mathbb{I} \Cross \vec{F}$ scaled with $\Sigma_J$. The analytical eigensystem of $J$ has already been described in \cite{Poya2023b} (i.e.\ via our $\vec{F}$-based formulation) and can be obtained in a straight-forward fashion using the same arrangement. In 3-dimensions all the 9 eigenvalues of the initial stiffness can be written concisely as
\begin{subequations}
\begin{align}
\bar{\lambda}_{1,2,3}^{\vec{\tilde{\mathcal{C}}}_{p_2}} = \Sigma_J \left( \sqrt[3]{J + \sqrt{J^2 - \frac{(\vec{F}:\vec{F})^3}{27}}} +  \sqrt[3]{J - \sqrt{J^2 - \frac{(\vec{F}:\vec{F})^3}{27}}} \right), \\
\bar{\lambda}_4^{\vec{\tilde{\mathcal{C}}}_{p_2}} = - \Sigma_J \lambda_1, \;
\bar{\lambda}_5^{\vec{\tilde{\mathcal{C}}}_{p_2}} = - \Sigma_J \lambda_2, \;
\bar{\lambda}_6^{\vec{\tilde{\mathcal{C}}}_{p_2}} = - \Sigma_J \lambda_3, \;
\bar{\lambda}_7^{\vec{\tilde{\mathcal{C}}}_{p_2}} =   \Sigma_J \lambda_1, \;
\bar{\lambda}_8^{\vec{\tilde{\mathcal{C}}}_{p_2}} =   \Sigma_J \lambda_2, \;
\bar{\lambda}_9^{\vec{\tilde{\mathcal{C}}}_{p_2}} =  \Sigma_J \lambda_3.
\end{align}
\label{eq:eigscp23D}%
\end{subequations}
where the first 3 eigenvalues are the 3 roots of the cubic expression above which always compute to real numbers\footnote{%
The 3 cubic roots of a complex number $z = x + yi$ is given by
\begin{align*}
r_k = \sqrt[3]{|z|} \left( \text{cos}(\phi_k) + \text{sin}(\phi_k) i \right), \qquad \phi_k = \frac{1}{3} \text{atan2}(y, x) + \frac{2 \pi k}{3}, \quad k=1,2,3.
\end{align*}%
}.
%
\begin{remark}
In terms of singular-value formulation of \citet{Poya2023b} the term $\Sigma_J \mathbb{I} \Cross \vec{F}$ also contributes in the $\vec{H}_W$ operator of \eqref{eq:invarHw} which can be written as
\begin{align}
\vec{H}_{W}^{\vec{\tilde{\mathcal{C}}}_{p_2}}  = \Sigma_J
\begin{bmatrix}
0& \lambda_3 & \lambda_2 \\
&  0 & \lambda_1 \\
sym &  & 0
\end{bmatrix}.
\label{eq:JHw}
\end{align}
The first 3 eigenvalues of $\Sigma_J \mathbb{I} \Cross \vec{F}$ in \eqref{eq:eigscp23D} are indeed the eigenvalues of above $\vec{H}_{W}^{\vec{\tilde{\mathcal{C}}}_{p_2}}$ matrix or alternatively the roots of its characteristic equation
\begin{align}
(\bar{\lambda}_i^{\vec{\tilde{\mathcal{C}}}_{p_2}})^3 -  \Sigma_J^2(\vec{F}:\vec{F})(\bar{\lambda}_i^{\vec{\tilde{\mathcal{C}}}_{p_2}}) + 2J \Sigma_J^3 = 0, \quad \mathrm{for} \quad i=1,2,3.
\end{align}
Using this approach, the eigenvectors of $\vec{H}_{W}^{\vec{\tilde{\mathcal{C}}}_{p_2}}$ are also needed which can be found analytically \cite{Poya2023b} exactly similar to \eqref{eq:invarHwPSD} and \eqref{eq:invarHwCPSD} for PSD projection
\begin{align}
\vec{H}_{W}^{\vec{\tilde{\mathcal{C}}}_{p_2}^{\texttt{PSD}}} = \sum_{i=1}^d \mathrm{max}(\bar{\lambda}^{ \vec{H}_{W}^{\vec{\tilde{\mathcal{C}}}_{p_2}} }_i, 0) \; [{\vec{e}}^{ \vec{H}_{W}^{\vec{\tilde{\mathcal{C}}}_{p_2}} }]_i \otimes [{\vec{e}}^{ \vec{H}_{W}^{\vec{\tilde{\mathcal{C}}}_{p_2}} }]_i,
\label{eq:invarHwcp2PSD}
\end{align}
where $\bar{\lambda}^{ \vec{H}_{W}^{\vec{\tilde{\mathcal{C}}}_{p_2}} }_i$ are the eigenvalues and $[\vec{{e}}^{ \vec{H}_{W}^{\vec{\tilde{\mathcal{C}}}_{p_2}} }]_i$ are the eigenvectors of the tensor $\vec{H}_{W}^{\vec{\tilde{\mathcal{C}}}_{p_2}}$.
\end{remark}
In 2-dimensions much simpler expressions are found for the eigenvalues of ${\vec{\tilde{\mathcal{C}}}_{p_2}}$. There are 4 with multiplicity of 2 and the first 2 are eigenvalues of $\vec{H}_{W}^{\vec{\tilde{\mathcal{C}}}_{p_2}} = \Sigma_J [0, 1; 1, 0]$, (with eigenvectors $ [\vec{{e}}^{ \vec{H}_{W}^{\vec{\tilde{\mathcal{C}}}_{p_2}} }]_1 = \frac{1}{2} [1,1]^T$ and  $ [\vec{{e}}^{ \vec{H}_{W}^{\vec{\tilde{\mathcal{C}}}_{p_2}} }]_2 = \frac{1}{2} [-1,1]^T$)
\begin{align}
\bar{\lambda}_1^{\vec{\tilde{\mathcal{C}}}_{p_2}} = \Sigma_J, \quad
\bar{\lambda}_2^{\vec{\tilde{\mathcal{C}}}_{p_2}} = -\Sigma_J, \quad
\bar{\lambda}_3^{\vec{\tilde{\mathcal{C}}}_{p_2}} = -\Sigma_J, \quad
\bar{\lambda}_4^{\vec{\tilde{\mathcal{C}}}_{p_2}} = \Sigma_J.
\label{eq:eigscp22D}
\end{align}%
The term ${\vec{\tilde{\mathcal{C}}}_{p_2}}$ can finally be reconstructed from its eigen-decomposition in 2 and 3 dimensions, respectively by noting the eigenmatrices presented for the $\vec{F}$-based formulation as
\begin{subequations}
\begin{align}
\vec{\tilde{\mathcal{C}}}_{p_2} &= %
\sum_{i=1}^2 \sum_{j=1}^2 \bar{\lambda}_i^{\vec{\tilde{\mathcal{C}}}_{p_2}}  [\vec{{e}}^{ \vec{H}_{W}^{\vec{\tilde{\mathcal{C}}}_{p_2}} }]_i^j [{\vec{D}}]_j \otimes  [\vec{{e}}^{ \vec{H}_{W}^{\vec{\tilde{\mathcal{C}}}_{p_2}} }]_i^j [{\vec{D}}]_j +%
\bar{\lambda}_{3}^{\vec{\tilde{\mathcal{C}}}_{p_2}} [{\vec{L}}]_1 \otimes  [{\vec{L}}]_1 + %
\bar{\lambda}_{4}^{\vec{\tilde{\mathcal{C}}}_{p_2}}  [{\vec{T}}]_1 \otimes [{\vec{T}}]_1, \quad \text{in 2D}, \\
\vec{\tilde{\mathcal{C}}}_{p_2} &= %
\sum_{i=1}^3 \sum_{j=1}^3 \bar{\lambda}_i^{\vec{\tilde{\mathcal{C}}}_{p_2}}  [\vec{{e}}^{ \vec{H}_{W}^{\vec{\tilde{\mathcal{C}}}_{p_2}} }]_i^j [{\vec{D}}]_j \otimes  [\vec{{e}}^{ \vec{H}_{W}^{\vec{\tilde{\mathcal{C}}}_{p_2}} }]_i^j [{\vec{D}}]_j +%
\sum_{i=4}^6 \bar{\lambda}_{i}^{\vec{\tilde{\mathcal{C}}}_{p_2}} [{\vec{L}}]_i \otimes  [{\vec{L}}]_i + %
\sum_{i=7}^9 \bar{\lambda}_{i}^{\vec{\tilde{\mathcal{C}}}_{p_2}}  [{\vec{T}}]_i \otimes [{\vec{T}}]_i, \quad \text{in 3D}.
\label{eq:cp2}
\end{align}
\end{subequations}%
Alternatively, the PSD (stabilised) version of $\vec{\tilde{\mathcal{C}}}_{p_2}$ can be constructed in 2 and 3 dimensions, respectively as
\small{%
\begin{subequations}
\begin{align}
\vec{\tilde{\mathcal{C}}}_{p_2}^{\texttt{PSD}} = %
\sum_{i=1}^2 \sum_{j=1}^2 \text{max}(\bar{\lambda}_i^{\vec{\tilde{\mathcal{C}}}_{p_2}}, 0)  [\vec{{e}}^{ \vec{H}_{W}^{\vec{\tilde{\mathcal{C}}}_{p_2}} }]_i^j [{\vec{D}}]_j \otimes  [\vec{{e}}^{ \vec{H}_{W}^{\vec{\tilde{\mathcal{C}}}_{p_2}} }]_i^j [{\vec{D}}]_j \nonumber \\
+
\text{max}(\bar{\lambda}_{3}^{\vec{\tilde{\mathcal{C}}}_{p_2}},0) [{\vec{L}}]_1 \otimes  [{\vec{L}}]_1 + %
\text{max}(\bar{\lambda}_{4}^{\vec{\tilde{\mathcal{C}}}_{p_2}},0)  [{\vec{T}}]_1 \otimes [{\vec{T}}]_1, \quad \text{in 2D}, \\
\vec{\tilde{\mathcal{C}}}_{p_2}^{\texttt{PSD}} = %
\sum_{i=1}^3 \sum_{j=1}^3 \text{max}(\bar{\lambda}_{i}^{\vec{\tilde{\mathcal{C}}}_{p_2}},0)  [\vec{{e}}^{ \vec{H}_{W}^{\vec{\tilde{\mathcal{C}}}_{p_2}} }]_i^j [{\vec{D}}]_j \otimes  [\vec{{e}}^{ \vec{H}_{W}^{\vec{\tilde{\mathcal{C}}}_{p_2}} }]_i^j [{\vec{D}}]_j \nonumber \\
+%
\sum_{i=4}^6 \text{max}(\bar{\lambda}_{i}^{\vec{\tilde{\mathcal{C}}}_{p_2}},0) [{\vec{L}}]_i \otimes  [{\vec{L}}]_i + %
\sum_{i=7}^9 \text{max}(\bar{\lambda}_{i}^{\vec{\tilde{\mathcal{C}}}_{p_2}},0)  [{\vec{T}}]_i \otimes [{\vec{T}}]_i, \quad \text{in 3D}.
\label{eq:cp2PSD}
\end{align}%
\end{subequations}%
}
To obtain the eigensystem of the first term in the initial stiffness namely
\begin{align}
\vec{\tilde{\mathcal{C}}}_{p_1}  &= \mathbb{I} \Cross \vec{\Sigma}_{\vec{H}},
\end{align}
it is important to note that, in polyconvex elasticity, the term $\vec{\Sigma}_{\vec{H}}$ is also independent variable whose information can not be retrieved from the geometry i.e.\ the deformation gradient. This is also the reason why it is possible to treat the two terms in the initial stiffness separately as they do not span the same eigen-space. Hence, intuitively we need the Singular-Value-Decomposition (SVD) of $\vec{\Sigma}_{\vec{H}}$ first
\begin{align}
\vec{\Sigma}_{\vec{H}} = \tilde{\vec{U}}\vec{\Lambda}_{\vec{\Sigma}_{\vec{H}}}\tilde{\vec{V}}^T,
\label{eq:svdSH}
\end{align}
where $\tilde{\vec{U}}$ and $\tilde{\vec{V}}$ are orthogonal tensors namely left and right singular-matrices, respectively and the tensor $\vec{\Lambda}_{\vec{\Sigma}_{\vec{H}}}$ encodes the singular-values of $\vec{\Sigma}_{\vec{H}}$ i.e.\ the principal work-conjugate stresses namely $\lambda_1^{\vec{\Sigma}_{\vec{H}}} \geq \lambda_2^{\vec{\Sigma}_{\vec{H}}} \geq \lambda_3^{\vec{\Sigma}_{\vec{H}}} \geq 0$ such that $\lambda_i^{\vec{\Sigma}_{\vec{H}}} = \Lambda_{{\Sigma}_{{\vec{H}}_{ii}}}$. It follows that, the eigensystem of $\vec{\tilde{\mathcal{C}}}_{p_1} = \mathbb{I} \Cross \vec{\Sigma}_{\vec{H}}$ is indeed the eigensystem of Hessian of $\mathrm{det}(\vec{\Sigma}_{\vec{H}}) = \tilde{J}$ which in 3-dimensions can be written as
\begin{subequations}
\begin{align}
\bar{\lambda}_{1,2,3}^{\vec{\tilde{\mathcal{C}}}_{p_1}} =  \sqrt[3]{\tilde{J} + \sqrt{\tilde{J}^2 - \frac{(\vec{\Sigma}_{\vec{H}}:\vec{\Sigma}_{\vec{H}})^3}{27}}} +  \sqrt[3]{\tilde{J}  - \sqrt{\tilde{J} ^2 - \frac{(\vec{\Sigma}_{\vec{H}}:\vec{\Sigma}_{\vec{H}})^3}{27}}},  \\
\bar{\lambda}_4^{\vec{\tilde{\mathcal{C}}}_{p_1}} = -  \lambda_1^{\vec{\Sigma}_{\vec{H}}}, \;
\bar{\lambda}_5^{\vec{\tilde{\mathcal{C}}}_{p_1}} = -  \lambda_2^{\vec{\Sigma}_{\vec{H}}}, \;
\bar{\lambda}_6^{\vec{\tilde{\mathcal{C}}}_{p_1}} = -  \lambda_3^{\vec{\Sigma}_{\vec{H}}}, \;
\bar{\lambda}_7^{\vec{\tilde{\mathcal{C}}}_{p_1}} =    \lambda_1^{\vec{\Sigma}_{\vec{H}}}, \;
\bar{\lambda}_8^{\vec{\tilde{\mathcal{C}}}_{p_1}} =    \lambda_2^{\vec{\Sigma}_{\vec{H}}}, \;
\bar{\lambda}_9^{\vec{\tilde{\mathcal{C}}}_{p_1}} =    \lambda_3^{\vec{\Sigma}_{\vec{H}}}.
\end{align}%
\label{eq:eigscp13D}%
\end{subequations}%
Once again, in terms of singular-value formulation of \citet{Poya2023b} the term $\mathbb{I} \Cross \vec{\Sigma_H}$  contributes a $d \times d$ whose eigenvalues are indeed the first 3 above eigenvalues
\begin{align}
\vec{H}_{W}^{\vec{\tilde{\mathcal{C}}}_{p_1}}  =
\begin{bmatrix}
0& \lambda_3^{\vec{\Sigma_H}} & \lambda_2^{\vec{\Sigma_H}} \\
&  0 & \lambda_1^{\vec{\Sigma_H}}  \\
sym &  & 0
\end{bmatrix}, \qquad
\vec{H}_{W}^{\vec{\tilde{\mathcal{C}}}_{p_1}^{\texttt{PSD}}} = \sum_{i=1}^d \mathrm{max}(\bar{\lambda}^{ \vec{H}_{W}^{\vec{\tilde{\mathcal{C}}}_{p_1}} }_i, 0) \; [{\vec{e}}^{ \vec{H}_{W}^{\vec{\tilde{\mathcal{C}}}_{p_1}} }]_i \otimes [{\vec{e}}^{ \vec{H}_{W}^{\vec{\tilde{\mathcal{C}}}_{p_1}} }]_i,
\label{eq:invarHwcp2PSD}
\end{align}
where $\bar{\lambda}^{ \vec{H}_{W}^{\vec{\tilde{\mathcal{C}}}_{p_1}} }_i$ are the eigenvalues and $[\vec{\tilde{e}}^{ \vec{H}_{W}^{\vec{\tilde{\mathcal{C}}}_{p_1}} }]_i$ are the eigenvectors of the tensor $\vec{H}_{W}^{\vec{\tilde{\mathcal{C}}}_{p_1}}$.

In 2-dimensions much simpler expressions are found for the eigenvalues of ${\vec{\tilde{\mathcal{C}}}_{p_1}}$. There are 4 with multiplicity of 2 and the first 2 are eigenvalues of $\vec{H}_{W}^{\vec{\tilde{\mathcal{C}}}_{p_1}} = [0, 1; 1, 0]$, , (with eigenvectors $ [\vec{{e}}^{ \vec{H}_{W}^{\vec{\tilde{\mathcal{C}}}_{p_1}} }]_1 = \frac{1}{2} [1,1]^T$ and  $ [\vec{{e}}^{ \vec{H}_{W}^{\vec{\tilde{\mathcal{C}}}_{p_1}} }]_2 = \frac{1}{2} [-1,1]^T$)
\begin{align}
\bar{\lambda}_1^{\vec{\tilde{\mathcal{C}}}_{p_1}} = 1, \quad
\bar{\lambda}_2^{\vec{\tilde{\mathcal{C}}}_{p_1}} = -1, \quad
\bar{\lambda}_3^{\vec{\tilde{\mathcal{C}}}_{p_1}} = -1, \quad
\bar{\lambda}_4^{\vec{\tilde{\mathcal{C}}}_{p_1}} = 1.
\label{eq:eigscp12D}
\end{align}%
Finally, the eigenmatrices have exactly similar structure as before namely \emph{scaling}, \emph{flip}, \emph{twist} but emanating from the SVD of $\vec{\Sigma_H}$ which in 2-dimensions can be written as
\begin{subequations}
\begin{align}
[\tilde{\vec{D}}]_1 &= \tilde{\vec{U}}
\begin{bmatrix}
1 & 0 \\
0 & 0
\end{bmatrix}
\tilde{\vec{V}}^{T}, &
[\tilde{\vec{D}}]_2 &= \tilde{\vec{U}}
\begin{bmatrix}
0 & 0 \\
0 & 1
\end{bmatrix}
\tilde{\vec{V}}^{T},\\
[\tilde{\vec{L}}]_1 &= \frac{1}{\sqrt{2}}\tilde{\vec{U}}
\begin{bmatrix}
0 & 1 \\
1 & 0
\end{bmatrix}
\tilde{\vec{V}}^{T}, &
[\tilde{\vec{T}}]_1 &= \frac{1}{\sqrt{2}}\tilde{\vec{U}}
\begin{bmatrix}
0 & -1 \\
1 & 0
\end{bmatrix}
\tilde{\vec{V}}^{T},
\end{align}%
\label{eq:fliptwisttensors2dSH}%
\end{subequations}
and similarly in 3-dimensions
\begin{subequations}
\begin{align}
[\vec{\tilde{D}}]_1 &= \tilde{\vec{U}}
\begin{bmatrix}
1 & 0 &  0 \\
0 & 0 &  0 \\
0 & 0 &  0
\end{bmatrix}
\tilde{\vec{V}}^{T}, \quad
[\vec{\tilde{L}}]_1 = \frac{1}{\sqrt{2}}\tilde{\vec{U}}
\begin{bmatrix}
0 & 0 &  0 \\
0 & 0 &  1 \\
0 & 1 &  0
\end{bmatrix}
\tilde{\vec{V}}^{T}, \quad
[\bar{\vec{T}}]_1 = \frac{1}{\sqrt{2}}\tilde{\vec{U}}
\begin{bmatrix}
0 & 0 &  0 \\
0 & 0 & -1 \\
0 & 1 &  0
\end{bmatrix}
\tilde{\vec{V}}^{T}, \\
[\vec{\tilde{D}}]_2 &= \tilde{\vec{U}}
\begin{bmatrix}
0 & 0 &  0 \\
0 & 1 &  0 \\
0 & 0 &  0
\end{bmatrix}
\tilde{\vec{V}}^{T}, \quad
[\vec{\tilde{L}}]_2 = \frac{1}{\sqrt{2}}\tilde{\vec{U}}
\begin{bmatrix}
0 & 0 &  1 \\
0 & 0 &  0 \\
1 & 0 &  0
\end{bmatrix}
\tilde{\vec{V}}^{T},  \quad
[\tilde{\vec{T}}]_2 = \frac{1}{\sqrt{2}}\tilde{\vec{U}}
\begin{bmatrix}
0 & 0 & -1 \\
0 & 0 &  0 \\
1 & 0 &  0
\end{bmatrix}
\tilde{\vec{V}}^{T},  \\
[\vec{\tilde{D}}]_3 &= \tilde{\vec{U}}
\begin{bmatrix}
0 & 0 &  0 \\
0 & 0 &  0 \\
0 & 0 &  1
\end{bmatrix}
\tilde{\vec{V}}^{T}, \quad
[\vec{\tilde{L}}]_3 = \frac{1}{\sqrt{2}}\tilde{\vec{U}}
\begin{bmatrix}
0 &  1 &  0 \\
1 &  0 &  0 \\
0 &  0 &  0
\end{bmatrix}
\tilde{\vec{V}}^{T}, \quad
[\tilde{\vec{T}}]_3 = \frac{1}{\sqrt{2}}\tilde{\vec{U}}
\begin{bmatrix}
0 & -1 &  0 \\
1 &  0 &  0 \\
0 &  0 &  0
\end{bmatrix}
\tilde{\vec{V}}^{T}.%
\end{align}%
\label{eq:fliptwisttensors3dSH}%
\end{subequations}%
The term ${\vec{\tilde{\mathcal{C}}}_{p_1}}$ can then be reconstructed from be its eigen-decomposition in 2 and 3 dimensions respectively as
\begin{subequations}
\begin{align}
\vec{\tilde{\mathcal{C}}}_{p_1} = %
\sum_{i=1}^2 \sum_{j=1}^2 \bar{\lambda}_i^{\vec{\tilde{\mathcal{C}}}_{p_1}} [{\vec{e}}^{ \vec{H}_{W}^{\vec{\tilde{\mathcal{C}}}_{p_1}} }]_i^j [\tilde{\vec{D}}]_j \otimes   [{\vec{e}}^{ \vec{H}_{W}^{\vec{\tilde{\mathcal{C}}}_{p_1}} }]_i^j [\tilde{\vec{D}}]_j
\bar{\lambda}_{3}^{\vec{\tilde{\mathcal{C}}}_{p_1}} [\tilde{\vec{L}}]_1 \otimes  [\tilde{\vec{L}}]_1 + %
\bar{\lambda}_{4}^{\vec{\tilde{\mathcal{C}}}_{p_1}}  [\tilde{\vec{T}}]_1 \otimes [\tilde{\vec{T}}]_1, \quad \text{in 2D}, \\
\vec{\tilde{\mathcal{C}}}_{p_1} = %
\sum_{i=1}^3  \sum_{j=1}^3 \bar{\lambda}_{i}^{\vec{\tilde{\mathcal{C}}}_{p_1}} [{\vec{e}}^{ \vec{H}_{W}^{\vec{\tilde{\mathcal{C}}}_{p_1}} }]_i^j [\tilde{\vec{D}}]_j \otimes   [{\vec{e}}^{ \vec{H}_{W}^{\vec{\tilde{\mathcal{C}}}_{p_1}} }]_i^j [\tilde{\vec{D}}]_j +%
\sum_{i=4}^6 \bar{\lambda}_{i}^{\vec{\tilde{\mathcal{C}}}_{p_1}} [\tilde{\vec{L}}]_i \otimes  [\tilde{\vec{L}}]_i + %
\sum_{i=7}^9 \bar{\lambda}_{i}^{\vec{\tilde{\mathcal{C}}}_{p_1}}  [\tilde{\vec{T}}]_i \otimes [\tilde{\vec{T}}]_i, \quad \text{in 3D}.
\label{eq:cp1}
\end{align}%
\end{subequations}%
Alternatively, the PSD (stabilised) version of $\vec{\tilde{\mathcal{C}}}_{p_1}$ can be constructed as
\small{%
\begin{subequations}
\begin{align}
\vec{\tilde{\mathcal{C}}}_{p_1}^{\texttt{PSD}} =
\sum_{i=1}^2  \sum_{j=1}^2 \text{max}(\bar{\lambda}_i^{\vec{\tilde{\mathcal{C}}}_{p_1}}, 0)  [{\vec{e}}^{ \vec{H}_{W}^{\vec{\tilde{\mathcal{C}}}_{p_1}} }]_i^j [\tilde{\vec{D}}]_j \otimes   [{\vec{e}}^{ \vec{H}_{W}^{\vec{\tilde{\mathcal{C}}}_{p_1}} }]_i^j [\tilde{\vec{D}}]_j \nonumber \\
+%
\text{max}(\bar{\lambda}_{3}^{\vec{\hat{\mathcal{C}}}_{p_1}},0)  [\tilde{\vec{L}}]_1 \otimes  [\tilde{\vec{L}}]_1 + %
\text{max}(\bar{\lambda}_{4}^{\vec{\hat{\mathcal{C}}}_{p_1}},0)  [\tilde{\vec{T}}]_1 \otimes [\tilde{\vec{T}}]_1, \quad \text{in 2D}, \\
\vec{\tilde{\mathcal{C}}}_{p_1}^{\texttt{PSD}} = %
\sum_{i=1}^3  \sum_{j=1}^3 \text{max}(\tilde{\lambda}_{i}^{\vec{\hat{\mathcal{C}}}_{p_1}},0)  [{\vec{e}}^{ \vec{H}_{W}^{\vec{\tilde{\mathcal{C}}}_{p_1}} }]_i^j [\tilde{\vec{D}}]_j \otimes   [{\vec{e}}^{ \vec{H}_{W}^{\vec{\tilde{\mathcal{C}}}_{p_1}} }]_i^j [\tilde{\vec{D}}]_j \nonumber \\
+%
\sum_{i=4}^6 \text{max}(\tilde{\lambda}_{i}^{\vec{\hat{\mathcal{C}}}_{p_1}},0)  [\tilde{\vec{L}}]_i \otimes  [\tilde{\vec{L}}]_i + %
\sum_{i=7}^9 \text{max}(\tilde{\lambda}_{i}^{\vec{\hat{\mathcal{C}}}_{p_1}},0)  [\tilde{\vec{T}}]_i \otimes [\tilde{\vec{T}}]_i, \quad \text{in 3D}.
\label{eq:cp1PSD}%
\end{align}%
\end{subequations}%
}

\begin{remark}
In the context of displacement-based variational formulation $\vec{\Sigma}_{\vec{H}}$ and $\vec{F}$ are co-axial and hence SVD of $\vec{\Sigma}_{\vec{H}}$ is not necessary but the structure of ${\vec{\tilde{\mathcal{C}}}_{p_1}}$ remains the same. In fact, for displacement-based variational formulation the eigensystem of ${\vec{\tilde{\mathcal{C}}}_{p_1}}$ is structurally exactly the same and simply the eigenmatrices change from $[\tilde{\vec{D}}]_i$, $[\tilde{\vec{L}}]_i$ and $[\tilde{\vec{T}}]_i$ to $[\vec{D}]_i$, $[\vec{L}]_i$ and $[\vec{T}]_i$. The singular-values of $\vec{\Sigma}_{\vec{H}}$ namely $\Lambda_{\Sigma_{H_{ii}}}=\lambda_i^{\vec{\Sigma_H}}$ from \eqref{eq:svdSH} needed in \eqref{eq:eigscp13D}-\eqref{eq:eigscp12D} are also straightforward to compute without performing an SVD as they are cross terms. In 2-dimensions they are simply
\begin{align}
\lambda_1^{\vec{\Sigma_H}} = \frac{\partial \tilde{W}}{\partial \lambda_2}, \quad
\lambda_2^{\vec{\Sigma_H}}= \frac{\partial \tilde{W}}{\partial \lambda_1}
\end{align}
and in 3-dimensions
\begin{align}
\lambda_1^{\vec{\Sigma_H}} = \frac{\partial \tilde{W}}{\partial (\lambda_2\lambda_3)}, \quad
\lambda_2^{\vec{\Sigma_H}} = \frac{\partial \tilde{W}}{\partial (\lambda_1\lambda_3)}, \quad
\lambda_3^{\vec{\Sigma_H}} = \frac{\partial \tilde{W}}{\partial (\lambda_1\lambda_2)}.
\end{align}
Finally, in the case of displacement-based variational formulation it is possible to add the eigenvalues of ${\vec{\tilde{\mathcal{C}}}_{p_1}}$ and ${\vec{\tilde{\mathcal{C}}}_{p_2}}$ together (since their eigenmatrices are the same) to form the total initial stiffness ${\vec{\tilde{\mathcal{C}}}_{p}}$ in 2 and 3 dimensions, respectively as
\begin{subequations}
\begin{align}
\vec{\tilde{\mathcal{C}}}_{p} &= %
\sum_{i=1}^2 \sum_{j=1}^2 \bar{\lambda}_i^{\vec{\tilde{\mathcal{C}}}_{p}}   [\vec{{e}}^{ \vec{H}_{W}^{\vec{\tilde{\mathcal{C}}}_{p}} }]_i^j [{\vec{D}}]_j \otimes  [\vec{{e}}^{ \vec{H}_{W}^{\vec{\tilde{\mathcal{C}}}_{p}} }]_i^j [{\vec{D}}]_j +%
\bar{\lambda}_{3}^{\vec{\tilde{\mathcal{C}}}_{p}} [{\vec{L}}]_1 \otimes  [{\vec{L}}]_1 + %
\bar{\lambda}_{4}^{\vec{\tilde{\mathcal{C}}}_{p}}  [{\vec{T}}]_1 \otimes [{\vec{T}}]_1, \quad \text{in 2D}, \\
\vec{\tilde{\mathcal{C}}}_{p} &= %
\sum_{i=1}^3 \sum_{j=1}^3 \bar{\lambda}_{i}^{\vec{\tilde{\mathcal{C}}}_{p}}  [\vec{{e}}^{ \vec{H}_{W}^{\vec{\tilde{\mathcal{C}}}_{p}} }]_i^j [{\vec{D}}]_j \otimes  [\vec{{e}}^{ \vec{H}_{W}^{\vec{\tilde{\mathcal{C}}}_{p}} }]_i^j [{\vec{D}}]_j +%
\sum_{i=4}^6 \bar{\lambda}_{i}^{\vec{\tilde{\mathcal{C}}}_{p}} [{\vec{L}}]_i \otimes  [{\vec{L}}]_i + %
\sum_{i=7}^9 \bar{\lambda}_{i}^{\vec{\tilde{\mathcal{C}}}_{p}}  [{\vec{T}}]_i \otimes [{\vec{T}}]_i, \quad \text{in 3D}.
\label{eq:cp}
\end{align}%
\end{subequations}%
Alternatively, the PSD (stabilised) version of $\vec{\tilde{\mathcal{C}}}_{p}$ can be constructed as
\small{%
\begin{subequations}
\begin{align}
\vec{\tilde{\mathcal{C}}}_{p}^{\texttt{PSD}} =
\sum_{i=1}^2  \sum_{j=1}^2 \mathrm{max}(\bar{\lambda}_i^{\vec{\tilde{\mathcal{C}}}_{p}}, 0)  [\vec{{e}}^{ \vec{H}_{W}^{\vec{\tilde{\mathcal{C}}}_{p}} }]_i^j [{\vec{D}}]_j \otimes  [\vec{{e}}^{ \vec{H}_{W}^{\vec{\tilde{\mathcal{C}}}_{p}} }]_i^j [{\vec{D}}]_j \nonumber \\
+%
\mathrm{max}(\tilde{\lambda}_{3}^{\vec{\tilde{\mathcal{C}}}_{p}},0)  [{\vec{L}}]_1 \otimes  [{\vec{L}}]_1 + %
\mathrm{max}(\tilde{\lambda}_{4}^{\vec{\tilde{\mathcal{C}}}_{p}},0)  [{\vec{T}}]_1 \otimes [{\vec{T}}]_1, \quad \text{in 2D}, \\
\vec{\tilde{\mathcal{C}}}_{p}^{\texttt{PSD}} = %
\sum_{i=1}^3  \sum_{j=1}^3 \mathrm{max}(\bar{\lambda}_{i}^{\vec{\tilde{\mathcal{C}}}_{p}},0)   [\vec{{e}}^{ \vec{H}_{W}^{\vec{\tilde{\mathcal{C}}}_{p}} }]_i^j [{\vec{D}}]_j \otimes  [\vec{{e}}^{ \vec{H}_{W}^{\vec{\tilde{\mathcal{C}}}_{p}} }]_i^j [{\vec{D}}]_j \nonumber \\
+%
\sum_{i=4}^6 \mathrm{max}(\bar{\lambda}_{i}^{\vec{\tilde{\mathcal{C}}}_{p}},0)  [{\vec{L}}]_i \otimes  [{\vec{L}}]_i + %
\sum_{i=7}^9 \mathrm{max}(\bar{\lambda}_{i}^{\vec{\tilde{\mathcal{C}}}_{p}},0)  [{\vec{T}}]_i \otimes  [{\vec{T}}]_i, \quad \text{in 3D}.
\label{eq:cpPSD}%
\end{align}%
\end{subequations}%
}
where $ \bar{\lambda}_{i}^{\vec{\tilde{\mathcal{C}}}_{p}} =  \bar{\lambda}_{i}^{\vec{\tilde{\mathcal{C}}}_{p_1}} +  \bar{\lambda}_{i}^{\vec{\tilde{\mathcal{C}}}_{p_2}}$ and $\vec{H}_{W}^{\vec{\tilde{\mathcal{C}}}_{p}} = \vec{H}_{W}^{\vec{\tilde{\mathcal{C}}}_{p_1}} + \vec{H}_{W}^{\vec{\tilde{\mathcal{C}}}_{p_2}} $ with $ [\vec{{e}}^{ \vec{H}_{W}^{\vec{\tilde{\mathcal{C}}}_{p}} }]_i$ its corresponding eigenvectors.
\end{remark}

\begin{remark}
It is also possible to consider yet another isotropic integrity basis or set of invariants based on the symmetric/objective strain measures for large strain isotropic polyconvex elasticity \cite{BonetPolyconvex15,Kraus2019}. This in particular, entails the Cauchy-Green strain tensor $\vec{C}$, its cofactor $\vec{G}$, and its determinant $C$ (presented earlier in Remark~\autoref{rem:cinvar}). To this end, we can express the polyconvex strain energy $\tilde{W}$ as
\begin{align}
e(\vec{\nabla}_0\vec{x}) = \tilde{W}(\vec{F}, \vec{H}, J) = \overline{\overline{W}}(\vec{C}, \vec{G}, C).
\end{align}
The procedure to find the analytic eigensystem for this this set of invariants follows exactly the same as that of polyconvex formulation in terms of $\mathcal{A_{FHJ}}$ bases. In particular, the initial stiffness operator emerging from this formulations is $ \mathbb{I} \Cross \vec{\Sigma_G} + \Sigma_C\, \mathbb{I} \Cross \vec{C}$ which is similar to \eqref{eq:lambCFHJ}.
\end{remark}

\subsection{Large strain isotropic elasticity using stretch tensor invariants}
One of the most recent formulations for large strain isotropic elasticity presented in \citet{BSmith19} uses the ``lower order" stretch tensor ($\vec{U}$) invariants (compared to Cauchy-Green invariants) as a new integrity basis defined as
\begin{align}
e(\vec{\nabla}_0 \vec{x}) = \phi(I_U, II_U, III_U)
\end{align}
where the set of invariants is defined as
\begin{align}
I_U = \mathrm{tr}(\vec{U}) = \sum_{i=1}^d \lambda_i, \quad II_U = \vec{U}:\vec{U} = \sum_{i=1}^d \lambda_i^2, \quad III_U = J = \prod_{i=1}^d \lambda_i.
\end{align}
The analytic eigensystem of these invariants have already surfaced in \cite{BSmith19} and \cite{Kim22a,Poya2022a} and in fact were the inspiration behind the work in \citet{Poya2023b}. We can however, find analytical eigensystems emanating from this integrity basis with the algebra developed thus far. Specifically noting that, consistent linearisation of $\phi$ yields
\begin{subequations}
\begin{align}
\frac{\partial \phi}{\partial \vec{F}} &= \frac{\partial \phi}{\partial I_U}\vec{R} + 2 \frac{\partial \phi}{\partial II_U}\vec{F} + \frac{\partial \phi}{\partial III_U} \vec{H}, \\
\frac{\partial^2 \phi}{\partial \vec{C} \partial \vec{C}} &=
\underbrace{%
\begin{bmatrix}
\vec{R}, & 2\vec{F}, & \vec{H}
\end{bmatrix}
[\vec{H}_{W}^{\phi}]
\begin{bmatrix}
\vec{R} \\ 2\vec{F} \\ \vec{H}
\end{bmatrix}%
}_{\vec{\mathcal{C}}_k^{\phi}}
+ \underbrace{%
\left( \frac{\partial \phi}{\partial I_U} \mathbb{T}_{\vec{R}} + 2 \frac{\partial \phi}{\partial II_U} \mathbb{I} +  \frac{\partial \phi}{\partial III_U} \left( \mathbb{I} \Cross \vec{F} \right)  \right)
}_{\vec{\mathcal{C}}_p^{\phi}}.
\label{eq:lambCU}
\end{align}%
\end{subequations}
One clear advantage of these invariants is their ability to express rotation gradients in a spectrally-decomposed form. In the above expression $\mathbb{T}_{\vec{R}}$ is the rotation gradient which just is a scaled \emph{twist} tensor as presented in \cite{BSmith19,Poya2023b} and $\vec{H}_{W}^{\phi}$ represents the symmetric (but not necessarily positive definite) $d\times d$ Hessian operator
\begin{align}
\vec{H}_{W}^{\phi} =
\begin{bmatrix}
\dfrac{\partial^2 \phi}{\partial I_U\partial I_U} & \dfrac{\partial^2 \phi}{\partial I_U\partial II_U} & \dfrac{\partial^2 \phi}{\partial I_U\partial III_U} \\
 & \dfrac{\partial^2 \phi}{\partial II_U\partial II_U} & \dfrac{\partial^2 \phi}{\partial II_U\partial III_U} \\
sym  & & \dfrac{\partial^2 \phi}{\partial III_U\partial III_U}
\end{bmatrix}.
\end{align}%
Similar to \eqref{eq:lambC2}, a sufficient condition for convexity is then positive semi-definiteness of \eqref{eq:lambCU} via the constitutive ${\vec{\mathcal{C}}}_k^{\phi}$ and initial ${\vec{\mathcal{C}}}_p^{\phi}$ contributions. The constitutive part once again, comprises of a $d\times d$ Hessian ($\vec{H}_{W}^{\phi}$) whose eigen-decomposition is simple to find \cite{Poya2023b,BSmith19,Kim22a}. The eigensystem of the initial stiffness once again contains $\mathbb{I} \Cross \vec{F}$ which represents the Hessian of $J$ and was described earlier.

\section{Stabilised large strain transversely isotropic elasticity} \label{sec:5}
In this section we briefly consider the case of large strain transversely isotropic elasticity presented earlier in \S\ref{sec:3}. It should be emphasised in advance that, transversely isotropic invariants of type considered in \S\ref{sec:3} are in general convex by themselves and do not require tangent stabilisation. However, functions of such invariants may not necessarily be convex. Furthermore, as will be shown in \S\ref{sec:6} depending on the variational formulation transversely isotropic invariants still require tangent stabilisation. In such cases, if necessary, their analytic eigensystem will have to be obtained on case by case basis. In the next subsections we consider different formulations for large strain transversely isotropic energies and show that analytic eigensystems can be found for common transerverly isotropic energies without relying on SVD informations or SVD differentials unlike some prior work \cite{yufeng15,yufeng21}.
\subsection{Large strain transversely isotropic elasticity based on Cauchy-Green invariants}
Let us consider the two transversely invariants introduced in  \S\ref{sec:3}.
\begin{align}
J_4 =  \mathrm{tr}(\vec{C}(\vec{M} \otimes \vec{M})), \quad
J_5 = \mathrm{tr}(\vec{C}^2(\vec{M} \otimes \vec{M})),
\end{align}%
which are certainly expressed in terms of $\vec{C}$ due to the anisotropic restriction imposed. Consistent linearisation of these invariants yield the first derivatives as
\begin{subequations}
\begin{align}
\frac{\partial J_4}{\partial \vec{C}} &= \vec{M} \otimes \vec{M}, &
\frac{\partial J_5}{\partial \vec{C}} &= 2 \vec{C} (\vec{M} \otimes \vec{M}) \\
\frac{\partial^2 J_4}{\partial \vec{C}\partial \vec{C}} &= \vec{0}, &
\frac{\partial^2 J_5}{\partial \vec{C}\partial \vec{C}} &= 2 [\mathbb{I}]_{ILJM} [\vec{M} \otimes \vec{M}]_{JK}
\end{align} %
\label{eq:lambCJ4J5}%
\end{subequations}
From \eqref{eq:lambCJ4J5} it is clear that both of these invariants are convex in $\vec{C}$ in that, $J_4$ does not contribute in the tangent elasticity operator and $J_5$ has a constant tangent. Nevertheless, for our subsequent development, it is useful to obtain an analytical formula for the eigen-decomposition of tangent  of $J_5$. This can be found in a straight-forward fashion
\begin{align}
\frac{\partial^2 J_5}{\partial \vec{C}\partial \vec{C}} = \sum_{i=1}^d 2 [\vec{E}]_i \otimes [\vec{E}]_i
\end{align}
revealing that there are $d$ equal and constant eigenvalues (all 2s) and the eigenmatrices namely the second order tensors $[\vec{E}]_i$ contain the transversely isotropic fibre orientation or rather the structural tensor $\vec{M}$ as their i$th$ row which in 2-dimensions can be written as
\begin{align}
 [\vec{E}]_1 =
 \begin{bmatrix}
\vec{M}^T \\
\vec{0}_{1\times2}
 \end{bmatrix}, \quad
  [\vec{E}]_2 =
 \begin{bmatrix}
\vec{0}_{1\times2} \\
\vec{M}^T
 \end{bmatrix},
\end{align}
and similarly in 3-dimensions
\begin{align}
 [\vec{E}]_1 =
 \begin{bmatrix}
\vec{M}^T \\
\vec{0}_{1\times3} \\
\vec{0}_{1\times3}
 \end{bmatrix}, \quad
  [\vec{E}]_2 =
 \begin{bmatrix}
\vec{0}_{1\times3} \\
\vec{M}^T \\
\vec{0}_{1\times3}
 \end{bmatrix}, \quad
  [\vec{E}]_3 =
 \begin{bmatrix}
\vec{0}_{1\times3} \\
\vec{0}_{1\times3} \\
\vec{M}^T
 \end{bmatrix}.
\end{align}
Finally, we consider a generalised class of transversely isotropic energy expressed in terms of these two invariants namely, $\bar{W}^{ti}(J_4, J_5)$. Consistent linearisation of the energy yields
\begin{subequations}
\begin{align}
\frac{\partial \bar{W}^{ti} }{\partial \vec{C} } &= \frac{\partial \bar{W}_{ti} }{\partial J_4} \left( \vec{M} \otimes \vec{M} \right) + 2\frac{\partial \bar{W}^{ti} }{\partial J_5} \vec{C} \left( \vec{M} \otimes \vec{M} \right), \\
\frac{\partial^2 \bar{W}^{ti} }{\partial \vec{C} \partial \vec{C} } &=
\begin{bmatrix}
\left( \vec{M} \otimes \vec{M} \right), & 2\vec{C}\left( \vec{M} \otimes \vec{M} \right)
\end{bmatrix}
[\vec{H}_{\bar{W}^{ti}}]
\begin{bmatrix}
\left( \vec{M} \otimes \vec{M} \right) \\ 2\vec{C}\left( \vec{M} \otimes \vec{M} \right)
\end{bmatrix}
+ 2\frac{\partial \bar{W}^{ti} }{\partial J_5} [\mathbb{I}]_{ILJM} [\vec{M} \otimes \vec{M}]_{JK}
\end{align}%
\end{subequations}
where $\vec{H}_{\bar{W}^{ti}}$ represents the $2 \times 2$ symmetric (but necessarily positive definite) Hessian operator
\begin{align}
\vec{H}_{\bar{W}^{ti}} =
\begin{bmatrix}
\dfrac{\partial^2 \bar{W}^{ti} }{\partial J_4 \partial J_4} & \dfrac{\partial^2 \bar{W}_{ti} }{\partial J_4 \partial J_5} \\
sym & \dfrac{\partial^2 \bar{W}^{ti} }{\partial J_5 \partial J_5}
\end{bmatrix},
\end{align}%
whose eigen-decomposition can be easily obtained analytically using the formula presented in \cite{Poya2023b}.

\subsection{Large strain transversely isotropic elasticity based on polyconvex invariants}
Following \citet{GilPolyconvex22}, similar to $J_4$ and $J_5$ we can introduce their transversely isotropic polyconvex counterparts $I_4$ and $I_5$ as
 \begin{align}
I_4 =  J_4 = (\vec{FM}) \cdot (\vec{FM}), \qquad
I_5 =  (\vec{HM}) \cdot (\vec{HM}), \qquad
\end{align}%
Similarly, if we consider a generalised class of transversely isotropic energy expressed in terms of these two invariants namely, $\tilde{W}^{ti}(I_4, I_5)$. Keeping in mind that, $\vec{F}$ and $\vec{H}$ are independent polyconvex variables, consistent linearisation of the energy yields \cite{GilPolyconvex22,Kim19a}
\begin{subequations}
\begin{align}
\frac{\partial \tilde{W}^{ti} }{\partial \vec{F} } &= 2 \vec{F} \left( \vec{M} \otimes \vec{M} \right), &
\frac{\partial \tilde{W}^{ti} }{\partial \vec{H} } &= 2 \vec{H} \left( \vec{M} \otimes \vec{M} \right), \\
\frac{\partial^2 \tilde{W}^{ti} }{\partial \vec{F} \partial \vec{F} } &=  2[\mathbb{I}]_{ILJM} [\vec{M} \otimes \vec{M}]_{JK}, &
\frac{\partial^2 \tilde{W}^{ti} }{\partial \vec{H} \partial \vec{H} } &= 2[\mathbb{I}]_{ILJM} [\vec{M} \otimes \vec{M}]_{JK}, \qquad
\end{align}%
\end{subequations}
which are all PSD and further do not contribute in the initial stiffness operator in \eqref{eq:lambCFHJ}. As we will see in \S\ref{sec:6} however, depending on the variational formulation they still require tangent stabilisation. Noting the Hessian operator in \eqref{eq:invarHwFHJ} we can collapse the terms relating to $\vec{F}$ and $\vec{H}$ in \eqref{eq:lambCFHJ} we obtain the constitutive tangent operator in a decomposed form
\begin{align}
\vec{\tilde{\mathcal{C}}}_k^{ti} =  \sum_{i=1}^d 2 [\vec{E}]_i \otimes [\vec{E}]_i + \left( \mathbb{I} \Cross \vec{F}  \right) \left(  \sum_{i=1}^d 2 [\vec{E}]_i \otimes [\vec{E}]_i  \right)    \left( \mathbb{I} \Cross \vec{F} \right).
\end{align}%
Before turning to the next section, it is worth noting that, all the algebra developed to handle analytic construction of tangents and their subsequent stabilisation is not computationally demanding, does not necessarily add extra overhead compared to implementations of classical elasticity and can be implemented extremely efficiently via static tensor types such as \cite{Poya201735} as presented in \cite{Poya2023b,Poya2022a}.

\section{Discretisation-aware load stepping}\label{sec:8}
In solid mechanics, usually coercive strain energies are used for elastic deformations \cite{Hartmann2003a,Poya2022a,Poya2023b} in order to ensure the assumption of no interpenetration of matter. Coercive strain energies are barrier energies that implicitly encode positivity of Jacobian ($J$). In other words, they disallow element inversion. However, under extreme scenarios element inversion still occurs since, in discrete setting, minimisation schemes such as Newton-Raphson minimise the deformation over the entire computational mesh in an ``average" sense. This phenomenon becomes more pronounced with tangent stabilised elasticity as it unleashes extreme deformations way beyond  conventional Newton-Raphson with full tangents/Hessians. A detailed analysis of element inversion and the possible remedies in the context of Newton-Raphson can be found in \citet{Poya2022a}. Hence, often a specialised and rather sophisticated numerical treatment is required to circumvent this issue and perform inversion-free simulations. One way of achieving this is through a line search scheme. However, conventional line search schemes that merely satisfy Wolfe's condition are also incapable of guaranteeing element inversion when deformations are too large \cite{JSmith15}. Inspired by the work of \cite{JSmith15, MinchenLi2020a}, we can devise a discretisation-aware line search scheme that explicitly checks for geometric validity. However, unlike these earlier works that apply only to simplicial meshes we provide an approach for general polyhedral finite elements of arbitrary order. Assuming that the variational formulations and finite element discretisation have been carried out (see \ref{sec:6} and \ref{sec:7}) the update rule for Newton-Raphson procedure equipped with line search can be written as
\begin{align}
\vec{x}_{k+1} = \vec{x}_{k} + \alpha \Delta\vec{u},
\end{align}
where $k$ is the iteration counter and $\alpha$ the line search parameter or the step size. If we write this equation for a single quadrature point ($q$) of a finite element, we obtain
\begin{align}
\vec{x}_{k+1}^q = \vec{x}_{k}^q + \alpha^q \Delta\vec{u}^q.
\end{align}
Taking derivatives yields
\begin{align}
\vec{\nabla}_0 \vec{x}_{k+1}^q = \vec{\nabla}_0 \left( \vec{x}_{k}^q + \alpha^q \Delta\vec{u}^e \right)
= \vec{\nabla}_0 \vec{x}_{k}^q + \alpha^q  \vec{\nabla}_0 \Delta\vec{u}^q
\end{align}
Noting that, $\vec{F}_{k} = \vec{\nabla}_0 \vec{x}_{k}^q$ is the deformation gradient at previous iteration (i.e. iteration $k$) and $\vec{F}_{k+1} = \vec{\nabla}_0 \vec{x}_{k+1}^q$ the deformation gradient at current iteration (i.e. iteration $k+1$) and further denoting $ \bar{\vec{F}} = \vec{\nabla}_0 \Delta\vec{u}^q$ and its determinant as $\text{det}(\Delta\vec{u}^q) = \bar{J}$ we have
\begin{align}
\vec{F}_{k+1} = \vec{F}_{k} + \alpha^q \bar{\vec{F}}.
\end{align}
The goal now is to find an $\alpha^q$ that drives the current iteration Jacobian to zero i.e.\ $J_{k+1}=\mathrm{det}(\vec{F}_{k+1})=0$. In other words, the problem of finding the right step size can be posed as a maximisation problem on per-quadrature basis as
\begin{align}
(\alpha^q)^\star = \argmax_{\alpha^q} \bigg\{ \mathrm{det} \left( \vec{F}_{k} + \alpha^q \bar{\vec{F}} \right) \bigg\},
\end{align}
where $(\alpha^q)^\star$ is the optimal step size. In the context of solid mechanics, this maximisation problem is well-defined as the starting point is always within the feasible region (initial configuration/mesh (or the one at previous iteration) is always assumed inversion-free). Fortunately, the above functional involves either $2 \times 2$ or $3\times3$ matrices for which (i.e.\ for the determinant of sums) simple closed-form expressions exist, in particular
\begin{subequations}
\begin{align}
\mathrm{det} \left( \vec{F}_{k} + \alpha^q \bar{\vec{F}} \right) &= J_k + \alpha^q \left( \bar{J} + \vec{H}_k : \bar{\vec{F}} \right), && \text{in\; 2D} \\
\mathrm{det} \left( \vec{F}_{k} + \alpha^q \bar{\vec{F}} \right) &= J_k + \alpha^q \left( \bar{J} + \vec{H}_k : \bar{\vec{F}} \right) + (\alpha^q)^2 \left(\bar{\vec{H}} : \vec{F}_k \right), && \text{in\; 3D}
\end{align}%
\end{subequations}%
where $J_k=\mathrm{det}(\vec{F}_k)$ and $\vec{H}_k = J_k\vec{F}_k^{-T} = \frac{1}{2} \vec{F}_k \Cross  \vec{F}_k$ are the Jacobian and the cofactor at previous iteration, respectively. The above expressions can be solved for $\alpha^q$ directly as they are linear and quadratic functions. The maximum affordable step sizes ($\alpha^q$s) are computed for all quadrature points and their minimum over the entire computational mesh is chosen as the safe global step size ($\alpha$). Once determined, $\alpha$ is further passed as the maximum allowed step size to a standard line search technique to further check for Wolfe's condition. The line search algorithm might further decrease $\alpha$. Note that, when determining $\alpha^q$ in 3-dimension, the quadratic equation above can have complex roots. Fortunately, unlike the arc-length methods, one is not spoilt for choice here. A complex root simply implies that $J_{k+1} = 0$ or collapsed configuration is infeasible. Such values including negative values of $\alpha$ are simply discarded. It is worth mentioning that, in sharp contrast to \cite{JSmith15} and \cite{MinchenLi2020a} techniques which are applicable to only linear/simplicial meshes our technique can be applied to general polyhedral finite elements and further requires only a linear functional in 2-dimensions and a quadratic in 3-dimensions (as opposed to quadratic in 2-dimensions and cubic in 3-dimensions).

In engineering analysis, continuation is a well-established practice as often it is not possible to apply the full load in one step within a nonlinear solution \cite{RIKS1979529,CRISFIELD198155, Wriggers-Book}. Tangent stabilised elasticity however, facilitates applying external loads (Neumann boundary conditions) in a single step by further relying on the above discretisation-aware line search scheme. For imposed displacements via Dirichlet boundary conditions however, situations can arise where applying the full imposed displacements in a single step might cause element-inversion and mesh invalidity. In such cases, it is understandable to rely on increments. Fortunately, an automatic way to determine the number of load increments (or rather the load factor) can be established using the same approach followed for the line search scheme. Recall that, the geometry update via imposed Dirichlet boundary condition at any given increment of the continuation can be written
\begin{align}
\vec{x}^{n} = \vec{x}^{n-1} + \Delta \lambda_{d} \vec{u}_p,
\end{align}
where $n$ is the increment counter, $\Delta \lambda_{d}$ the desired user-defined load factor increment (which is constant with equal load increments) and $\vec{u}_p$ the total prescribed displacement. Instead of working with a constant $ \Delta \lambda_{d}$ which might be infeasible, similar to the above line search procedure, we can choose to explicitly solve for the load factor increment $ \Delta \lambda^{n}$ by re-writing the above equation on a quadrature point level as
\begin{align}
(\vec{x}^{n})^q = (\vec{x}^{n-1})^q + (\Delta \lambda^{n})^q \vec{u}_p^q,
\end{align}
where the solution to the right load factor increment can be posed as a maximisation problem resulting in a similar set of equations
\begin{subequations}
\begin{align}
\mathrm{det} \left( \vec{F}^{n} + (\Delta \lambda^{n})^q \bar{\vec{F}} \right) &= J^n + (\Delta \lambda^{n})^q \left( \bar{J} + \vec{H}^n : \bar{\vec{F}} \right), && \text{in\; 2D} \\
\mathrm{det} \left( \vec{F}^{n} +(\Delta \lambda^{n})^q \bar{\vec{F}} \right) &= J^n + (\Delta \lambda^{n})^q \left( \bar{J} + \vec{H}^n : \bar{\vec{F}} \right) + ((\Delta \lambda^{n})^q)^2 \left(\bar{\vec{H}} : \vec{F}^n \right), && \text{in\; 3D}
\end{align}%
\end{subequations}%
where $\vec{F}^{n} = \vec{\nabla}_0 (\vec{x}^{n})^q$, $J^n=\mathrm{det}(\vec{F}^n)$ and $\vec{H}^n = J^n(\vec{F}^n){^{-T}} = \frac{1}{2} \vec{F}^n \Cross \vec{F}^n$ are the deformation gradient, Jacobian and the cofactor at previous increment, respectively, and with slight abuse of notation $\bar{\vec{F}} = \vec{\nabla}_0 \vec{u}_p^q$. Like the \emph{non-consistent} arc-length method, the value of global $(\Delta \lambda^{n})^q$ is solved for explicitly but now by taking the minimum of their local per-quadrature points values. The right global $ \Delta \lambda^{n} $ then limits the amount of load/Dirichlet BC at any given increment to the maximum discretisation/inversion-safe load factor. Certainly, cost-savings can be applied here by only traversing over elements which are on the boundary/interfaces where the loading is applied. This discretisation-aware continuation technique coupled with Projected Newton-Raphson is shown in Algorithms \autoref{algo:pnr} and \autoref{algo:safeload}. For problems with no prestressing where assumed stored energy is continuously increasing during the course of minimisation, a simplified version of the line search presented in \cite{Poya2023b} can be articulated as shown in Algorithm \autoref{algo:linesearch}. \ \\ \

\begin{tcolorbox}[skin=enhanced, frame style={fill=black!60}, interior style={color=yellow!5}, fonttitle=\bfseries, title=Newton-Raphson with automatic safe load stepping]
\begin{algorithm}[H]
\label{algo:pnr}
\SetKwFunction{ComputeSafeLoadFactor}{ComputeSafeLoadFactor}
\SetKwFunction{ComputeSafeStepSize}{ComputeSafeStepSize}
\SetKwData{LineSearch}{LineSearch}
\SetKwData{maxIter}{maxIter}
\SetKwData{thresholdr}{thresholdr}
\SetKwData{thresholdc}{thresholdc}
\SetKwData{break}{break}
\SetKwData{or}{or}

\KwData{$\vec{X}, \Delta \lambda_d, \lambda_{\text{max}}, \thresholdr, \thresholdc$}

\KwResult{$\vec{x}$}

Initialise $n = 0$

\While{$\lambda < \lambda_{\mathrm{max}}$ }
{
	Initialise $k = 0$

	Assemble external forces $\textbf{F}$ using desired load factor $ \Delta \lambda_d$

	Impose incremental Dirichlet BC using desired load factor $ \Delta \lambda_d$

	$\Delta \lambda^{n} = $ \ComputeSafeLoadFactor($\vec{x}^n, \Delta \lambda_d, \lambda_{\text{max}}$)

	Re-assemble forces $\textbf{F}, \textbf{R}$ with safe load factor $\Delta \lambda^{n} $

    	\While{$||\vec{\mathrm{R}}|| > \thresholdr$ \or $k < $ \maxIter}
	{
		Assemble internal forces and tangent stiffness $\textbf{R}, \textbf{K}$

		Add external forces $\textbf{R} = \textbf{R} + \textbf{F}$

		Solve for displacements $\textbf{K} \Delta \vec{u} = -\textbf{R}$

		\If{$|\vec{\mathrm{R}} \cdot \Delta \vec{u}| < \thresholdc$}
		{
			\break
		}

		$\alpha = $ \ComputeSafeStepSize{$\vec{x}_{k}, \vec{\Delta} \vec{u}$ }

		$\alpha = $ \LineSearch($\vec{x}_{k}, \textbf{R}, \vec{\Delta} \vec{u}, \alpha$)

        		$\vec{x}_{k+1} = \vec{x}_{k} + \alpha \vec{\Delta} \vec{u}$

        		$k = k + 1$
    	}

	$\lambda = \lambda + \Delta \lambda^{n} $

	$n = n + 1$
}
\end{algorithm}
\end{tcolorbox}

\begin{tcolorbox}[skin=enhanced, frame style={fill=black!60}, interior style={color=yellow!5}, fonttitle=\bfseries, title=Compute safe load factor]
\begin{algorithm}[H]
\label{algo:safeload}
\SetKwData{break}{break}
\SetKwData{or}{or}
\SetKwData{in}{in}
\SetKwData{ndim}{ndim}
\SetKwData{DBLMAX}{DBLMAX}
\SetKwData{max}{max}

\KwData{$\vec{x}^n, \Delta \lambda_d, \lambda_{\text{max}}$}

\KwResult{$ \Delta \lambda^n$}

Initialise $ \Delta \lambda^n = $ \DBLMAX, $c=0.9$

\For{$e$ \in elements }
{
	\For{$q$ \in quadratures }
	{
		\uIf{\ndim $= 2$}
		{
			Solve $J^n + (\Delta \lambda^{n})^q \left( \bar{J} + \vec{H}^n : \bar{\vec{F}} \right)$ for $(\Delta \lambda^{n})^q$
		}
		\uElseIf{\ndim $= 3$}
		{
			Solve $J^n + (\Delta \lambda^{n})^q \left( \bar{J} + \vec{H}^n : \bar{\vec{F}} \right) + ((\Delta \lambda^{n})^q)^2 \left(\bar{\vec{H}} : \vec{F}^n \right)$ for $(\Delta \lambda^{n})^q $
		}

		\If{$\Delta \lambda^{n} > (\Delta \lambda^{n})^q $}
		{
			$\Delta \lambda^{n} = (\Delta \lambda^{n})^q $
		}
	}
}
$\Delta \lambda^{n} = c\Delta \lambda^{n} $
\end{algorithm}
\end{tcolorbox}

\begin{tcolorbox}[skin=enhanced, frame style={fill=black!60}, interior style={color=yellow!5}, fonttitle=\bfseries, title=Line search procedure]
\begin{algorithm}[H]
\label{algo:linesearch}
\SetKwFunction{LineSearch}{LineSearch}
\SetKwData{break}{break}
\SetKwData{or}{or}
\SetKwData{maxIter}{maxIter}

\KwData{$\vec{x}_k, \textbf{R}, \Delta \vec{u}, \alpha$}

\KwResult{$\alpha$}

Initialise $m = 0,\; \maxIter = 50,\; \alpha_{\text{min}} = 10^{-16},\; \rho = 0.5,\; c = 0.95$

Compute curvature at $\vec{x}_k$:  $\mathcal{R}_0 = \textbf{R} \cdot \Delta \vec{u}$

\While{$\alpha > \alpha_{\text{min}}$ \or $m < $ \maxIter}
{
        Assemble residual at $\vec{x}_k + \alpha \Delta \vec{u}$: $\textbf{R}_\alpha = \textbf{R}(\vec{x}_k + \alpha \Delta \vec{u})$

        Compute curvature at $\vec{x}_k + \alpha \Delta \vec{u}$:  $\mathcal{R}_\alpha = \textbf{R}_\alpha \cdot \Delta \vec{u}$

        \If{$\mathcal{R}_\alpha \leq c \mathcal{R}_0 $}
        {
		\break
        }

        $\alpha = \rho \alpha$

        $m = m + 1$
}
\end{algorithm}
\end{tcolorbox}

\section{Numerical examples}\label{sec:9}
\subsection{Comparison of all presented formulations of nonlinear elasticity in 2 and 3 dimensions}
The goal of this example is to
\begin{enumerate}
\item \emph{Compare all classic implementations of nonlinear elasticity with the new ones proposed in this work in particular, for $\vec{F}$-based, $\vec{C}$-based and $\vec{b}$-based formulations}
\item \emph{Show that, the proposed spectrally-decomposed tangent constructions are not ``approximation" and achieve the same quadratic Newton-Raphson convergence of classic implementations}
\end{enumerate}
To facilitate these comparisons, we have implemented the standard $\vec{F}$-based, $\vec{C}$-based and $\vec{b}$-based formulations of nonlinear elasticity too. This is followed by their new implementations based on spectrally-decomposed tangent construction. With tangent stabilised elasticity, we choose to work with coercive (i.e.\ barrier) polyconvex energies of quasi-conformal type [see Lemma C.7 in \citet{Hartmann2003a}]
\begin{align}
e(\vec{F}) = \mu_1 \left ( \frac{\vec{F}:\vec{F}}{3J^{2/3}} - 1 \right)  +  \mu_2 \left ( \frac{\vec{H}:\vec{H}}{3J^{2/3}} - 1 \right) + \kappa \left( J + J^{-1} - 2 \right),
\label{eq:amips_energy3d}
\end{align}
which is a nearly-incompressible Mooney-Rivlin type model. The reason for choosing coercive energies in the context of tangent stabilised elasticity was laid out in \cite{Poya2023b}, i.e. to guarantee inversion-free simulations. This is in addition to safe load stepping scheme presented earlier in \autoref{algo:pnr} and in fact complementary to it. In 2-dimensions, a well-known particularisation of this energy is available which we can write in terms of $\vec{F}$-based, $\vec{C}$-based and $\vec{b}$-based formulations as
\begin{subequations}
\begin{align}
e(\vec{F}) = \mu \left ( \frac{\vec{F}:\vec{F}}{2J} - 1 \right) + \kappa \left( J + J^{-1} - 2 \right), \\
\breve{W}(\vec{C}) = \mu \left ( \frac{ \mathrm{tr}(\vec{C}) }{2\sqrt{C}} - 1 \right) + \kappa \left( \sqrt{C} + {\sqrt{C}}^{-1} - 2 \right), \\
\dbar{W}(\vec{b}) = \mu \left ( \frac{ \mathrm{tr}(\vec{b}) }{2\sqrt{b}} - 1 \right) + \kappa \left( \sqrt{b} + {\sqrt{b}}^{-1} - 2 \right),
\end{align}%
\label{eq:amips_energy2d}%
\end{subequations}
The analytic eigensystem of this energy in 2-dimensions was already worked out in \cite{Poya2023b} for $\vec{F}$-based formulations and eigen-spectrum was extensively analysed; see also \cite{Poya2022a}. We show the analytic eigensystem for the $\vec{C}$-based counterpart which can be obtained easily using the algebra developed in the previous sections. The first two eigenvalues of constitutive tangent namely $\bar{\lambda}_1^{\vec{\breve{\mathcal{C}}}_k} $ and $\bar{\lambda}_2^{\vec{\breve{\mathcal{C}}}_k}$ are found from the Hessian matrix
\begin{align}
\vec{H}_{\bar{W}} =
\begin{bmatrix}
\dfrac{3\mathrm{tr}(\vec{C})\mu}{2\sqrt{C}\lambda_1^4} - \dfrac{2\mu}{\sqrt{C}\lambda_1^2}  - \dfrac{\kappa}{\lambda_1^4}  \left(\sqrt{C} - 3 \sqrt{C}^{-1}\right) &
\kappa \left( \dfrac{1}{\sqrt{C}}  + \dfrac{1}{ (\sqrt{C})^3 } \right) - \dfrac{\mathrm{tr}(\vec{C})\mu }{2(\sqrt{C})^3} \\
sym & \dfrac{3\mathrm{tr}(\vec{C})\mu}{2\sqrt{C}\lambda_2^4} - \dfrac{2\mu}{\sqrt{C}\lambda_2^2}  - \dfrac{\kappa}{\lambda_2^4}  \left(\sqrt{C} - 3\sqrt{C}^{-1}\right)
\end{bmatrix},
\end{align}
and the third one corresponds to flip mode which is
\begin{align}
\bar{\lambda}_3^{\vec{\breve{\mathcal{C}}}_k} =   \dfrac{\mathrm{tr}(\vec{C})\mu }{2(\sqrt{C})^3}  - \kappa \frac{ C  - 1 }{(\sqrt{C})^3},
\end{align}
and finally the initial/geometric stiffness eigenvalues are indeed the principal components of PK2
\begin{subequations}
\begin{align}
\Lambda_{S_{11}} &=  \frac{\mu}{\sqrt{C}} - \frac{\mathrm{tr}(\vec{C}) \mu}{2\sqrt{C} \lambda_1^2} +  \frac{\kappa}{\lambda_1^2} \left(\sqrt{C} - \sqrt{C}^{-1} \right), \\
\Lambda_{S_{22}} &=  \frac{\mu}{\sqrt{C}} - \frac{\mathrm{tr}(\vec{C}) \mu}{2\sqrt{C} \lambda_2^2} +  \frac{\kappa}{\lambda_2^2} \left(\sqrt{C} - \sqrt{C}^{-1}\right).
\end{align}%
\label{eq:amips_pk1lambs}%
\end{subequations}
Similarly, the eigensystem for $\vec{b}$-based formulation can be obtained using the push-forward relationships from  \eqref{eq:pushfd_Hw}, \eqref{eq:pushfd_flips} and \eqref{eq:pushfd_IS}. The 3-dimensional tangent eigensystem is longer but can be obtained as easily from the symbolic code provided in \autoref{fig:symcodeCbased}.

\begin{figure}
\centering
\includegraphics[scale=0.3]{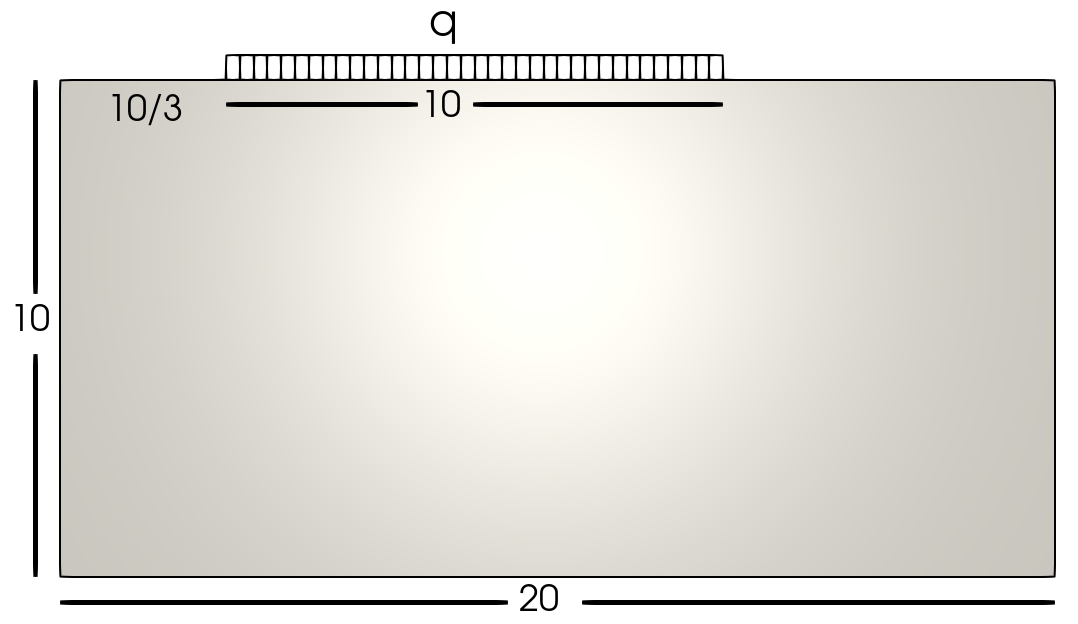} \\
\begin{tabular}{ccc}
\includegraphics[scale=0.14]{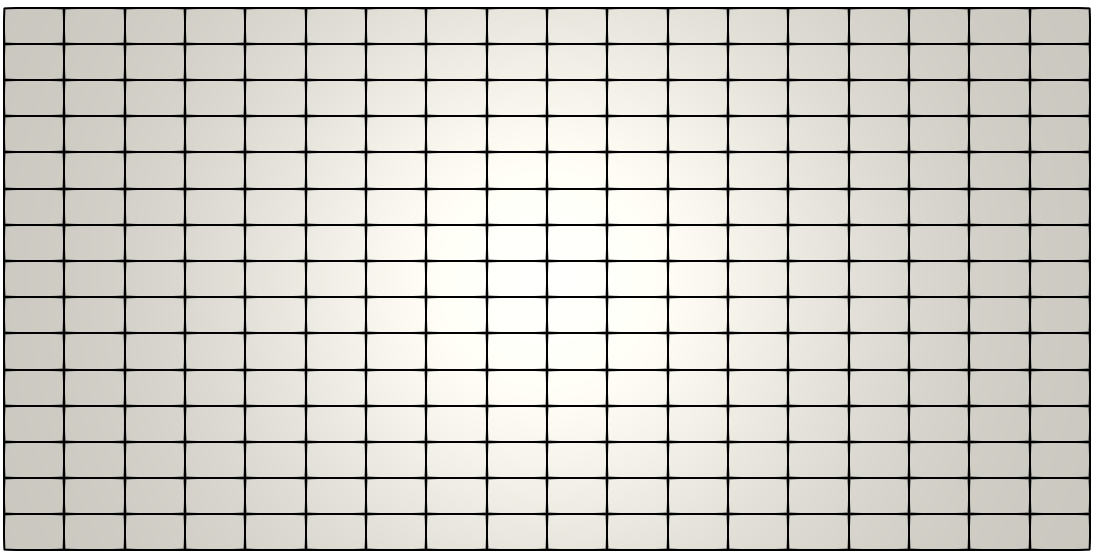} &
\includegraphics[scale=0.14]{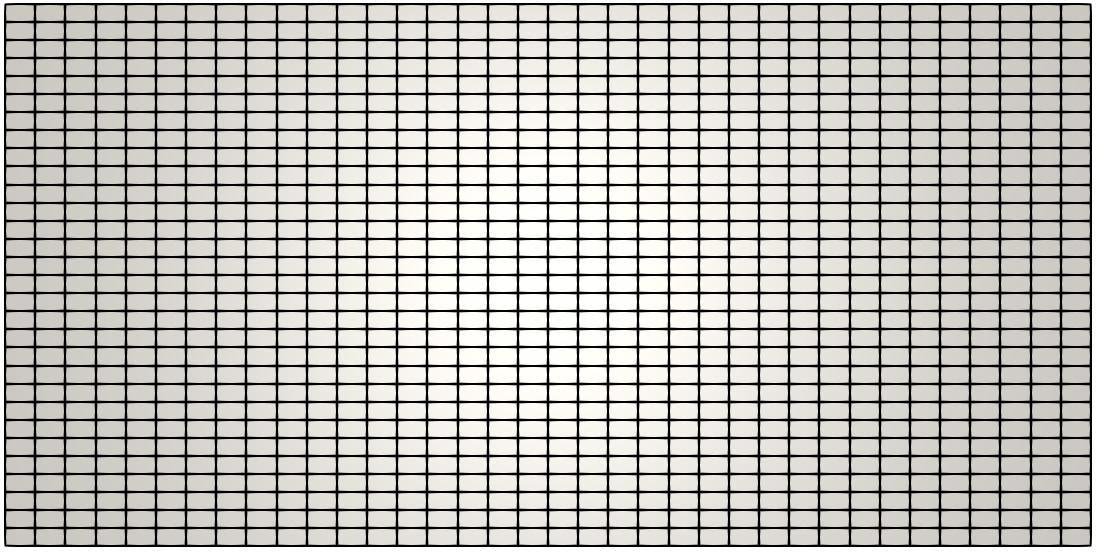} &
\includegraphics[scale=0.14]{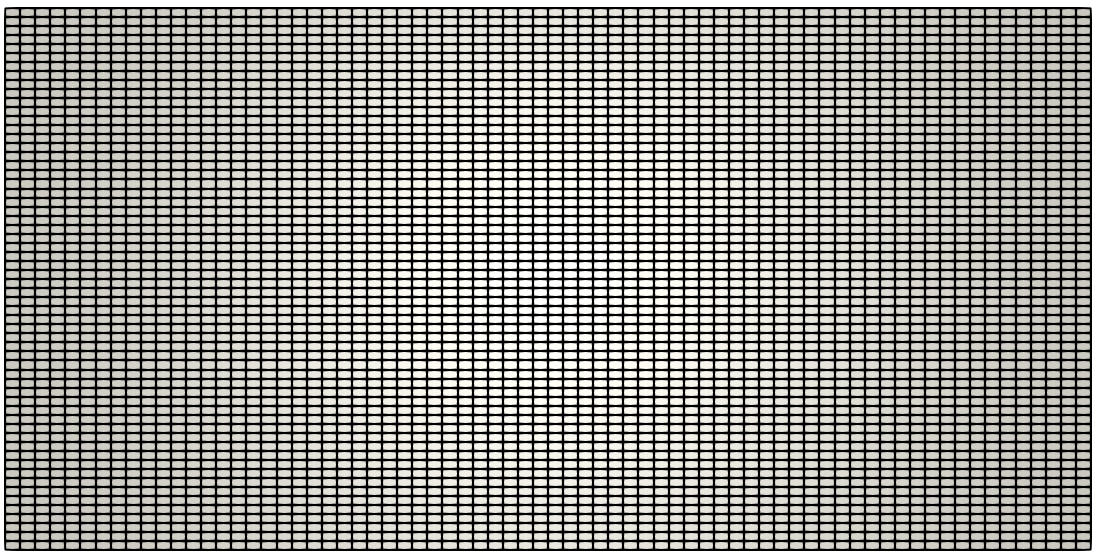} \\
a) Level 1 & b) Level 2 & c) Level 3
\end{tabular}
\caption{Geometry and loading setup for the rectangular block specimen and the computational mesh with 3 refinement levels.}
\label{fig:compblockgeom}%
\end{figure} %

For the first example, we consider compression of an inhomogeneous rectangular block using asymmetric loading as shown \autoref{fig:compblockgeom}. This example has been analysed in the past by others and is known to show a mild degree of localised failure \cite{Bieber2023a} due to geometric instability. For the purpose of verifications, we first start with bilinear finite element ansatz functions for quadrilaterals. Three refinement levels are considered as will be shown in \autoref{fig:compblockgeom}. Material properties used for the above coercive energy are: Shear modulus $\mu=1000$ and Poisson's ratio $\nu = 0.49$.

\begin{table}
\centering
\resizebox{\columnwidth}{!}{%
\begin{tabular}{| c | c | c | c | c | c | c |}
\hline
Formulation & Iter 1& Iter 2 & Iter 3 & Iter 4 & Iter 5 & Iter 6  \\ \hline
$\vec{F}$-based Classic &  1.00000e+00 &  1.31091e-01 &  3.58628e-02 &  7.82149e-04 &  4.75692e-07 &  6.46615e-11  \\ \hline
$\vec{F}$-based PS &  1.00000e+00 &  1.31091e-01 &  3.58628e-02 &  7.82149e-04 &  4.75673e-07 &  6.89650e-11  \\ \hline
$\vec{C}$-based Classic &  1.00000e+00 &  1.31091e-01 &  3.58628e-02 &  7.82149e-04 &  4.75706e-07 &  6.38668e-11  \\ \hline
$\vec{C}$-based PS &  1.00000e+00 &  1.31091e-01 &  3.58628e-02 &  7.82149e-04 &  4.75707e-07 &  6.35543e-11  \\ \hline
$\vec{b}$-based Classic &  1.00000e+00 &  1.31091e-01 &  3.58628e-02 &  7.82149e-04 &  4.75690e-07 &  5.56428e-11  \\ \hline
$\vec{b}$-based PS &  1.00000e+00 &  1.31091e-01 &  3.58628e-02 &  7.82149e-04 &  4.75701e-07 &  5.87015e-11  \\ \hline
\end{tabular}
}
\caption{Quadratic convergence of Newton-Raphson for the first increment for the 2-dimensional inhomogeneous compression of rectangular block using new (spectrally-decomposed tangent) and classic implementations of $\vec{F}$-based, $\vec{C}$-based and $\vec{b}$-based formulations. Results are in solid agreement validating all numerical implementations.}%
\label{tab:nrconv2d1}
\end{table}

\begin{table}
\centering
\resizebox{\columnwidth}{!}{%
\begin{tabular}{| c | c | c | c | c | c | c |}
\hline
Formulation & Iter 1& Iter 2 & Iter 3 & Iter 4 & Iter 5  \\ \hline
$\vec{F}$-based Classic &  1.00000e+00 &  2.38769e-02 &  2.30346e-03 &  1.07726e-06 &  6.96135e-11 \\ \hline
$\vec{F}$-based PS &  1.00000e+00 &  2.38769e-02 &  2.30346e-03 &  1.07722e-06 &  7.88531e-11 \\ \hline
$\vec{C}$-based Classic &  1.00000e+00 &  2.38769e-02 &  2.30346e-03 &  1.07725e-06 &  7.31756e-11 \\ \hline
$\vec{C}$-based PS &  1.00000e+00 &  2.38769e-02 &  2.30346e-03 &  1.07727e-06 &  7.85511e-11 \\ \hline
$\vec{b}$-based Classic &  1.00000e+00 &  2.38769e-02 &  2.30346e-03 &  1.07726e-06 &  8.63721e-11 \\ \hline
$\vec{b}$-based PS &  1.00000e+00 &  2.38769e-02 &  2.30346e-03 &  1.07730e-06 &  5.98287e-11 \\ \hline
\end{tabular}
}
\caption{Quadratic convergence of Newton-Raphson for last (5th) increment for 2-dimensional inhomogeneous compression of rectangular block using new (spectrally-decomposed tangent) and classic implementations of $\vec{F}$-based, $\vec{C}$-based and $\vec{b}$-based formulations. Results are in solid agreement validating all numerical implementations.}%
\label{tab:nrconv2d2}
\end{table}

\begin{figure}
\centering
\includegraphics[scale=0.12]{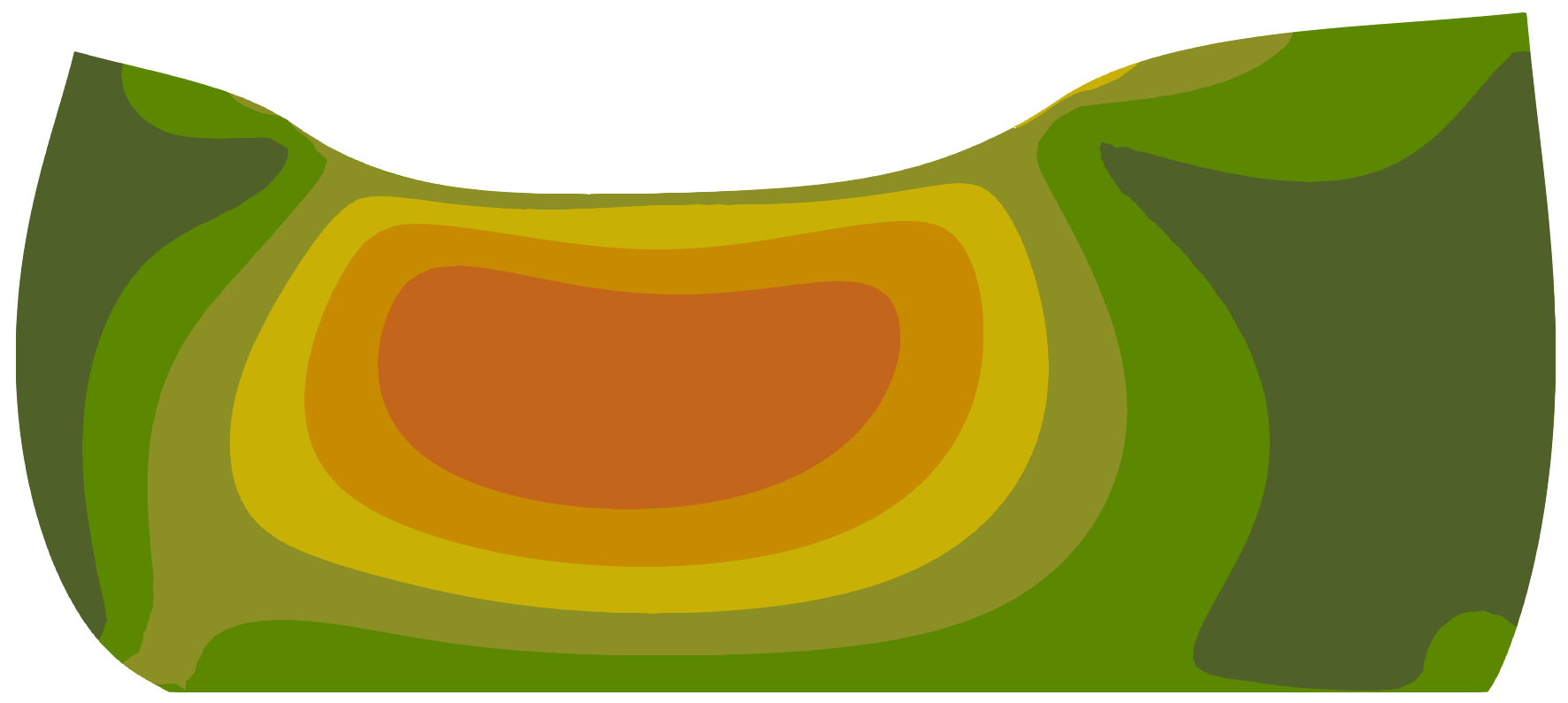}
\includegraphics[scale=0.17]{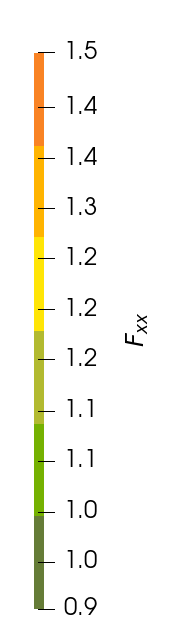}
\includegraphics[scale=0.12]{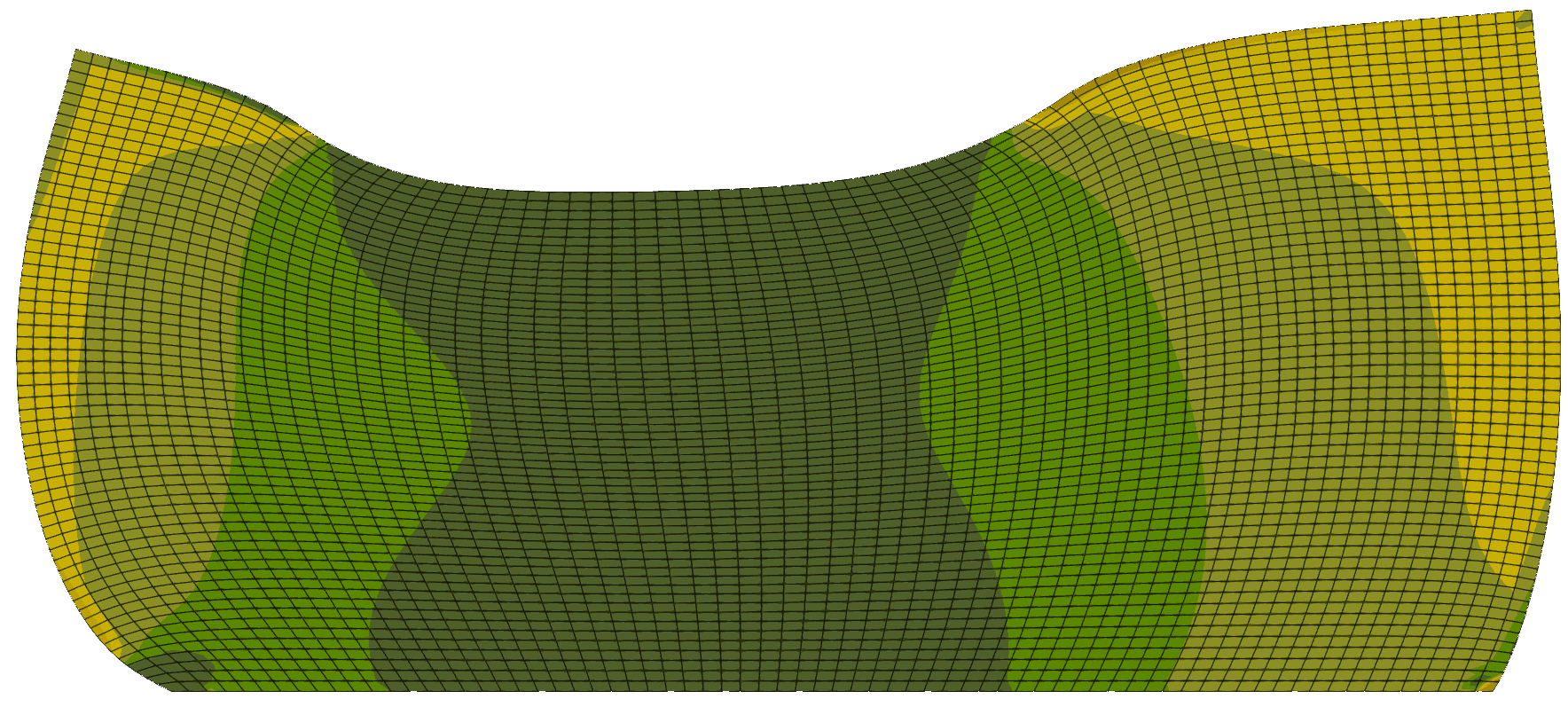}
\includegraphics[scale=0.17]{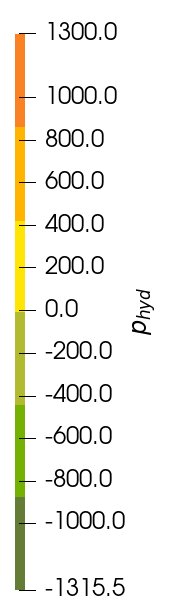} \\
\caption{$F_{xx}$ and hydrostatic pressure $p_{hyd}$ distribution for the 2-dimensional rectangular block from \autoref{tab:nrconv2d1} and \autoref{tab:nrconv2d2}. All implementations result in identical distribution hence, only $\vec{b}$-based result is shown here.}
\label{fig:compblock2d_nr_fxx_phyd}%
\end{figure}

We start the comparisons with applying a compressive load on the rectangular block and use the standard NR with 5 increments and check that all implementations achieve quadratic convergence by using a residual norm of $10^{-8}$ as stopping criterion. \autoref{tab:nrconv2d1} and \autoref{tab:nrconv2d2} show that this is certainly the case. All implementations with spectrally-decomposed tangents (denoted as PS) achieve quadratic convergence and are in excellent agreement with classic implementations. Further, in \autoref{fig:compblock2d_nr_fxx_phyd} we plot the distribution of derived quantities namely, the deformation gradient component $F_{xx}$ and the hydrostatic pressure $p_{hyd}$ and verify that these quantities match for all formulations. \autoref{fig:compblock2d_nr_fxx_phyd} also shows the finest mesh (Level 3) for the geometry shown in \autoref{fig:compblockgeom} with 4356 quadrilateral elements 4490 nodes.

\begin{figure}
\centering
\includegraphics[scale=0.16]{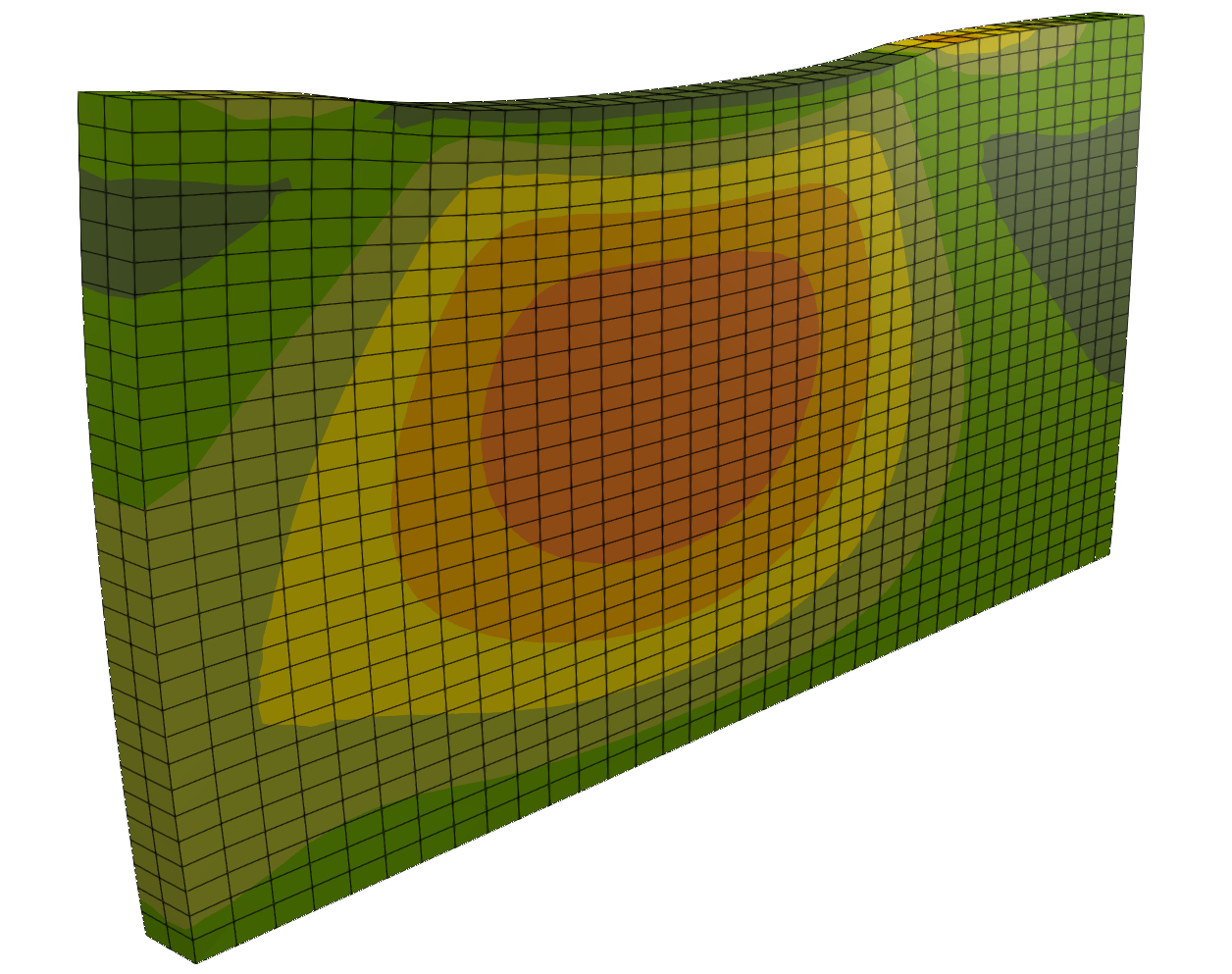}
\includegraphics[scale=0.2]{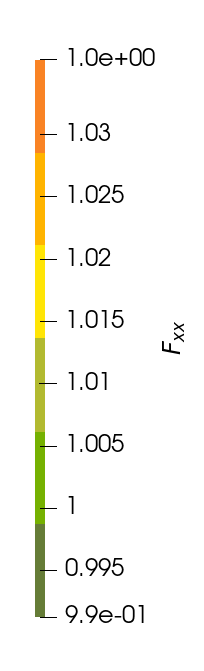}
\includegraphics[scale=0.16]{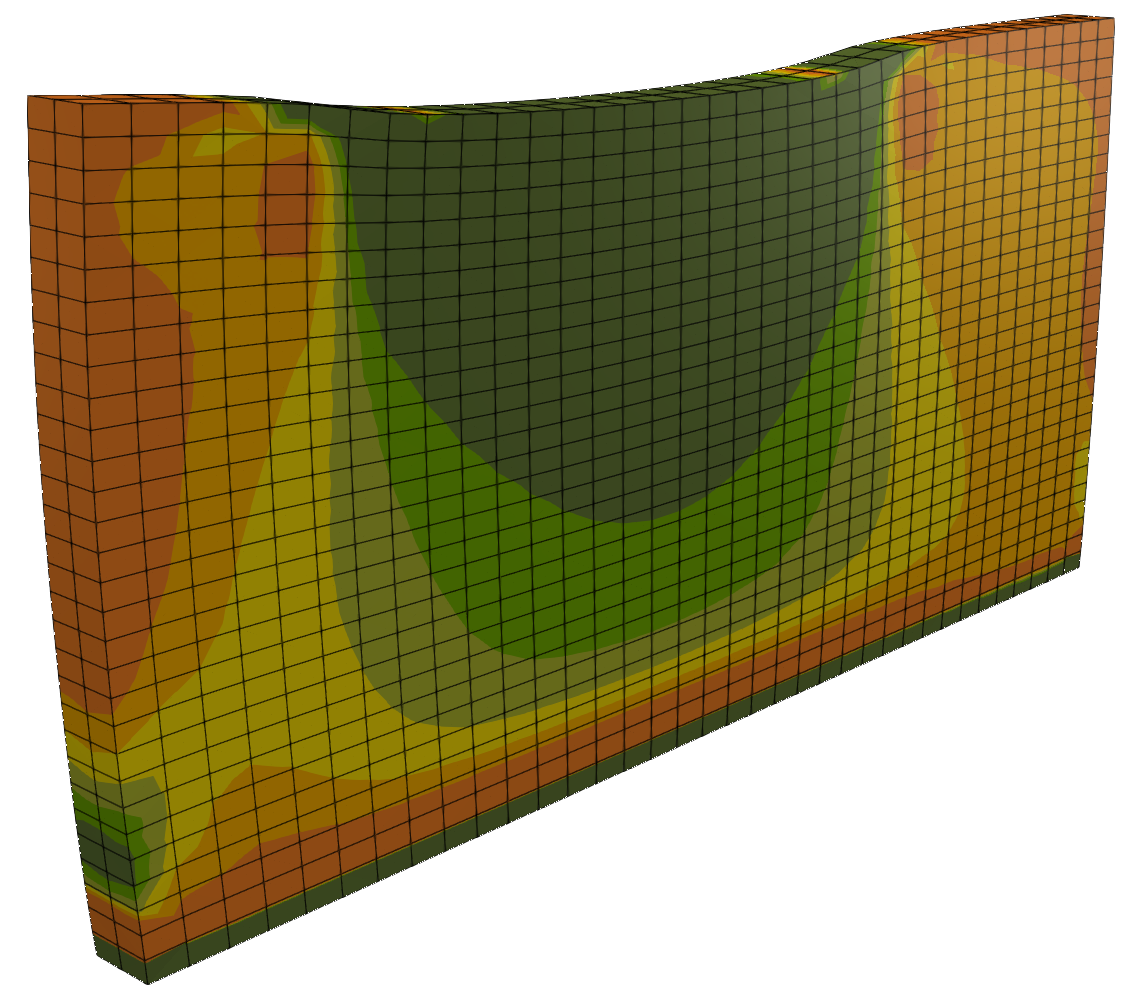}
\includegraphics[scale=0.2]{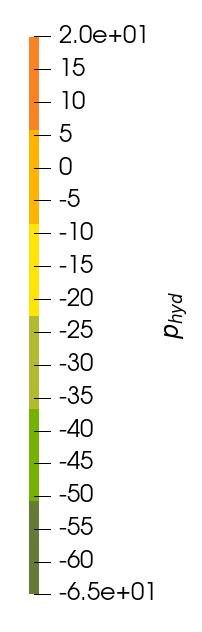}
\caption{$F_{xx}$ and hydrostatic pressure $p_{hyd}$ distribution for the 3-dimensional rectangular block from \autoref{tab:nrconv3d1} and \autoref{tab:nrconv3d2}. All implementations result in identical distribution hence, only $\vec{b}$-based result is shown here.}
\label{fig:compblock3d_nr_fxx_phyd}%
\end{figure}

We then use the (Level 2) mesh and extrude it in the $Z$ direction by unit length and re-perform the same experiment but now in 3-dimensions. Once again, \autoref{tab:nrconv3d1} and \autoref{tab:nrconv3d2} show that, all implementations with spectrally-decomposed tangents achieve quadratic convergence and are in excellent agreement with classic implementations. Once again, in \autoref{fig:compblock3d_nr_fxx_phyd} we plot the distribution of derived quantities namely, the deformation gradient component $F_{xx}$ and the hydrostatic pressure $p_{hyd}$ and verify that these quantities match for all formulations in 3-dimensions too.

\begin{figure}
\centering
\includegraphics[scale=0.16]{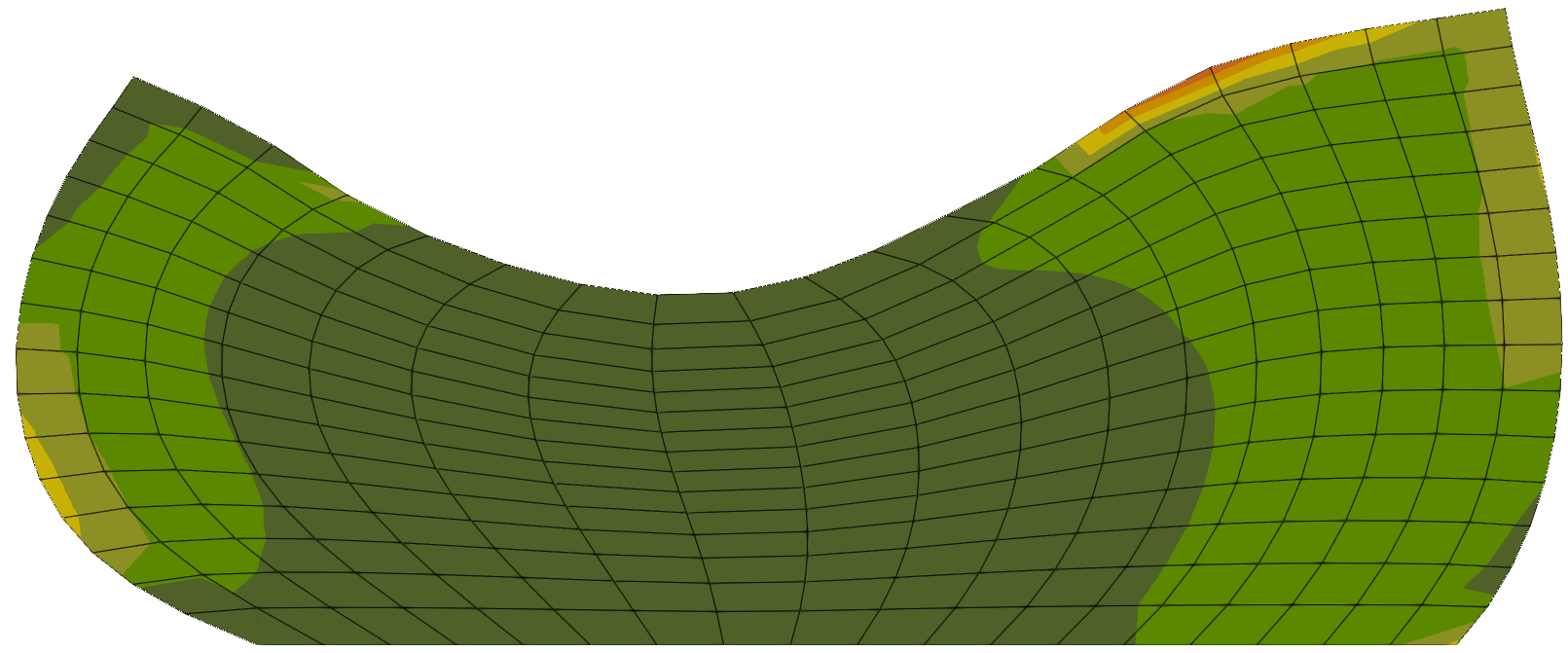}
\includegraphics[scale=0.19]{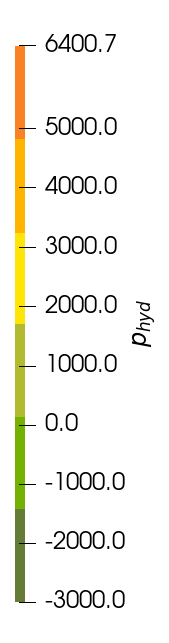}
\includegraphics[scale=0.35]{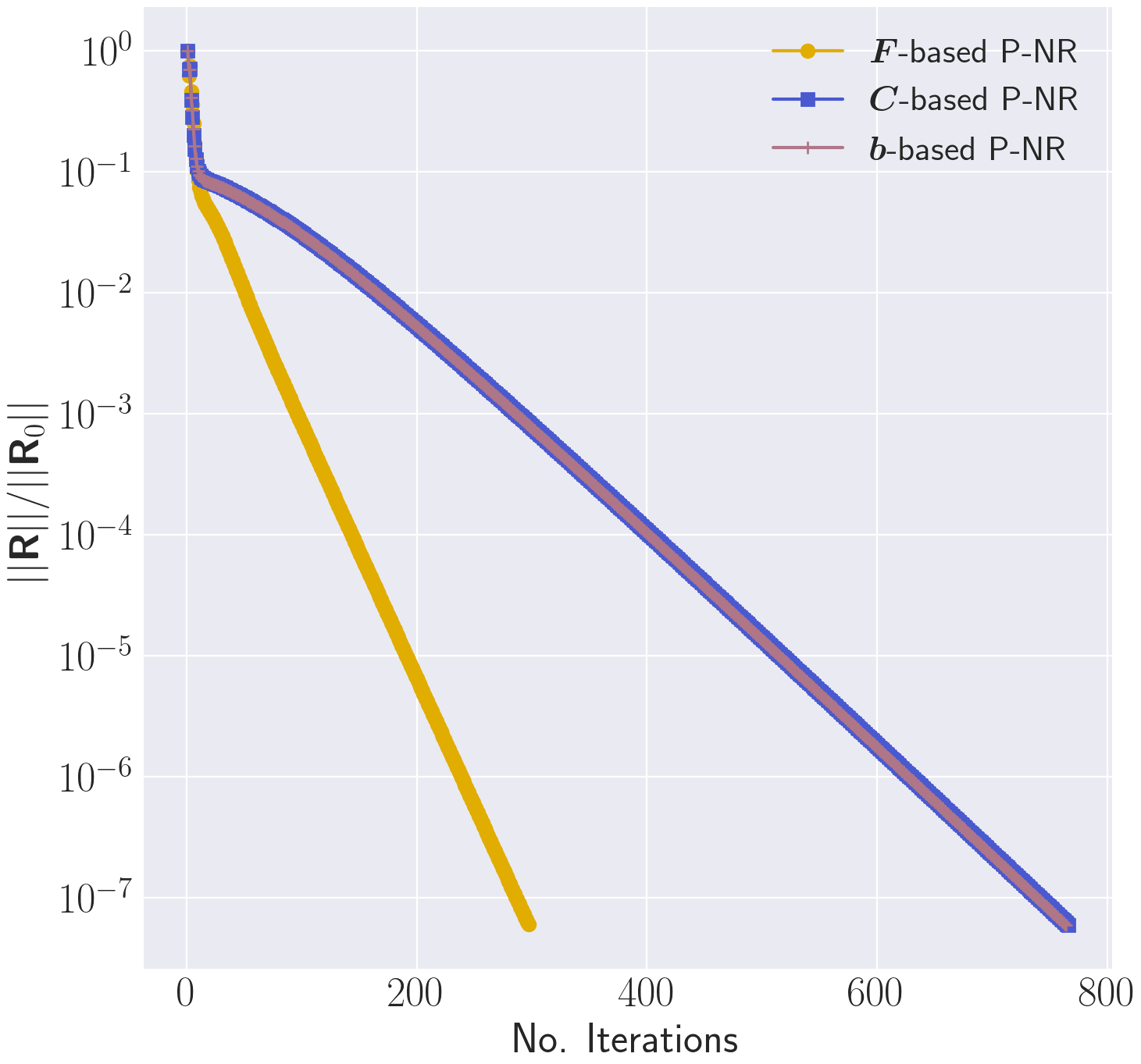}
\caption{Convergence of P-NR with a load factor for which NR with constant load factor fails (NR was tested up to 1000 increments). Figure confirms that, PSD projection does not work the same way between different formulations. In $\vec{C}$- and $\vec{b}$-based formulations constitutive and geometric components need to be projected separately resulting in a less accurate approximation of tangents. This is not the case with $\vec{F}$-based formulation as there the initial and constitutive operators are all ``collapsed" into a single tangent operator. The performance of  $\vec{C}$- and $\vec{b}$-based in general varies significantly depending on which eigen-modes are pruned.}
\label{fig:compblockpnr_convergence}%
\end{figure} %

\begin{figure}
\centering
\includegraphics[scale=1.2]{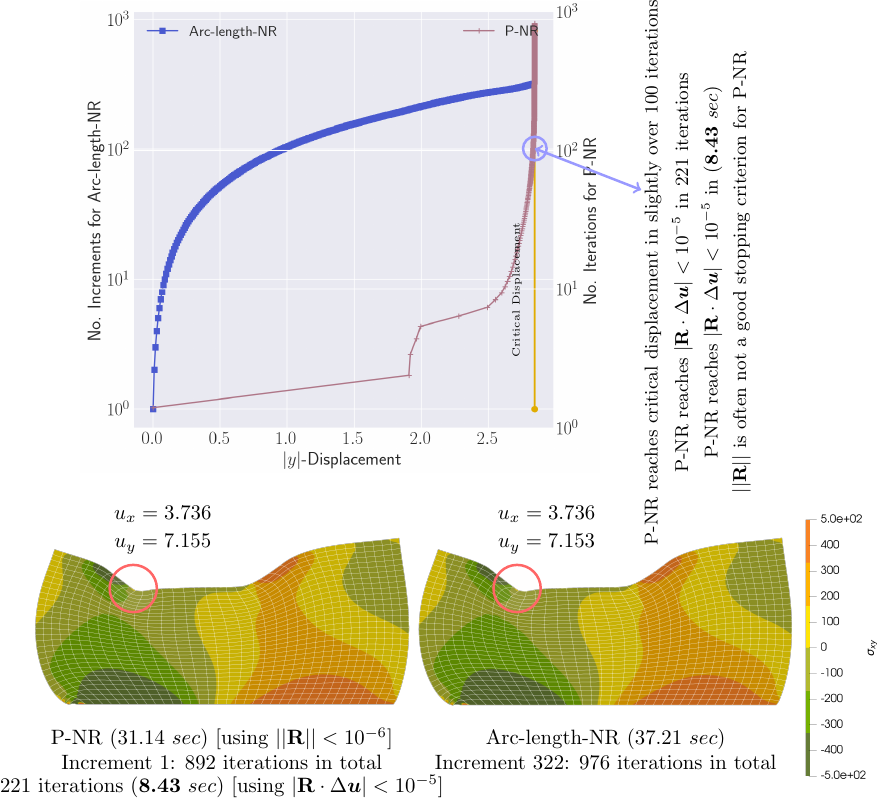}
\caption{\textbf{Side-by-side comparison Arc length-NR and P-NR}: P-NR uses the discretisation-aware line search from \autoref{algo:pnr} and reaches the critical displacement faster by over $\vec{4\times}$. Arc-length was manually tuned to reach the critical load factor in the least number of increments. Both results are in good agreement.}
\label{fig:compblockpnr_al_vs_pn}%
\end{figure} %

\begin{table}
\centering
\resizebox{\columnwidth}{!}{%
\begin{tabular}{| c | c | c | c | c | c | c |}
\hline
Formulation & Iter 1& Iter 2 & Iter 3 & Iter 4 & Iter 5  \\ \hline
$\vec{F}$-based Classic &  1.00000e+00 &  2.78701e-02 &  8.52802e-03 &  2.66989e-05 &  1.75363e-09 \\ \hline
$\vec{F}$-based PS &  1.00000e+00 &  2.78701e-02 &  8.52802e-03 &  2.66989e-05 &  1.74557e-09 \\ \hline
$\vec{C}$-based Classic &  1.00000e+00 &  2.78701e-02 &  8.52802e-03 &  2.66989e-05 &  1.75040e-09 \\ \hline
$\vec{C}$-based PS &  1.00000e+00 &  2.78701e-02 &  8.52802e-03 &  2.63867e-05 &  1.34560e-09 \\ \hline
$\vec{b}$-based Classic &  1.00000e+00 &  2.78701e-02 &  8.52802e-03 &  2.66989e-05 &  1.75855e-09 \\ \hline
$\vec{b}$-based PS &  1.00000e+00 &  2.78701e-02 &  8.52802e-03 &  2.63867e-05 &  1.34560e-09 \\ \hline
\end{tabular}
}
\caption{Quadratic convergence of Newton-Raphson for the first increment for the 3-dimensional inhomogeneous compression of  block using new (spectrally-decomposed tangent) and classic implementations of $\vec{F}$-based, $\vec{C}$-based and $\vec{b}$-based formulations. Results are in solid agreement validating all numerical implementations.}%
\label{tab:nrconv3d1}
\end{table}

\begin{table}
\centering
\resizebox{\columnwidth}{!}{%
\begin{tabular}{| c | c | c | c | c | c | c |}
\hline
Formulation & Iter 1& Iter 2 & Iter 3 & Iter 4 & Iter 5 & Iter 6  \\ \hline
$\vec{F}$-based Classic &  1.00000e+00 &  4.76957e-02 &  2.01982e-02 &  1.96365e-04 &  9.63069e-08 &  2.78094e-11  \\ \hline
$\vec{F}$-based PS &  1.00000e+00 &  4.76957e-02 &  2.01982e-02 &  1.96365e-04 &  9.62955e-08 &  2.86818e-11  \\ \hline
$\vec{C}$-based Classic &  1.00000e+00 &  4.76957e-02 &  2.01982e-02 &  1.96365e-04 &  9.63116e-08 &  3.08771e-11  \\ \hline
$\vec{C}$-based PS &  1.00000e+00 &  4.76957e-02 &  2.01721e-02 &  1.89533e-04 &  7.22232e-08 &  1.61624e-10  \\ \hline
$\vec{b}$-based Classic &  1.00000e+00 &  4.76957e-02 &  2.01982e-02 &  1.96365e-04 &  9.63003e-08 &  2.70922e-11  \\ \hline
$\vec{b}$-based Classic &  1.00000e+00 &  4.76957e-02 &  2.01723e-02 &  1.91047e-04  &  8.68032e-08 &  3.11638e-10   \\ \hline
\end{tabular}
}
\caption{Quadratic convergence of Newton-Raphson for the last (5th) increment for the 3-dimensional inhomogeneous compression of block using new (spectrally-decomposed tangent) and classic implementations of $\vec{F}$-based, $\vec{C}$-based and $\vec{b}$-based formulations. Results are in solid agreement validating all numerical implementations.}%
\label{tab:nrconv3d2}
\end{table}

\subsection{Verification of tangent stabilised elasitcity}
The goal of this example is to:
\begin{enumerate}
\item Verify the result of tangent stabilised elasticity with P-NR with classical NR and arc-length NR
\item Validate the result and accuracy of tangent stabilised elasticity with external finite element suite
\end{enumerate}

While, as shown in the previous example, the spectrally-decomposed tangent constructions achieve identical convergence to classic implementations, they come with an added benefit of analytically and easily being stabilised. Hence, we now shift the focus to clearly showcase the convergence characteristics of different formulations with P-NR using a single increment. To this end, we use the already described rectangular block example under compression with (Level 1) mesh and increase the loading enough for standard NR (with constant load factor) to fail all the way up to 1000 increments and instead use P-NR to perform the simulation. \autoref{fig:compblockpnr_convergence} shows the deformed configuration and distribution of $F_{xx}$ obtained with P-NR. The convergence of P-NR is shown in \autoref{fig:compblockpnr_convergence} which as can be observed is certainly not quadratic. However, it is worth noting that, standard NR fails for this case irrespective of number of load increments hence, this is more of a stress test for P-NR. If load increments and P-NR are combined the performance is certainly better as shown in \cite{Poya2023b}. Interestingly, as shown earlier, while spectrally-decomposed tangent constructions for all formulations produce the same convergence patterns the situation is not the same when tangents are stabilised. The answer to this is straightforward. In $\vec{C}$- and $\vec{b}$-based formulations constitutive and geometric components need to be projected separately resulting in a more-than-necessary stabilisation and hence, lesser accurate approximation of tangents. This is not the case with $\vec{F}$-based formulation, as there, the initial and constitutive tangents are all ``collapsed" into one resulting in smaller stabilisation. The performance of $\vec{C}$- and $\vec{b}$-based in general varies significantly from example to example depending on the loading, geometry and material models and depending on which eigen-modes are pruned. The $\vec{F}$-based P-NR on the other hand, has more of a uniform convergence pattern and remains the most competitive.

As noted in \cite{Bieber2023a}, the localised failure profile is only obtained with Q2 elements for this problem. To this end, we use the (Level 2) mesh and increase the finite element polynomial degree to 2 and redo the experiment with both arc length NR and P-NR. \autoref{fig:compblockpnr_al_vs_pn} shows a mild kink (highlighted in red circles) using both arc length NR and P-NR. The result of arc length NR and P-NR are in good agreements. Arc length-NR ultimately reaches a load factor of $\lambda \approx 1.1$ similar to \cite{Bieber2023a} which is then used as a total load for P-NR. It can be observed that P-NR has a much faster convergence towards the solution through the course of iterations and reaches the expected result within a crude accuracy in slightly over a 100 iterations. For a fixed residual tolerance of $10^{-6}$ P-NR was still slightly faster than arc length (both over 30 seconds). Alternatively, if a curvature based stopping criterion is used as shown in Algorithm \autoref{algo:pnr}, 221 iterations of P-NR is sufficient to reach a tolerance of $10^{-5}$, making it $4.4\times$ faster than the arc-length-NR. The authors experience suggest that, for P-NR style algorithms residual based tolerances are often too tight implying that the stopping criterion of standard NR must be revisited for P-NR since through the course of iterations displacements drop rapidly. This makes curvature based stopping criterion more suited for P-NR. This example uses the discretisation-aware line search and the $\vec{F}$-based formulation. The arc-length method uses a cylindrical constraint with secant path correction \cite{FENG199543,FENG1996479}. Arc-length was repeatedly tuned (with different user-defined starting load factor and number of increments) to reach the critical displacement in the least number of increments. P-NR needed no tuning.

\begin{figure}
\centering
\includegraphics[scale=0.8]{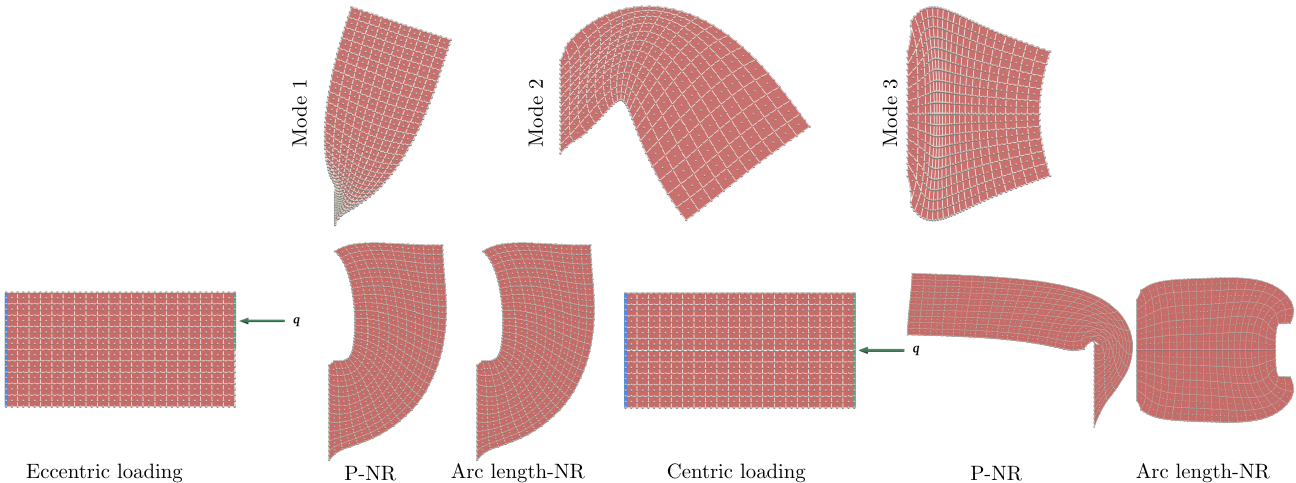}
\caption{Dominant buckling modes (top) and quasi-static simulations (bottom) of bar clamped on the left and under the action of centric (bottom left) and eccentric (bottom right) loading. Under eccentric loading, both P-NR and arc length-NR accurately capture the dominant buckling mode. Under centric loading, arc length-NR produces a different profile that corresponds to the third dominant mode of the bar, while P-NR captures the same dominant buckling mode by effectively eliminating bifurcation, resulting in a more deterministic simulation.}
\label{fig:verif_PNR}
\end{figure}

The examples presented above demonstrate that P-NR captures the same deformation pattern as NR and arc length-NR with strong agreement. We now consider a simple example to emphasise the correctness of tangent stabilised elasticity in accurately capturing stable configurations. \autoref{fig:verif_PNR} shows example of a bar with aspect ratio 2:1 clamped on the left side, discretised with Q2 mesh under two loading scenarios namely, centric, and eccentric loading. The first three deformation modes are first obtained by performing linear modal analysis by evaluating the previously described nearly-incompressible Mooney-Rivlin energy (with $\mu=1000$ and $\nu = 0.49$) at the origin. These deformation modes are shown in \autoref{fig:verif_PNR}(a,b,c). When subjected to eccentric loading, both P-NR and arc length-NR accurately capture the well-known dominant buckling mode. However, under centric loading, arc length-NR produces a different profile that corresponds to the third dominant mode of the bar, while P-NR achieves the same buckling profile. With NR, in the presence of multiple potential solutions, the bar may randomly buckle in one direction or remain compressed in the central region. P-NR, on the other hand, effectively eliminates such bifurcations through stabilisation, resulting in a more deterministic simulation that captures the most dominant mode. This stabilising effect was also observed in a study by \citet{Poya2023b} (refer to Figure 31), where P-NR consistently yielded the same buckling profile regardless of refinements in h (and p) and variations in finite element formulations (displacement vs mixed), whereas NR produced different profiles.

We finally consider a standard benchmark problem recommended by the National Agency for Finite Element Methods and Standards (NAFEMS) \cite{nafems1990} namely the $Z$-section cantilever under torsional loading shown in \autoref{fig:zsection_impdisp}. We perform this to validate the results of the new formulations with other established finite element suites. The cantilever is 10$m$ long, the flanges are 1$m$ wide, the web 2$m$ wide and the thickness is uniform, 0.1$m$. The benchmark entails subjecting the structure to torsional moment of 2MN$m$, generated by the two forces uniformly distributed across the flanges, $S = 0.6$MN; see specification in \cite{Krysl2022a}. The Young's modulus is taken as 210GPa, and the Poisson ratio is 0.3. The target quantity is the mid-surface axial stress ($\sigma_{xx}$) at point A (at the edge of a flange, 2.5 m from the clamped end) of value - 108MPa (compressive) \cite{nafems1990}. After calibrating material parameters, using the material model in \eqref{eq:amips_energy3d} using Q2 hexahedral elements with 20 elements along the length, 10 across the flange and 1 across the thickness, we obtain $\sigma_{xx}^A = -107.113$MPa (for all formulations), closely matching the reference value and hence validating the implementation and accuracy of tangent stabilised elasticity. Although not shown, we further refined the mesh with 100 elements along the length (keeping the refinement fixed in other directions) and compared this value with Abaqus results reported in \cite{Krysl2022a}. We obtain, a value of $\sigma_{xx}^A = -111.048$MPa vs Abaqus's shear flexible quadratic element (S8R) $\sigma_{xx}^A = -110.767$MPa and shear rigid quadratic element (S8R5) $\sigma_{xx}^A = -111.103$MPa further validating our implementations.

\subsection{Automatic discretisation-safe load stepping}
\begin{figure}
\centering
\includegraphics[scale=0.09]{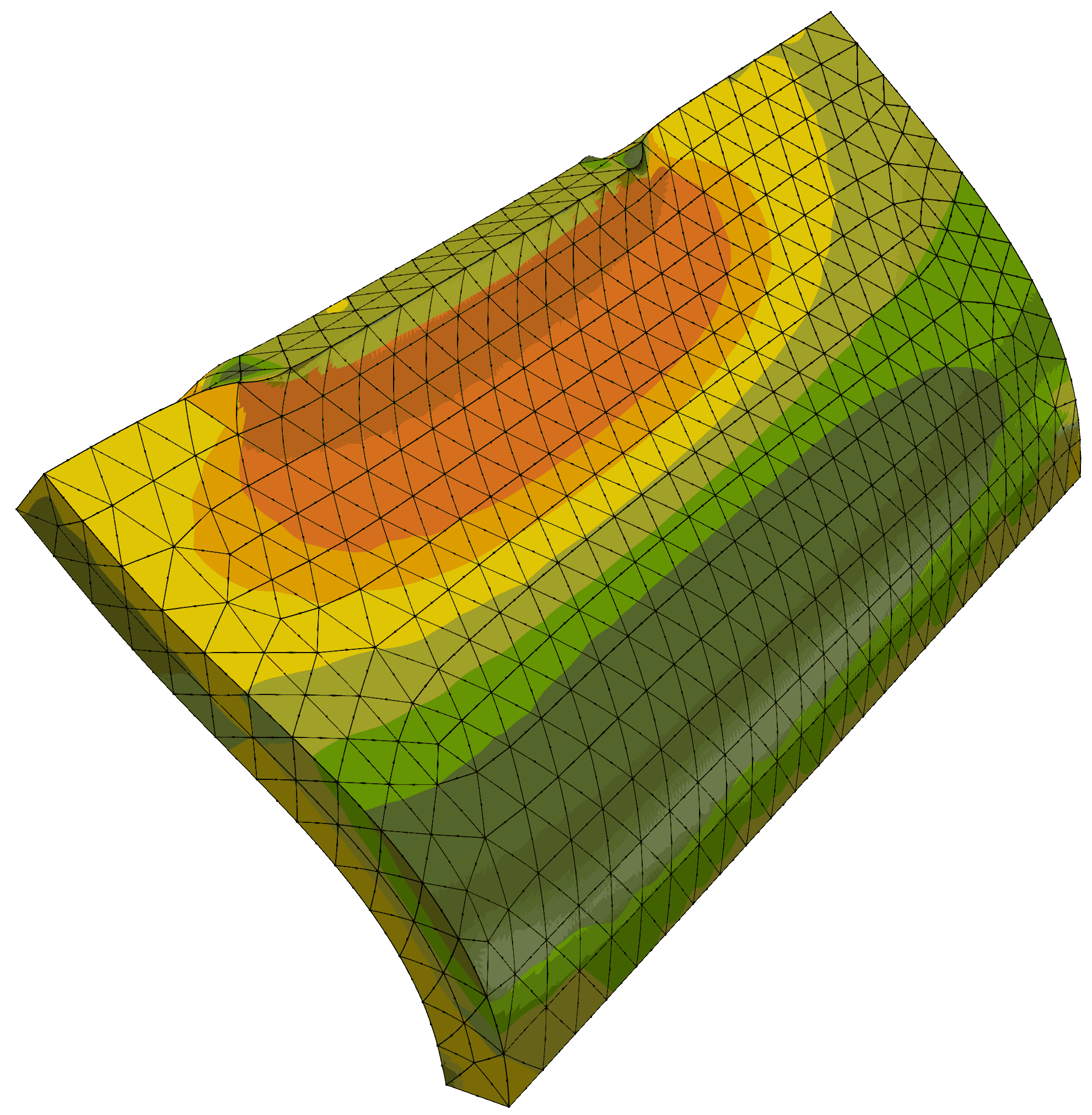}
\includegraphics[scale=0.09]{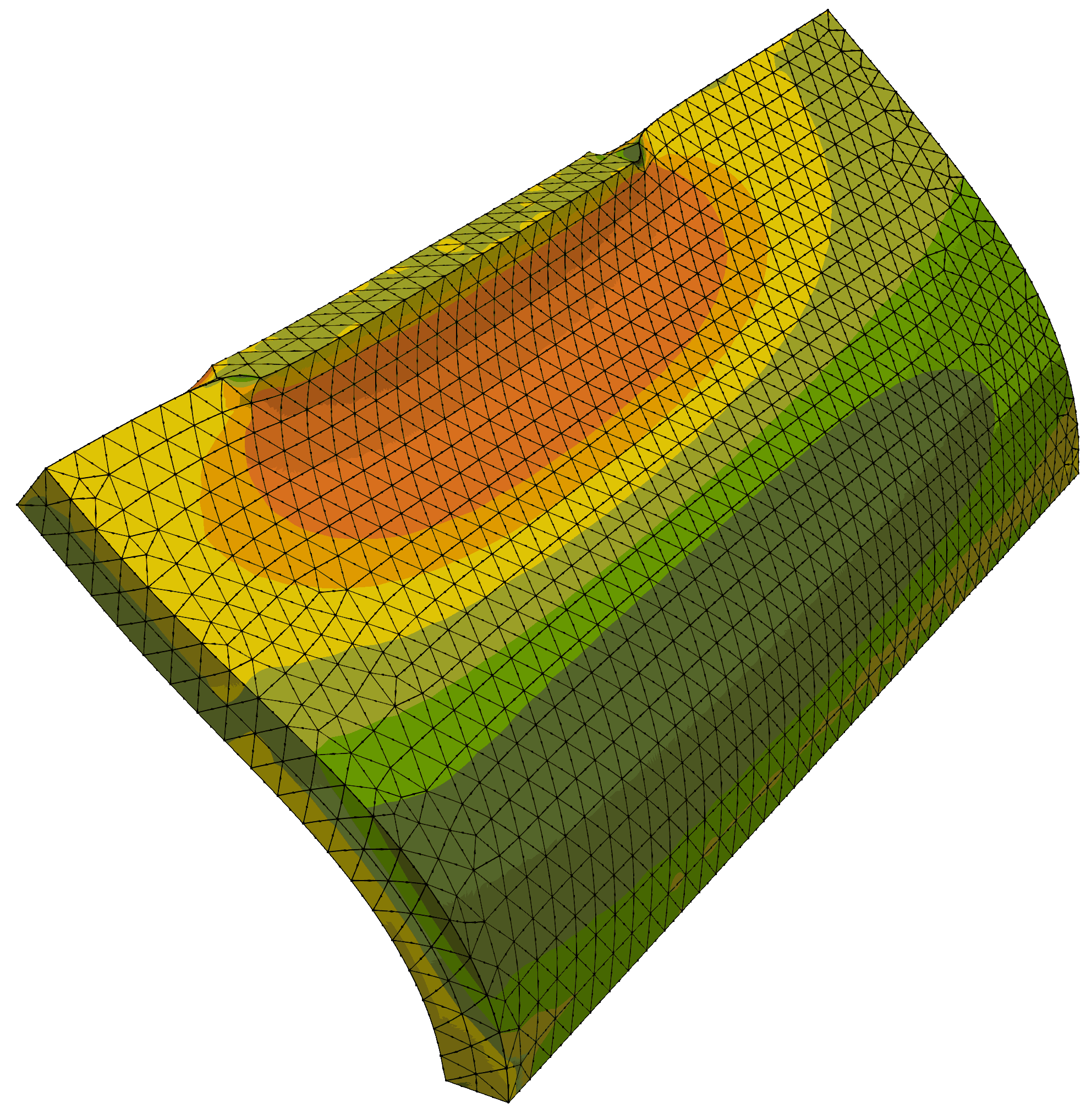}
\includegraphics[scale=0.14]{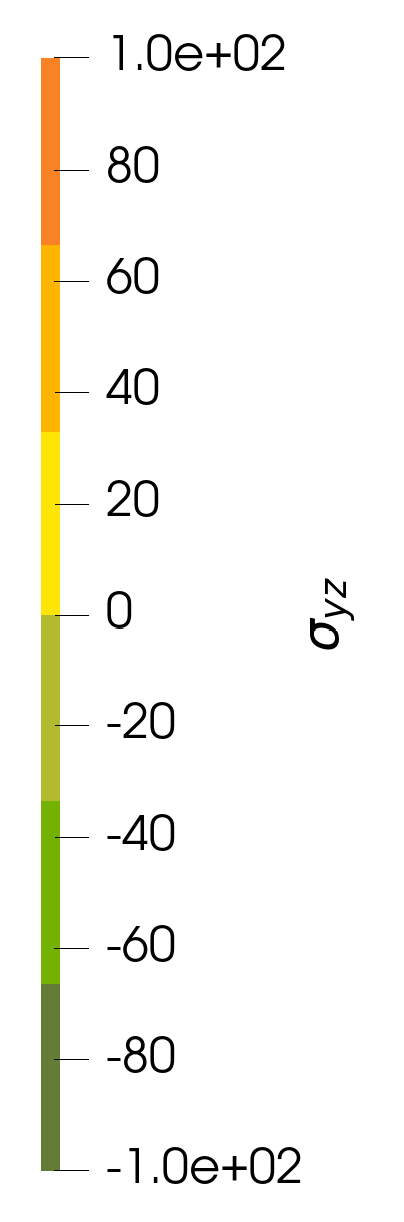} \\
\includegraphics[scale=0.4]{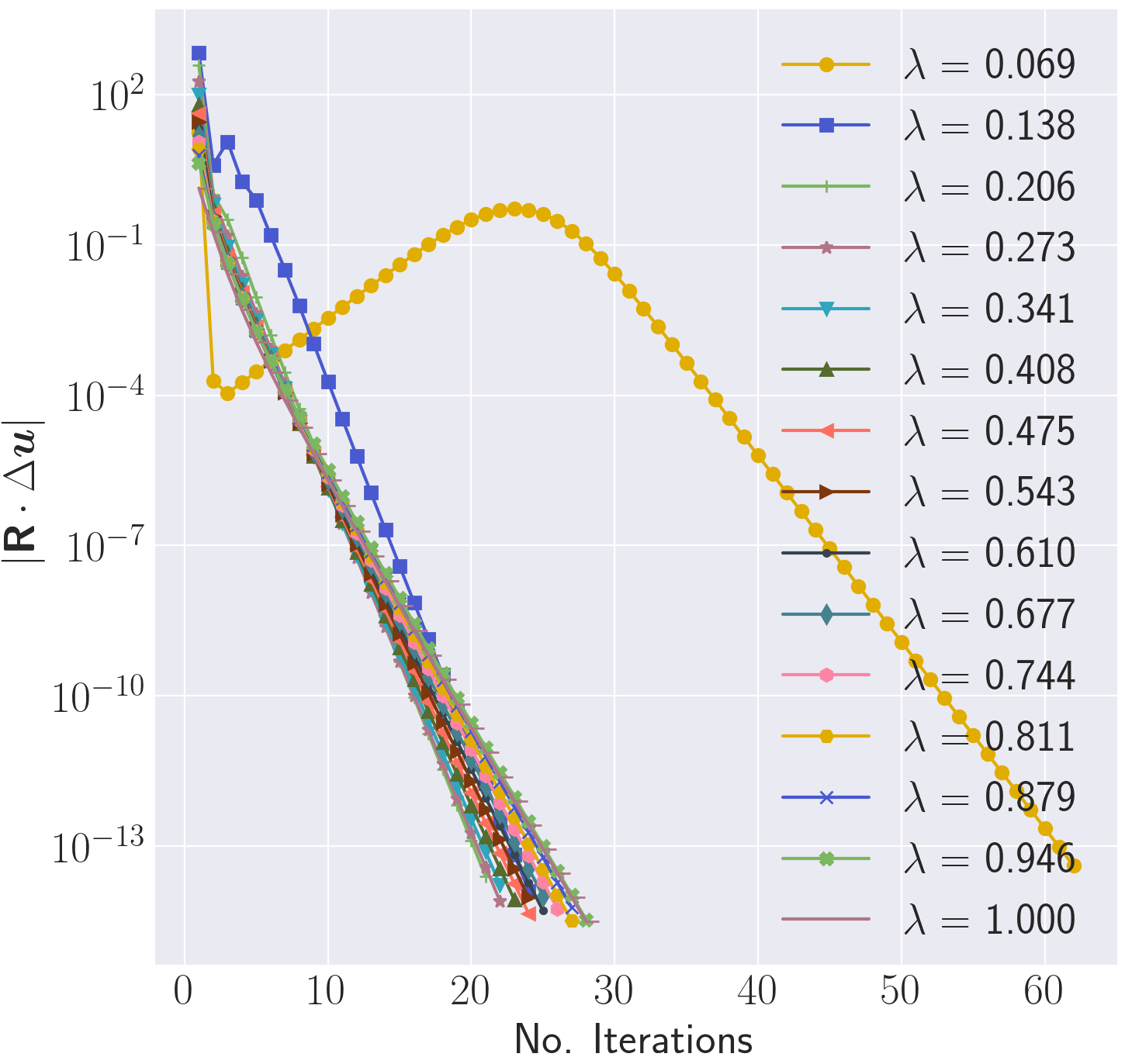}
\includegraphics[scale=0.4]{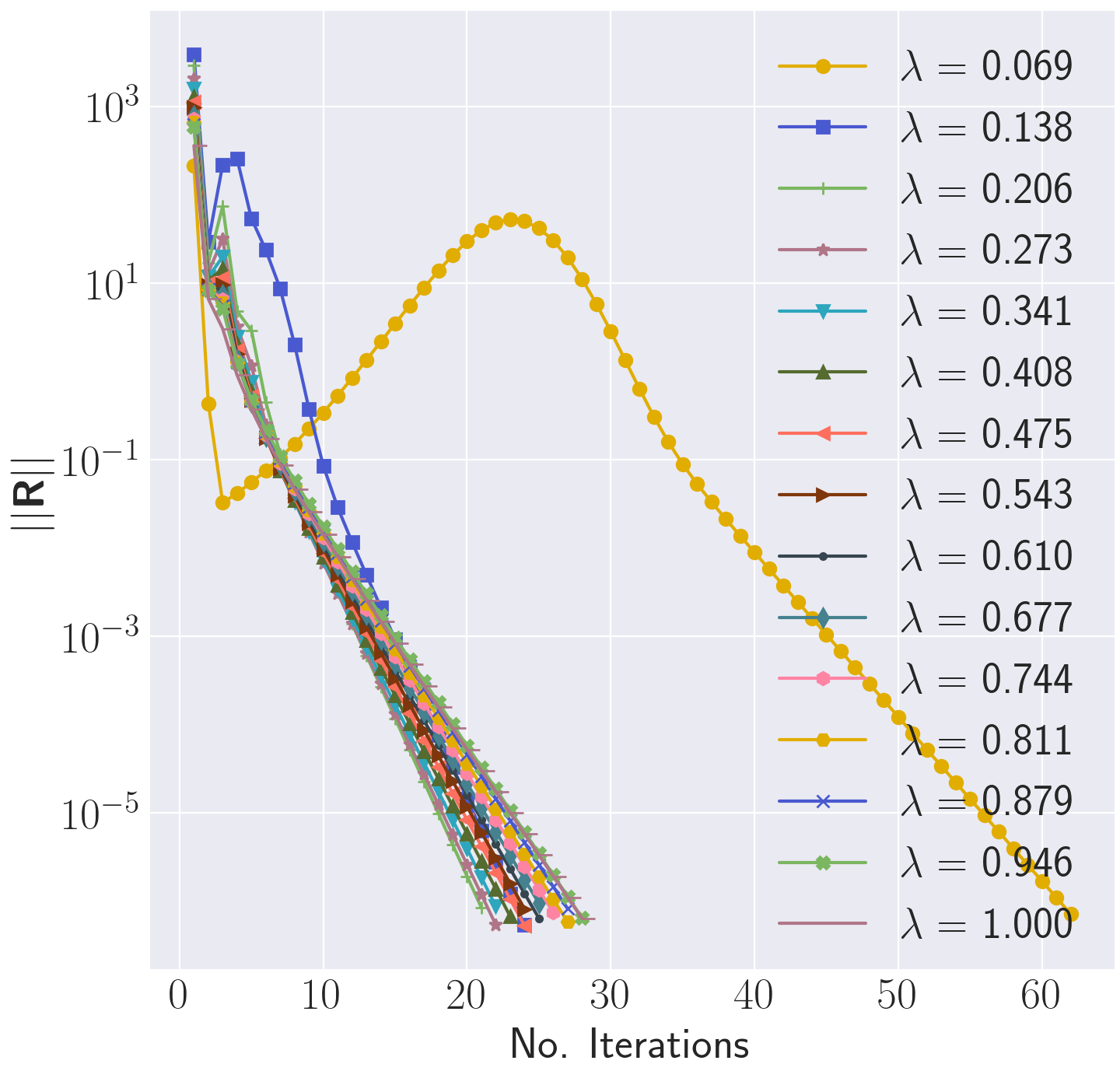} \\
\includegraphics[scale=0.4]{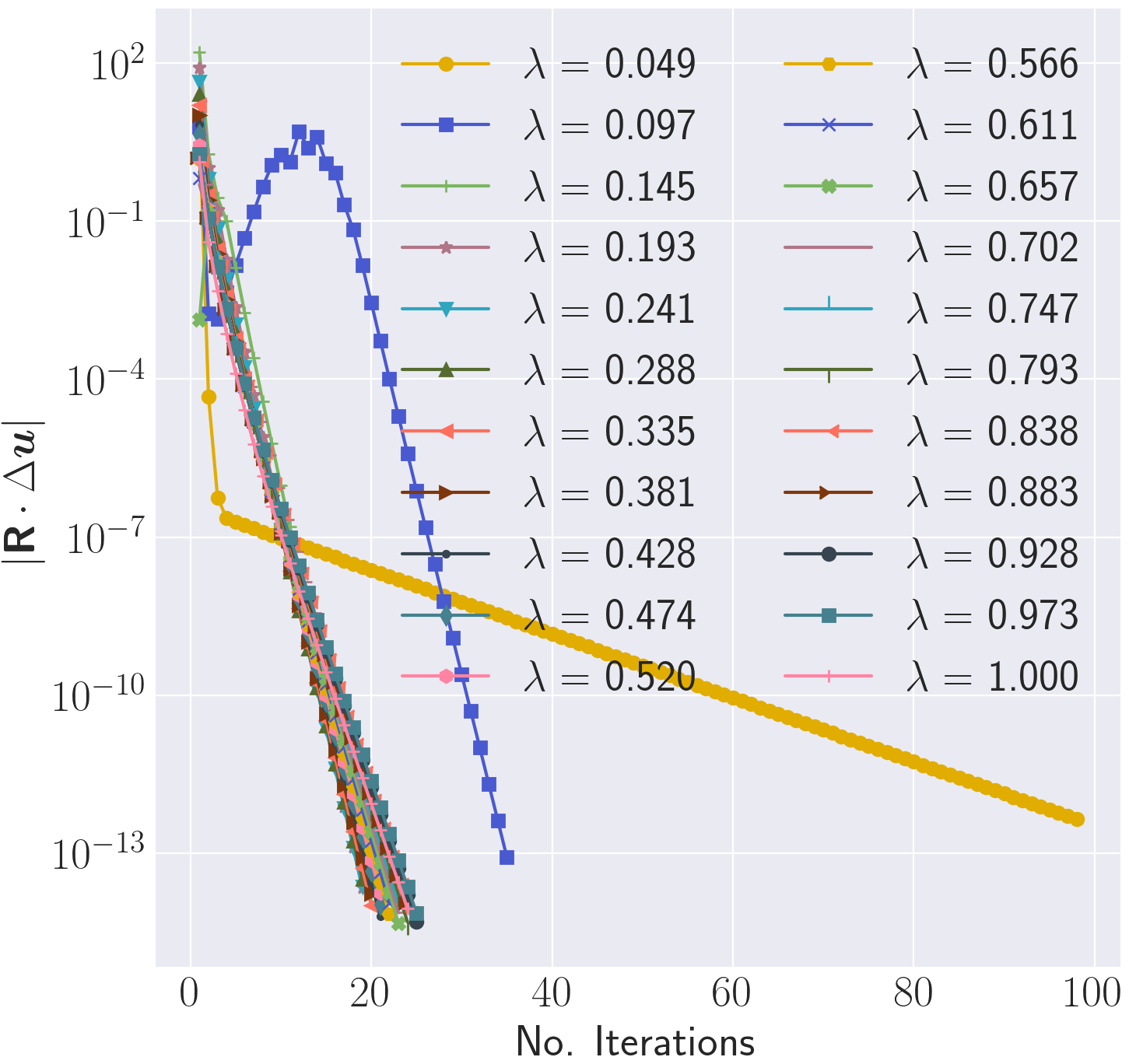}
\includegraphics[scale=0.4]{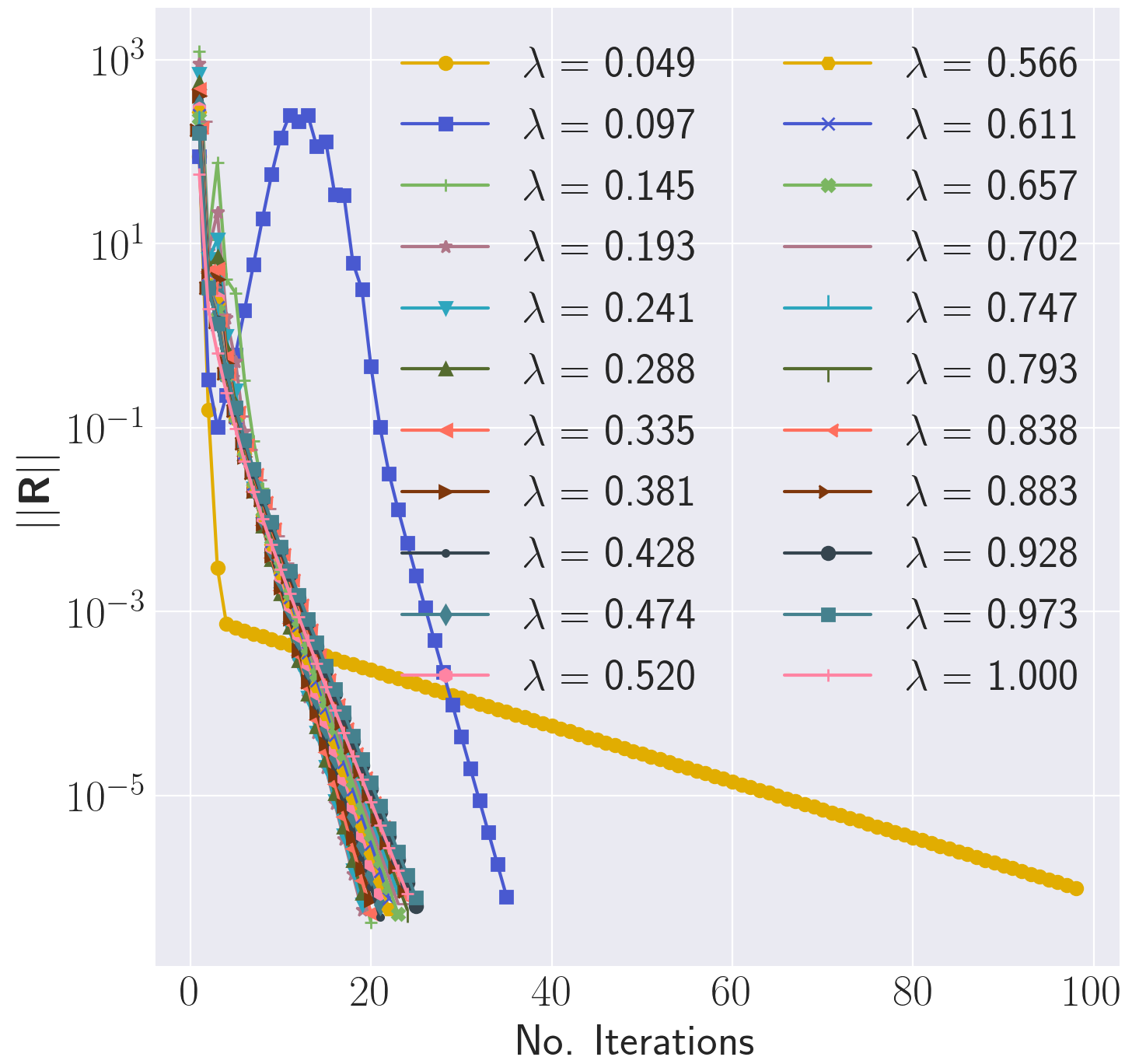}
\caption{\textbf{Automatic discretisation-aware load stepping with P-NR}: A block (with two $p=2$ tetrahedral meshes) is pressed with imposed displacements applied in full that creates element inversion. Discretisation-aware load stepping, described in \ref{algo:pnr} and \ref{algo:safeload}, ensures that the loading is automatically adjusted to maximum inversion-safe displacements per increment. Further, load incrementation improves the convergence of P-NR on a per-increment basis. Two bottom rows show per-increment convergence (with accumulative load factor $\lambda$ in legends) of curvature $|\textbf{R} \cdot \Delta \vec{u}|$, and residual $||\textbf{R}||$ for coarse and fine meshes, respectively. Curvature drops rapidly within 15 iterations.}
\label{fig:compblock_impdisp}
\end{figure}

The goal of this example is to showcase the robustness of the discretisation-aware load stepping and line search from Algorithms \ref{algo:pnr}, and \ref{algo:safeload}. For this, we consider the case of displacement-driven loading. \autoref{fig:compblock_impdisp} shows the same rectangular block considered early but now with two $p=2$ tetrahedral meshes that is pressed with imposed displacements applied in one increment. Such excess displacement is too large to be applied in one increment and causes element inversion near the boundary. As can be seen, the proposed discretisation-aware load stepping, described in Algorithm \ref{algo:pnr} and \ref{algo:safeload}, ensures that the loading is automatically adjusted to maximum inversion-safe displacements per increment. Further, it is noticeable that automatic load incrementation also improves the convergence of P-NR on a per-increment basis. From the bottom rows in \autoref{fig:compblock_impdisp} we can observe once again that, curvature drops much faster compared to pure residual since displacements disappear. An engineering accuracy is reached in all increments within 15 iterations. Top row in \autoref{fig:compblock_impdisp} shows distribution of Cauchy stress $\sigma_{yz}$ for both refinements. 

\begin{figure}
\centering
\includegraphics[scale=0.12]{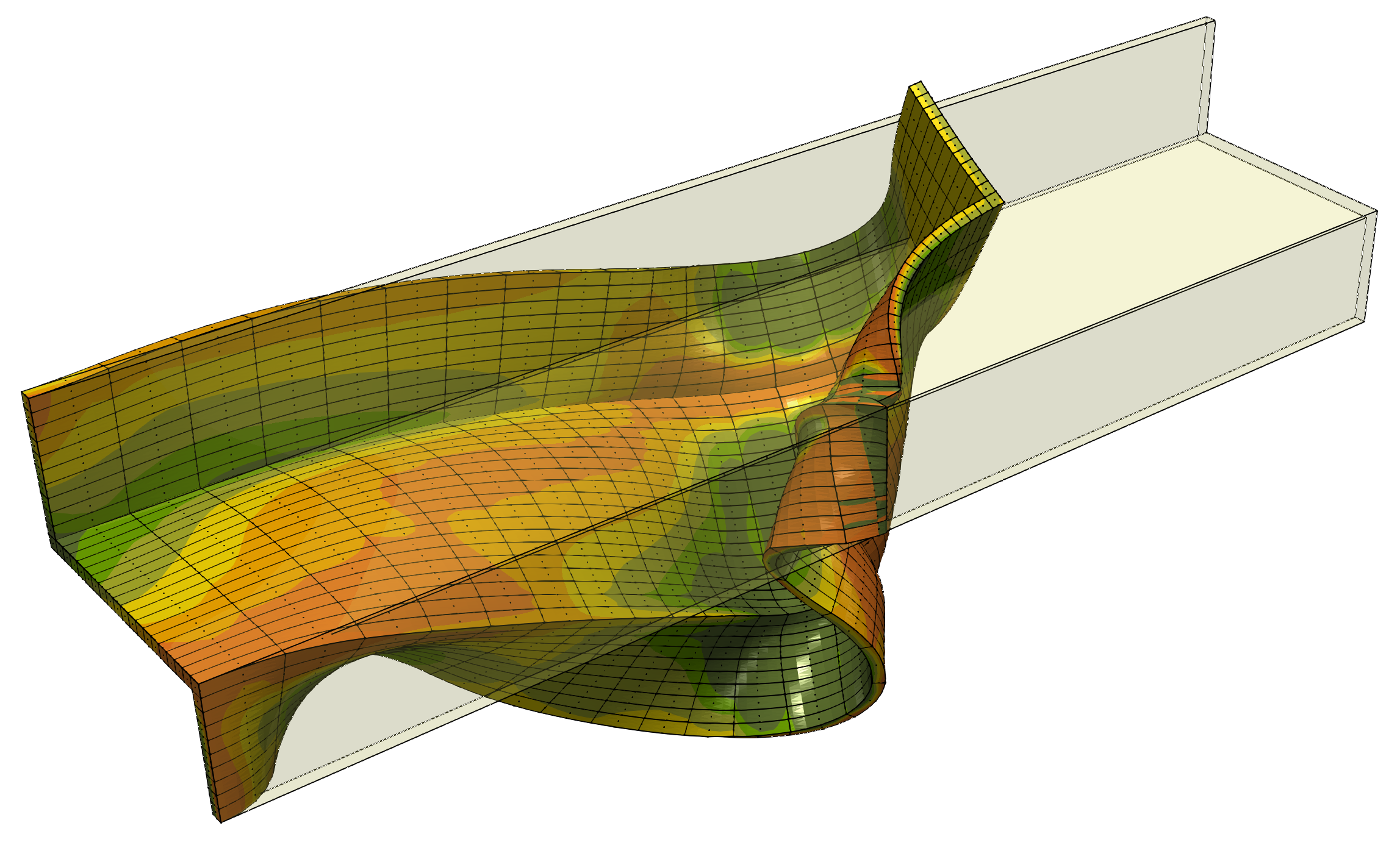}
\includegraphics[scale=0.12]{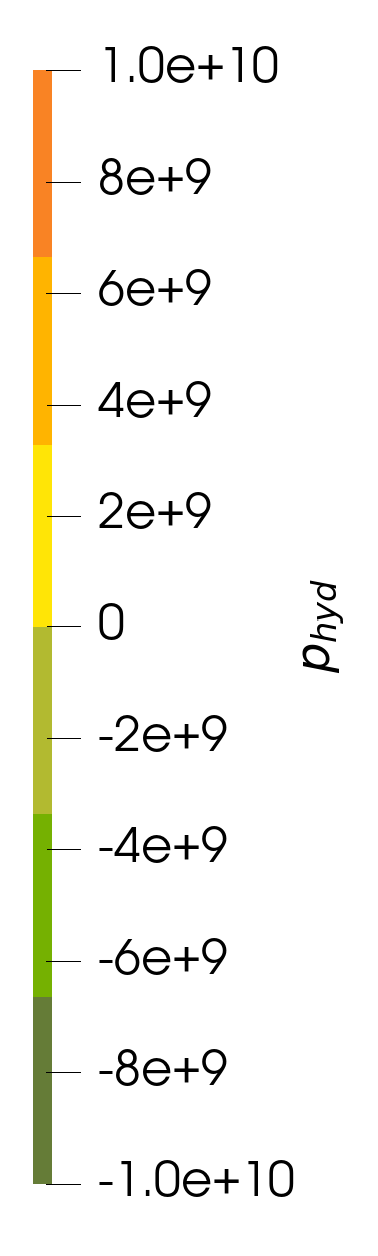}
\includegraphics[scale=0.32]{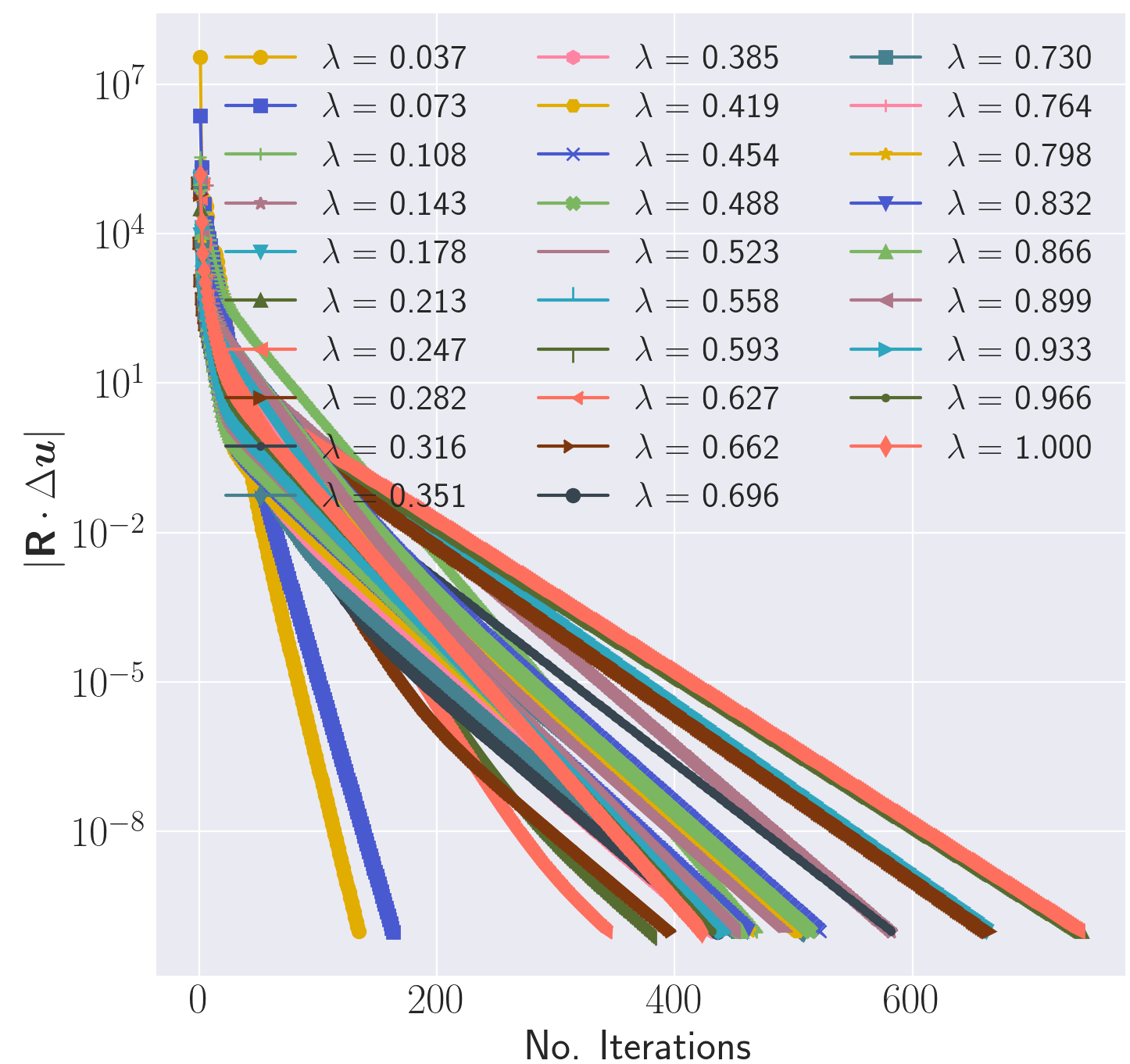}
\caption{\textbf{Guaranteed inversion-free deformation} of a Z-section cantilever through imposed displacements on the lower flange by half the section length using tangent stabilised $\vec{C}$-based formulation. The level of displacements causes structural collapse and hence element inversion during simulation when imposed in one increment. Discretisation-aware load stepping, described in \ref{algo:pnr} and \ref{algo:safeload}, ensures that the loading is automatically adjusted to maximum inversion-safe displacements per increment. Despite the level of deformation and high element aspect ratio, the final result is inversion-free Q2 elements i.e. $J>0$ when sampled at 27, 64 and 125 quadrature point. The right figure shows the maximum load factors determined by the algorithm and their corresponding convergence.}
\label{fig:zsection_impdisp}
\end{figure}

To showcase the robustness of discretisation-aware load stepping, we finally, consider an extreme hypothetical loading scenario where the NAFEM $Z$-section cantilever is pressed to half its length via displacements, imposed on the lower flange. This is too extreme of a displacement which causes buckling, structural collapse and element inversion if the load increments are not sufficiently small. As can be seen in \autoref{fig:zsection_impdisp}, the discretisation-aware load stepping, described in Algorithm \ref{algo:pnr} and \ref{algo:safeload}, ensures that the loading is automatically adjusted to maximum inversion-safe displacements per increment. The final result is inversion-free Q2 elements i.e. $J>0$ when sampled with 27, 64, 125 point quadrature rules for hexahedral elements. The right plot in \autoref{fig:zsection_impdisp} shows the per-increment convergence of P-NR for all load factors determined by the algorithm. Convergence plots in \autoref{fig:zsection_impdisp} was obtained with $\vec{C}$-based formulation.

\subsection{Comparison of arc length with P-NR under extreme loading scenarios and structural collapse}
\begin{figure}
\centering
\includegraphics[scale=1.2]{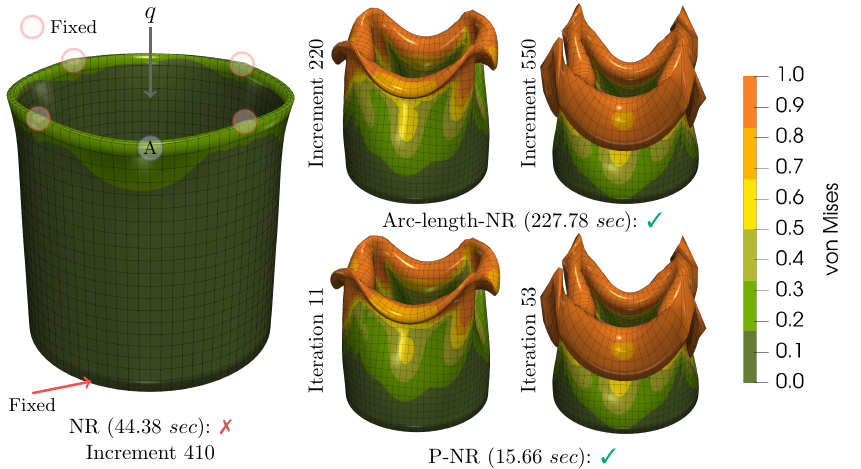}
\caption{Crash simulation of a cylindrical column using different methods: Newton-Raphson [NR] (1000 increments), Arc length based Newton-Raphson [Arc length NR] (550 increments: 1655 iterations in total), and Tangent stabilised/Projected Newton-Raphson [P-NR] (1 increment: 53 iterations in total). Standard NR failed regardless of increments, while Arc length NR was 14.5 times slower than P-NR. Arc length NR required parameter adjustments to work correctly. P-NR used the load factor achieved by Arc length NR. Similar final results were obtained for both methods.}
\label{fig:cylcrash}
\end{figure}

\begin{figure}
\centering
\includegraphics[scale=0.42]{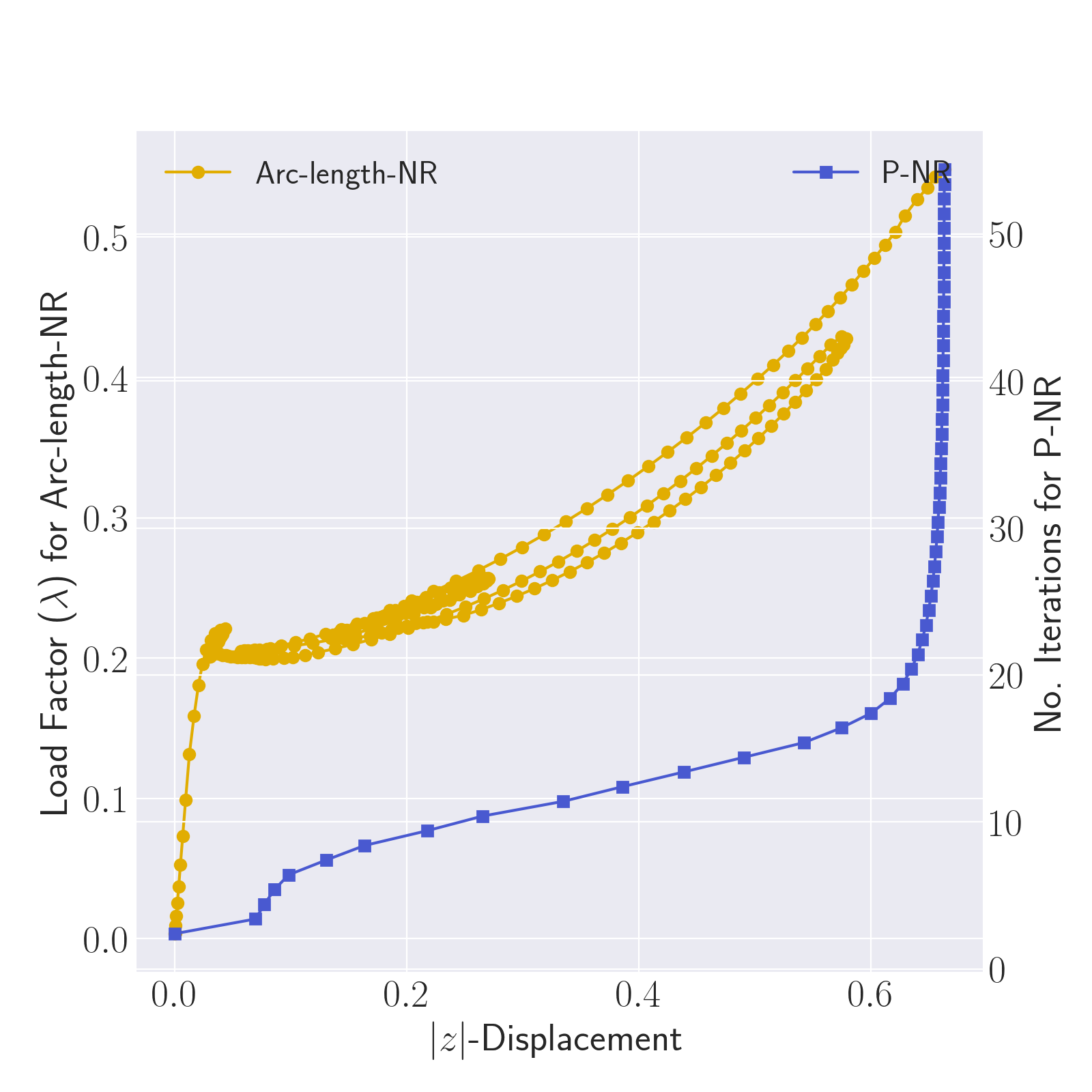}
\includegraphics[scale=0.42]{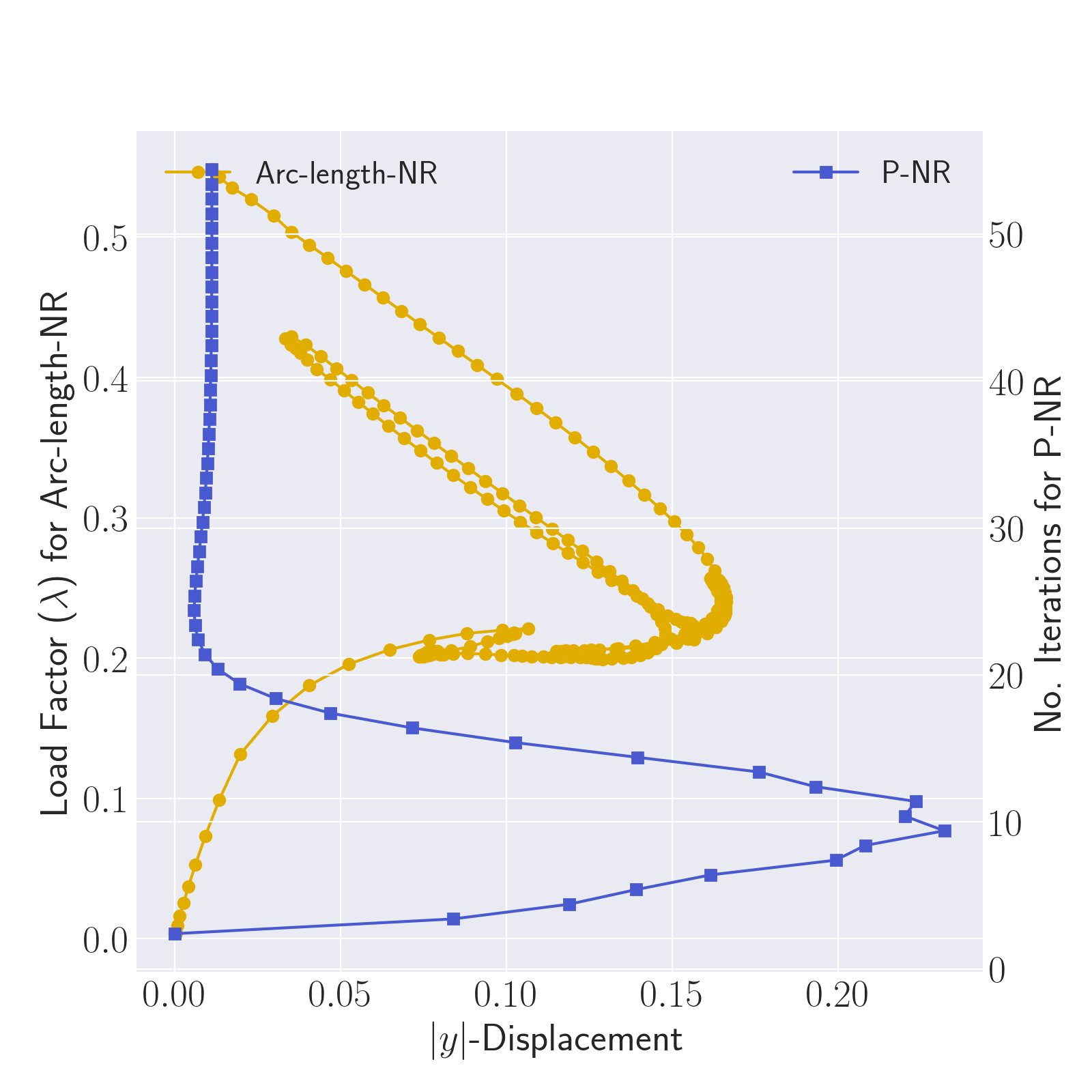}
\caption{Crash simulation of a cylindrical column from \autoref{fig:cylcrash}: Load-deflection curve for point $A$ in \autoref{fig:cylcrash} for arc length vs convergence of $\vec{F}$-based P-NR with iterations. The final $y$ and $z$ displacements for point $A$ obtained with P-NR match well with arc-length NR.}
\label{fig:cylcrashconv_al_vs_pn}
\end{figure}

The goal of this example is to:
\begin{enumerate}
\item Perform a thorough comparison of P-NR with the arc length method in 3-dimensional settings and under extreme loading scenarios and highlight the differences in performance and characteristics of the two approaches
\item Discuss performance of different tangent stabilisation techniques including geometrically stabilised polyconvexity and their merit/usefulness in solving such extreme problems
\end{enumerate}
To this end, we choose a cylindrical shell under compression as shown in \autoref{fig:cylcrash}. The base and four symmetrical points at the top of cylinder are fixed. A nominal load of $q=-0.224$ is applied to the rest of the nodes on the top face and a load controlled arc length is used to monitor the percentage of applied load that the structure can take. The final load factor achieved $\lambda \approx 0.56$ is then given to P-NR with a single increment. The same nearly incompressible Mooney-Rivlin material with reference parameters (shear modulus $\mu=1$ and Poisson's ratio $\nu = 0.45$) is used to keep the rest of the setting simplified for comparisons. Further, for the purpose of demonstrating differences between the arc-length-NR and P-NR it suffices to choose a Q1 hexahedral mesh as the same concepts hold for higher order finite elements too.
\autoref{fig:cylcrash} shows the final deformed configuration obtained with NR with constant load factor, NR with arc-length and NR with tangent stabilised elasticity (P-NR) when the cylinder is crashed to the failure point. This example illustrates a few key points. For this example, NR with constant load factor fails regardless of the number of increments. Arc-length NR succeeds but requires a few attempts to get its parameters (i.e. the starting arc-length and number of increments) tuned correctly but even with best effort it is still \textbf{14.5}$\vec{\times}$ slower than P-NR. The load-deflection curves at point $A$ in \autoref{fig:cylcrash} obtained with arc-length NR is shown in \autoref{fig:cylcrashconv_al_vs_pn} for both $y$ and $z$ directions. On the same plots, the displacement vs iteration count of P-NR is shown where the final displacement of point $A$ agrees well with the arc-length method. The arc-length-NR employed here uses a cylindrical constraint with secant path correction and the key to its success is often in getting the parameters correct, as has been well-known in the literature \cite{FENG199543,FENG1996479,DESOUZANETO199981}. This certainly illustrates the lack of automation inherent in arc length based approaches. On the other hand, for this example, P-NR succeeds with a single load increment and 53 iterations while giving similar result to arc-length NR. Furthermore, it requires no tuning and extra parameter input. For many applications mentioned earlier such automation remains key.

\begin{figure}
\centering
\includegraphics[scale=0.42]{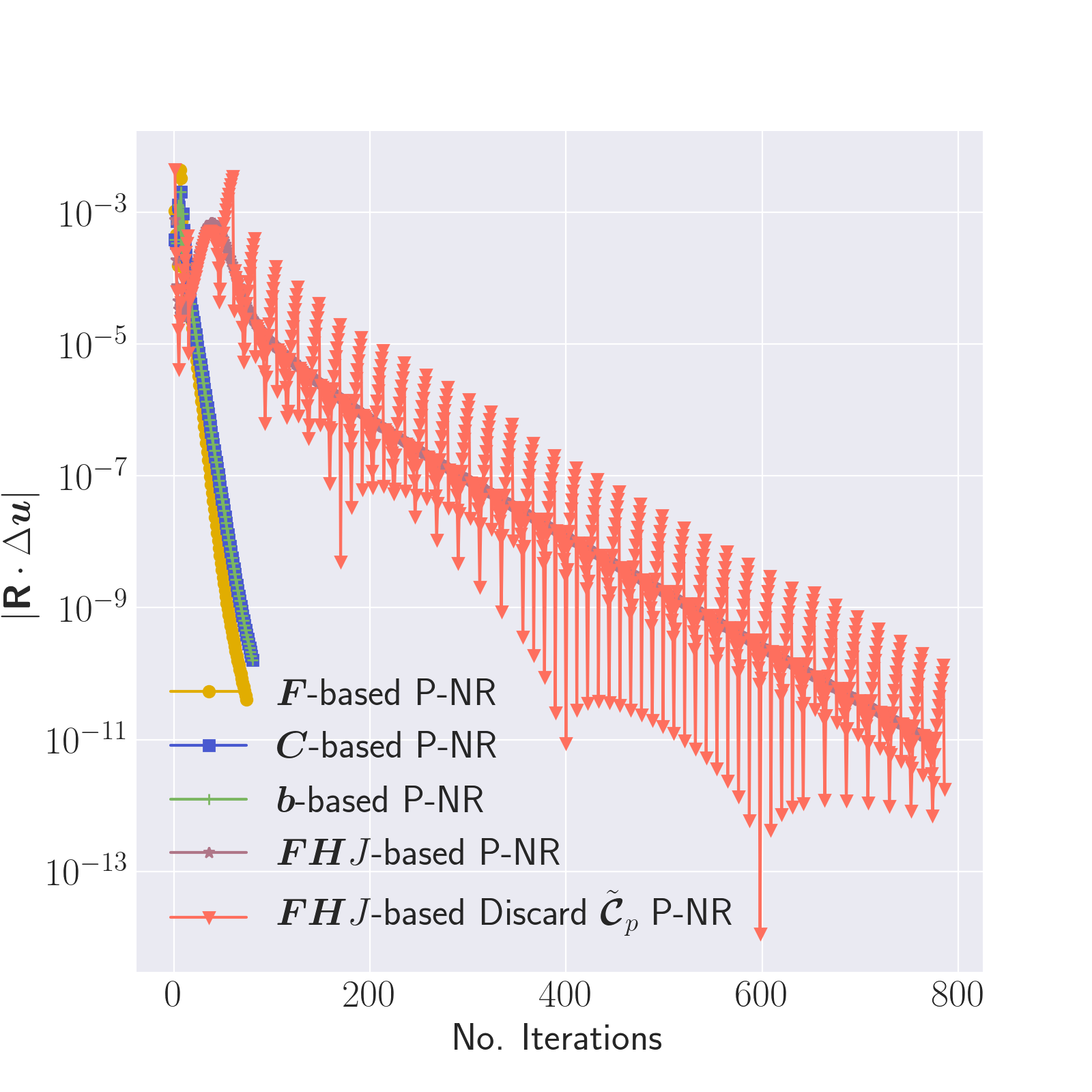}
\includegraphics[scale=0.42]{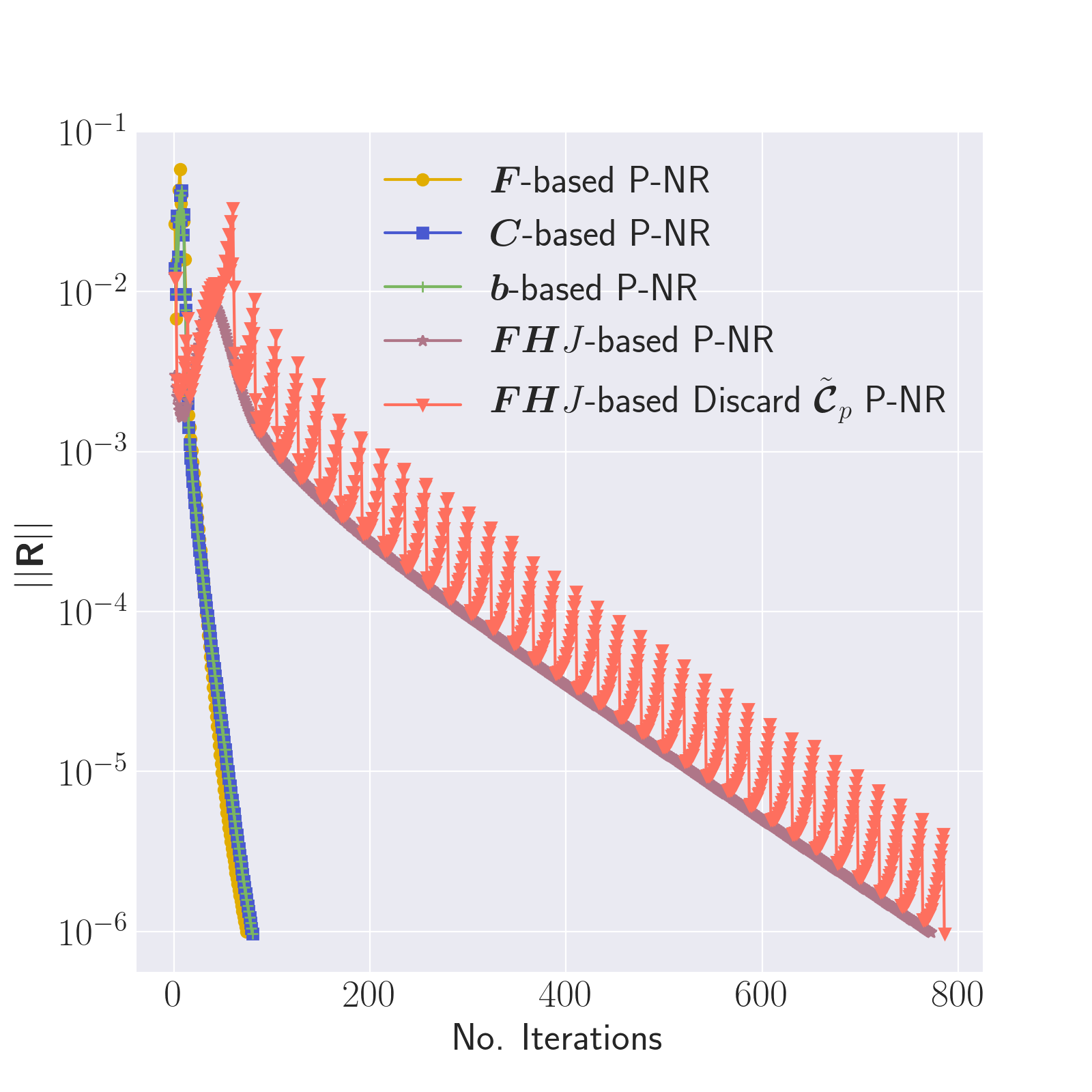}
\caption{Crash simulation of a cylindrical column from \autoref{fig:cylcrash} : Convergence of different tangent stabilisation techniques with P-NR. The performance of $\vec{C}$- and $\vec{b}$-based formulations is close to $\vec{F}$-based although $\vec{F}$-based formulation is still marginally faster and requires fewer iterations. PSD projection of the initial stiffness ($\vec{\tilde{\mathcal{C}}}_p$) alone in polyconvex elasticity leads to unnecessarily truncation of second-order information (i.e. Hessian terms) which results in inferior convergence pattern. Discarding this initial stiffness altogether results in oscillation in residual/gradient.}
\label{fig:cylcrashconv}
\end{figure}

Finally, we would like to assess the performance of different tangent stabilisation techniques presented for this complex problem. \autoref{fig:cylcrashconv} shows convergence of P-NR with different tangent stabilisation technique for both residual and curvature. First of all, it is clear that all techniques converge to the same solution without getting stuck. Unlike \autoref{fig:compblockpnr_convergence} however, we see that the $\vec{C}$- and $\vec{b}$-based formulations is now closer in performance to $\vec{F}$-based. As pointed out earlier, the convergence characteristics of these formulations are problem dependent although the $\vec{F}$-based formulation still converges quicker. The more interesting insight however comes from the geometrically stabilised elasticity formulation whose performance is considerably worse than the former 3 formulations. The explanation for this follows the same logic discussed earlier that is, while the initial stiffness ($\vec{\tilde{\mathcal{C}}}_p$) that emerges in polyconvex elasticity might not be (and often cannot be as clear from \ref{eq:amips_pk1lambs}, for instance) PSD, the total tangent stiffness ($\vec{\tilde{\mathcal{C}}}$) could very well be PSD owing to the contribution of always PSD constitutive tangent ($\vec{\tilde{\mathcal{C}}}_k$). Hence, PSD projection of the initial stiffness ($\vec{\tilde{\mathcal{C}}}_p$) alone truncates second-order information (i.e. Hessian terms) unnecessarily which results in damped and rather inferior convergence. Finally, it makes sense to include in the results the case of polyconvex elasticity by completely discarding the initial stiffness. This is a rather popular Gauss-Newton approximation in numerical optimisation. We see from the figures that discarding the initial stiffness results in significant oscillation in convergence however, the total number of iterations remains the same as that of geometrically stabilised polyconvex elasticity, for this problem. In sum, $\vec{F}$-based formulation remains the least intrusive and most competitive way of tangent stabilisation.


\section{Concluding remarks}\label{sec:10}
In this work, we have generalised the notion of tangent stabilised elasticity to virtually all known invariant formulations of nonlinear elasticity. We have shown that, closed-form eigen-decomposition of tangents is easily available irrespective of invariant formulation. In particular, we have worked out closed-form tangent eigensystems for Total Lagrangian deformation gradient ($\vec{F}$)-based and right Cauchy-Green ($\vec{C}$)-based as well as Updated Lagrangian left Cauchy-Green ($\vec{b}$)-based formulations and present their exact convexity conditions postulated in terms of their corresponding tangent and initial stiffness eigenvalues. In addition, we have introduced the notion of geometrically stabilised polyconvex large strain elasticity for models that are materially stable but exhibit geometric instabilities for whom we have constructed their initial stiffness in a spectrally-decomposed form analytically. Numerical experiments however show that compared to $\vec{F}$-based formulation, geometrically stabilised polyconvex elasticity might not be competitive as it leads to over-truncation of tangent elasticity. We have further extended the framework to the case of transverse isotropy where once again, closed-form tangent eigensystems are found for common transversely isotropic invariants. We have further presented a mixed variational formulation for models with arbitrarily directed inextensible fibres. Finally, we have proposed a discretisation-aware load-stepping together with a line search scheme for a robust implementation of tangent stabilised elasticity over general polyhedral meshes. Multiple examples have been presented that show the robustness and performance characteristics of tangent stabilised elasticity specifically when compared to the path-following approaches such as arch-length. In this regard Newton-Raphson with Projected (stabilised) tangents i.e. P-NR is shown to be competitive to arc-length requiring no manual user intervention. The automatic nature of specific P-NR flavours showcased in this work is due in part to the load stepping scheme that is sensitive to discretisation. Although we have introduced curvature as a different way to measure convergence, we do not claim that it is the best measure. Additionally, although P-NR seems highly robust, it may still encounter convergence issues. This can occur when the line search becomes too small, causing the simulations to become unresponsive. There may be other criteria for tracking convergence in P-NR that are superior. We note that, an analysis of performance of P-NR with implicit and explicit transient dynamic analysis algorithms would be indeed an interesting future piece of work. \\ \ \\
\textbf{Acknowledgement} \\
The first author thanks Brent Meranda manager of the Meshing Framework, Discretization, Simulation and Test Solutions, Siemens Digital Industries Software. The second author acknowledges the support of grant PID2022-141957OA-C22 funded by MCIN/AEI/10.13039/501100011033 and by ``RDF A way of making Europe". The second author also acknowledges the support provided by the Autonomous Community of the Region of Murcia, Spain through the programme for the development of scientific and technical research by competitive groups (21996/PI/22), included in the Regional Program for the Promotion of Scientific and Technical Research of Fundacion Seneca - Agencia de Ciencia y Tecnologia de la Region de Murcia. The third author acknowledges the financial support received through the UK Defence, Science and Technology Laboratory, grants: DSTLX1000160466R and RQ0000028640.
\\ \ \\

\appendix

\section{Variational formulations}\label{sec:6}
\subsection{Standard displacement-based variational principle}
The solution of large strain elastic problems is often expressed by means of the total energy minimisation variational principle as
\begin{align}
\Pi(\vec{x}^\star) = \min_{\vec{x}} \int_V W(\mathcal{A}_{\vec{x}}) \;\mathrm{d}V - \int_V \vec{f} _0 \cdot \vec{x} \;\mathrm{d}V - \int_{\partial_t V} \vec{t} _0 \cdot \vec{x} \;\mathrm{d}A,
\end{align}
where $\vec{x}^\star$ denotes the exact solution. The stationary condition of this functional leads to the principle of virtual work, commonly written as
\begin{align}
D\Pi[\delta \vec{u}] = \int_V \vec{P^x} : \delta  \vec{\nabla}_0 \vec{u}  \;\mathrm{d}V - \int_V \vec{f} _0 \cdot \delta \vec{u} \;\mathrm{d}V -  \int_{\partial_t V} \vec{t} _0 \cdot \delta \vec{u} \;\mathrm{d}A  = 0.
\end{align}
In this expression, the first Piola-Kirchhoff tensor $\vec{P^x}$ is evaluated in the standard fashion using \eqref{eq:lambpk1} in terms of the gradient of the deformation $\vec{\nabla}_0 \vec{x}$ and in our setting its singular-values. For better clarification of the notation, it is useful to introduce the definition of the geometrically compatible strain and stretch measures as
\begin{align}
\vec{F_x} = \hat{\vec{U_x}}\vec{\Lambda_x}\hat{\vec{V_x}}^T = \vec{\nabla}_0 \vec{x}, \quad
\lambda_{\vec{x}_i}  = \Lambda_{x_{ii}}, \quad \vec{P^x} =  \sum_{i=1}^d  \Sigma_{\lambda_i}^{\vec{x}} \frac{\partial \lambda_{\vec{x}_i}}{\partial \vec{F_x}} = \hat{\vec{U_x}} \vec{\Lambda}_{\vec{P^x}} \hat{\vec{V_x}}^T, \quad \Sigma^{x}_{\lambda_i} = \Sigma_{\lambda_i}(\lambda_{x_i}),
\label{eq:lambpk1x}%
\end{align}
where the superscript (subscript) $\vec{x}$ is used for stresses (strains) to indicate that they are evaluated in terms of geometry. An iterative Newton-Raphson process to converge towards the solution is usually established by solving a linearised system for the increment $\Delta \vec{u}$ as
\begin{align}
D^2\Pi[\delta\vec{u};\Delta \vec{u}] = -D\Pi(\vec{x})[\delta\vec{u}], \qquad \vec{x}_{k+1} = \vec{x}_k + \alpha \Delta \vec{u},
\end{align}
where $\alpha$ is the line search parameter and the second derivative of the total energy functional is given by
\begin{align}
D^2\Pi[\delta\vec{u};\Delta \vec{u}] = \int_V D^2 e[\delta \vec{u},\Delta \vec{u}]\;\text{d}V.
\end{align}
The tangent operator is evaluated taking  $\vec{F_x} = \vec{\nabla}_0 \vec{x}$ and $\lambda_{\vec{x}_i}$ as its singular-values.
\subsection{Mixed variational principle for isotropic elasticity}
A series of novel mixed variational principles for principal stretch based formulations were presented in \citet{Poya2023b} for general analysis of isotropic compressible and nearly incompressible elastica which for completeness we briefly summarise. For isotropic elasticity, a variational principle can be written in terms of the geometry and principal stretches as a constrained minimisation problem using a standard Lagrange multiplier approach
\begin{align}
&\Pi_M(\vec{x}^\star,\mathcal{A}^\star, \mathcal{S}^\star) = \min_{\vec{x}, \mathcal{A}} \bigg\{ \max_{\mathcal{S}}  \bigg\{ \int_V W(\mathcal{A}) \;\mathrm{d}V + \sum_{i=1}^d \int_V \Sigma_{\lambda_i}(\lambda_{\vec{x}_i} - \lambda_i) \;\mathrm{d}V - \int_V \vec{f} _0 \cdot \vec{x} \;\mathrm{d}V -  \int_{\partial_t V} \vec{t} _0 \cdot \vec{x} \;\mathrm{d}A  \bigg\}  \bigg\}.
\end{align}
This 7-field formulation belongs to the general class of Hu-Washizu type of mixed variational principles. Note that, the conjugate variables $\Sigma_{\lambda_i}$ in this expression, are simply Lagrange multipliers.

\subsection{Mixed variational principle for isotropic rigid and stiff models}
A variational principle can be easily established as particular case of the above one for the case of nearly rigid i.e.  stiff materials or the practically relevant case, the truly rigid formulation which do not have a deformable part
\begin{align}
\Pi_M^R(\vec{x}^\star, \mathcal{S}^\star) = \min_{\vec{x}} \bigg\{ \max_{\mathcal{S}}  \bigg\{ \sum_{i=1}^d \int_V \Sigma_{\lambda_i}(\lambda_{\vec{x}_i} - 1) -\frac{1}{2C_c} \Sigma_{\lambda_i}^2 \;\mathrm{d}V - \int_V \vec{f} _0 \cdot \vec{x} \;\mathrm{d}V -  \int_{\partial_t V} \vec{t} _0 \cdot \vec{x} \;\mathrm{d}A  \bigg\}  \bigg\},
\end{align}
where $C_c > 0$ is an additional penalty term to avoid saddle point problem and the formulation above falls under the general category of Perturbed Lagrangian formulation.

\subsection{Mixed variational principle for transversely isotropic elasticity with inextensible fibres}
\citet{Poya2023b}'s approach is essentially an $\vec{F}$-based formulation in terms of principal stretches. As seen in the above $\vec{\tilde{\mathcal{C}}}_k^{ti} $ it is already positive definite. Hence, it requires no special treatment. In general, following this approach all isotropic invariants are resolved using the SVD but no SVD is performed for transversely isotropic cases. Following \cite{Auricchio2017a, Wriggers2016a}, we introduce the following variational principle
\begin{align}
\Pi_M^I(\vec{x}^\star, \Sigma_{c_i}^\star) = \min_{\vec{x}} \bigg\{ \max_{\Sigma_{c_i}}  \bigg\{ \int_V {W}(\mathcal{A}_x) \;\mathrm{d}V + \sum_{i=1}^m \int_V \Sigma_{c_i}(I_4^i - 1) -\frac{1}{2C_i} \Sigma_{c_i}^2 \;\mathrm{d}V - \int_V \vec{f} _0 \cdot \vec{x} \;\mathrm{d}V -  \int_{\partial_t V} \vec{t} _0 \cdot \vec{x} \;\mathrm{d}A  \bigg\}  \bigg\},
\end{align}
where $m$ is the number of specified fibre directions and $I_4^i = \vec{F}[\vec{M}]^i\cdot \vec{F}[\vec{M}]^i$ denotes the inextensibility constraint condition in the $i$th direction with $[\vec{M}]^i$ the corresponding $i$th fibre direction. The stationary conditions of this hybrid functional is evaluated in the same fashion as previous formulation. For instance, the first derivative with respect to geometry gives the principle of virtual work as
\begin{align}
D_{1}\Pi_M^I[\delta \vec{u}] = \int_V \vec{P}^I :  \vec{\nabla}_0 \delta \vec{u}  \;\mathrm{d}V - \int_V \vec{f} _0 \cdot \delta \vec{u} \;\mathrm{d}V -  \int_{\partial_t V} \vec{t} _0 \cdot \delta \vec{u} \;\mathrm{d}A  = 0,
\end{align}
where the first Piola-Kirchoff stress tensor is now evaluated as
\begin{align}
\vec{P}^I =  \sum_{i=1}^d  \Sigma_{\lambda_{\vec{x}_i}} \frac{\partial \lambda_{\vec{x}_i}}{\partial \vec{F_x}} + 2 \sum_{i=1}^m  \Sigma_{c_i} \vec{F} \left( [\vec{M}]^i \otimes [\vec{M}]^i \right).
\label{eq:lambpk1i}%
\end{align}
The first derivative with respect to $\Sigma_{c_i}$ enforces the inextensibility constraint
\begin{align}
D_{2}\Pi_M^I[\delta \Sigma_{c_i}] &= \sum_{i=1}^m \int_V \delta \Sigma_{c_i}  \left( (I_4^i - 1) - \frac{1}{C_i}  \Sigma_{c_i} \right)  \;\mathrm{d}V.
\label{eq:Rsc}
\end{align}
The evaluation of second derivatives required for a Newton-Raphson process proceeds along the same lines. For instance, the second derivative with respect to geometry leads to the initial stiffness operator
\begin{align}
D^2_{1;1}\Pi_M^I[\delta \vec{u}; \Delta \vec{u}] &= \int_V \vec{\nabla}_0 \delta \vec{u} : \vec{\mathcal{C}}^I:  \vec{\nabla}_0 \Delta \vec{u}  \;\mathrm{d}V,
\label{eq:ISmixed3}
\end{align}
where the total tangent stiffness is now evaluated as
\begin{align}
\vec{\mathcal{C}}^I = \sum_{i=1}^d  \sum_{j=1}^d \frac{\partial^2 W}{\partial {\lambda_{\vec{x}_i}} {\lambda_{\vec{x}_j}} }  \frac{\partial {\lambda_{\vec{x}_i}} }{\partial \vec{F}} \otimes  \frac{\partial {\lambda_{\vec{x}_j}} }{\partial \vec{F}} + \sum_{i=1}^d  \Sigma_{\lambda_{\vec{x}_i}} \frac{\partial^2 \lambda_{\vec{x}_i}}{\partial \vec{F_x}\partial \vec{F_x}} + 2 \sum_{i=1}^m  \Sigma_{c_i} \left(  \sum_{j=1}^d [\vec{E}]^i_j \otimes [\vec{E}]^i_j \right),
\label{eq:lambCI}%
\end{align}
where once again, $[\vec{E}]^i_j$ represents the $j$th eigenmatrix for the $i$th fibre. Alternatively, the PSD (stabilised) version of the tangent can be constructed as
\begin{align}
\vec{\mathcal{C}}^{I^{\texttt{PSD}}} &= %
\begin{bmatrix}
\dfrac{\partial \lambda_{\vec{x}_1}}{\partial \vec{F}}, &
\cdots, &
\dfrac{\partial \lambda_{\vec{x}_d}}{\partial \vec{F}}
\end{bmatrix}
[\vec{H}^{\texttt{PSD}}_W]
\begin{bmatrix}
\dfrac{\partial \lambda_{\vec{x}_1}}{\partial \vec{F}} \\
\vdots \\
\dfrac{\partial \lambda_{\vec{x}_d}}{\partial \vec{F}}
\end{bmatrix} \nonumber \\
&+ \sum_{i=1}^k \text{max}(\bar{\lambda}_{x_i}^{\vec{\mathcal{C}_p}}, 0)  [\vec{L}]_i \otimes [\vec{L}]_i + \text{max}(\bar{\lambda}_{x_{i+k}}^{\vec{\mathcal{C}_p}}, 0)  [\vec{T}]_i \otimes [\vec{T}]_i \nonumber \\
&+ 2 \sum_{i=1}^m  \text{max}(\Sigma_{c_i}, 0) \left(  \sum_{j=1}^d [\vec{E}]^i_j \otimes [\vec{E}]^i_j \right),
\label{eq:lambCIPSD}%
\end{align}
where $\bar{\lambda}_{x_i}^{\vec{\mathcal{C}}}$ indicates the evaluation tangent operator with respect to geometry. Crucially, PSD projection (stabilisation) requires pruning the contribution of transversely isotropic invariant to tangent operator when the inextensibility constraint $\Sigma_{c_i}$ becomes negative. The second derivative with respect to $\Sigma_{c_i}$ has a standard Galerkin mass matrix format
\begin{align}
D^2_{2;2}\Pi_M^I[\delta \Sigma_{c_i}; \Delta \Sigma_{c_i}] &=  -\sum_{i=1}^m \frac{1}{C_i} \int_V \delta \Sigma_{c_i} \cdot \Delta \Sigma_{c_i}  \;\mathrm{d}V,
\label{eq:ISmixed3}
\end{align}
and finally, the cross terms emerging from derivatives of geometry and inextensibility constraint are obtained as
\begin{subequations}
\begin{align}
D^2_{1;2}\Pi_M^I[\delta \vec{u}; \Delta \Sigma_{c_i}] &= 2\sum_{i=1}^m \int_V \left( \vec{F}[\vec{M}]^i : \vec{\nabla}_0 \delta \vec{u} \right) \cdot \Delta \Sigma_{c_i} \;\mathrm{d}V, \\
D^2_{2;1}\Pi_M^I[\delta \Sigma_{c_i}; \Delta \vec{u}] &= 2\sum_{i=1}^m \int_V \delta \Sigma_{c_i} \cdot \left( \vec{F}[\vec{M}]^i : \vec{\nabla}_0 \Delta \vec{u} \right) \;\mathrm{d}V.
\end{align}%
\label{eq:ISmixed4}%
\end{subequations}%

\section{Finite element discretisation}\label{sec:7}
The implementation of the various variational principles described in the previous section is based on a finite element  partition of the domain into a set of elements. Inside each element the problem variables are interpolated in terms of a set of shape functions $N_a$ as
\begin{align}
\vec{x} = \sum_{a=1}^{n_x} \vec{x}_a N_a^x, \quad
\Sigma_{c_i} = \sum_{a=1}^{n_{\Sigma_{c_i}}} \Sigma_{c_{i_a}} N_a^{\Sigma_{c_i}},
\end{align}
where $a$ denotes the nodes or other degrees of freedom used in the interpolation of the above variables. In general, different interpolations are often used to describe different variables. However, the same interpolation space will invariably be used for strain-stress conjugate pairs; that is, $N_a^{\lambda_i} = N_a^{\Sigma_{\lambda_i}}$, etc. The virtual and incremental variables are also interpolated using the same spaces as
\begin{align}
\delta \vec{u} &= \sum_{a=1}^{n_x} \delta \vec{u}_a N_a^{x}, \quad
\delta \Sigma_{\lambda_i} = \sum_{a=1}^{n_{\Sigma_{\lambda_i}}} \delta \Sigma_{\lambda_{i_a}} N_a^{\Sigma_{c_i}}, \quad
\Delta \vec{u} = \sum_{a=1}^{n_x} \Delta \vec{u}_a N_a^{x}, \quad
\Delta \Sigma_{c_i} = \sum_{a=1}^{n_{\Sigma_{c_i}}} \Delta \Sigma_{c_{i_a}} N_a^{\Sigma_{c_i}}.
\end{align}
Finite element equations are derived by simply substituting the above  expressions into the functional expressions provided in the previous section. For instance, substituting the above interpolation for the virtual displacements into any of the virtual work statements given in the previous section leads to residual forces as
\begin{align}
D_1\Pi[\delta \vec{u}] = \sum_a \vec{R}_a^{\vec{x}} \cdot \delta \vec{u}, \qquad
\vec{R}_a^{\vec{x}} = \int_{V^e} \vec{P} \vec{\nabla}_0 N_a^{x} \; \text{d}V^e - \int_{V^e} \vec{f}_0 N_a^{x} \; \text{d}V^e - \int_{\partial_t V^e} \vec{t}_0 N_a^{x} \; \text{d}A^e,
\end{align}
where the first Piola-Kirchhoff stress tensor above will be evaluated in accordance with each of the formulations presented in the previous section and $V^e$ and $A^e$ denote the volume and area of element $e$ in the original configuration. We only present the finite element discretisation for the transversely isotropic elasticity with inextensible fibres as the discretisation for other formulations have already been presented in \citet{Poya2023b}. For instance, the geometric compatibility residuals $\vec{R}_a^{\Sigma_{c_i}}$ emerge from the discretisation of \eqref{eq:Rsc}.
\begin{align}
\vec{R}_a^{\Sigma_{c_i}} = \sum_{i=1}^m
 \int_{V^e} \left(I_4^i - 1 - \frac{\Sigma_{c_i}}{C_i} \right) N_a^{\Sigma_{c_i}}
\; \text{d}V^e.
\end{align}
In order to complete the finite element formulation it is necessary to derive equations for the components of the tangent matrix by discretising the tangent operators. For the case of mixed inextensible formulation, the resulting tangent operator expands to
\begin{align}
D^2\Pi_M[\delta \vec{u}, \delta  \Sigma_{c_i}; \Delta \vec{u}, \Delta  \Sigma_{c_i} ] &=
\begin{bmatrix}
\delta \vec{u} & \delta \Sigma_{c_i}
\end{bmatrix}
\begin{bmatrix}
\vec{K}_{\vec{xx}}                      & \vec{K}_{\vec{x} \Sigma_{c_i}} \\
\vec{K}_{ \Sigma_{c_i} \vec{x}}  & \vec{K}_{ \Sigma_{c_i} \Sigma_{c_i}}
\end{bmatrix}
\begin{bmatrix}
\Delta \vec{u} \\ \Delta  \Sigma_{c_i}
\end{bmatrix},
\end{align}
where $\vec{K}_{\vec{x} \Sigma_{c_i}}$ and $\vec{K}_{\Sigma_{c_i} \Sigma_{c_i}}$ emerge from the discretisation of \eqref{eq:ISmixed4} and \eqref{eq:ISmixed3}, respectively
\begin{subequations}
\begin{align}
\vec{K}_{\vec{x} \Sigma_{c_i}}^{ab} &= 2 \sum_{i=1}^m
 \int_{V^e} \left( \vec{F}[\vec{M}]^i : \vec{\nabla}_0 N_a^{\vec{x}}  \right) N_b^{\Sigma_{c_i}} \; \text{d}V^e, \\
\vec{K}_{\Sigma_{c_i} \Sigma_{c_i}}^{ab} &= - \sum_{i=1}^m \frac{1}{C_i} \int_{V^e} N_a^{\Sigma_{c_i}} \cdot N_b^{\Sigma_{c_i}} \; \text{d}V^e.
\end{align}
\end{subequations}
These equations can be implemented using a variety of finite element spaces but, not all choices will lead to effective or valid finite element formulations \cite{CHAPELLE1993537,AURICCHIO20051075,AuricchioChapter2017a,CERVERA20102559,BROOKS1982199,HUGHES198685,ELGUEDJ2014388,SchroderMixed17a}. Some choices of ansatz spaces are given in  \cite{Auricchio2017a, Wriggers2016a} for continuous interpolation of all fields and in \cite{Poya2023b} for discontinuous interpolation of secondary (dual) variables. 

\section{Analytic eigensystem of neo-Hookean type energies in Voigt notation}\label{app:voigt}
A practically relevant case of Voigt notation arises for neo-Hookean-style energies of type
\begin{align}
\breve{W} = \frac{\mu}{2} (\text{tr}(\vec{C}) - N) + f(J, \kappa),
\end{align}
where $\mu$ and $\kappa$ are Lam\`e parameters. For such energy functions the constitutive tangent takes the following form
\begin{align}
\breve{\vec{\mathcal{C}}}_k = \alpha C_{IJ}^{-1}C_{KL}^{-1} + \beta (C_{IK}^{-1}C_{JL}^{-1} + C_{IL}^{-1}C_{JK}^{-1})
\end{align}
where $\alpha$ and $\beta$ are in general variables depending on material parameters and $C = J^2$. The explicit constitutive tangent of these energies in 3-dimensions can be written as
\begin{align}
\breve{\vec{\mathcal{C}}}_k^{Voigt} =
\begin{bmatrix}
C_{1111} & C_{1122} & C_{1133} & 0 & 0 & 0 \\
& C_{2222} & C_{2233} & 0 & 0 & 0 \\
& & C_{3333} & 0 & 0 & 0 \\
& & & C_{1212} & 0 & 0 \\
& & & & C_{1313} & 0 \\
sym & & & & & C_{2323}
\end{bmatrix},
\end{align}
whose analytic eigensystem separates again into \emph{scaling} and \emph{flip} modes similar to \eqref{eq:lambCC2} but  now in simpler vector form due to Voigt notation
\begin{align}
\breve{\vec{\mathcal{C}}}_k^{Voigt} = \sum_{i=1}^3\sum_{j=1}^3 \bar{\lambda}_i^{ \breve{\vec{\mathcal{C}}}_k^{Voigt} } [\bar{\vec{e}}]_i^j [\vec{d}]_j \otimes  [\bar{\vec{e}}]_i^j [\vec{d}]_j +  \sum_{i=1}^3 \bar{\lambda}_{i+3}^{ \breve{\vec{\mathcal{C}}}_k^{Voigt} } [\vec{l}]_i  \otimes [\vec{l}]_i
\end{align}
where the  \emph{scaling} and \emph{flip} eigenvectors are simply unit vectors
\begin{align}
[\vec{d}]_1 = [1, 0, 0, 0, 0, 0]^T, \quad
[\vec{d}]_2 = [0, 1, 0, 0, 0, 0]^T, \quad
[\vec{d}]_3 = [0, 0, 1, 0, 0, 0]^T, \\
[\vec{l}]_1 = [0, 0, 0, 1, 0, 0]^T, \quad
[\vec{l}]_2 = [0, 0, 0, 0, 1, 0]^T, \quad
[\vec{l}]_3 = [0, 0, 0, 0, 0, 1]^T
\end{align}
and the last 3 eigenvalues are already decoupled and appear as
\begin{align}
\bar{\lambda}_4^{ \breve{\vec{\mathcal{C}}}_k^{Voigt} }  = C_{1212}, \quad
\bar{\lambda}_5^{ \breve{\vec{\mathcal{C}}}_k^{Voigt} }  = C_{1313}, \quad
\bar{\lambda}_6^{ \breve{\vec{\mathcal{C}}}_k^{Voigt} }  = C_{2323}.
\end{align}
For neo-Hookean style energies often $C_{1111} = C_{2222} = C_{3333}$, $C_{1122} = C_{1133} = C_{2233}$ and $C_{1212} = C_{1313} = C_{2323}$ and, the first 3 eigenvalues (and eigenvectors $[\bar{\vec{e}}]_i)$ are obtained from the eigen-decomposition of the upper $3\times 3$ block of $\breve{\vec{\mathcal{C}}}_k^{Voigt}$ which actually corresponds to $\vec{H}_{\bar{W}}$ from \eqref{eq:invarHwC}
\begin{align}
\bar{\lambda}_1^{ \breve{\vec{\mathcal{C}}}_k^{Voigt} }  = C_{1111} + 2 C_{1122}, \quad
\bar{\lambda}_2^{ \breve{\vec{\mathcal{C}}}_k^{Voigt} }  = C_{1111} - C_{1122}, \quad
\bar{\lambda}_3^{ \breve{\vec{\mathcal{C}}}_k^{Voigt} }  = C_{1111} - C_{1122}.
\end{align}
In 3-dimensional Voigt format the eigenvectors of $\vec{H}_{\bar{W}}$ can always be chosen as $[\bar{\vec{e}}]_1 = \frac{1}{\sqrt{3}} [1, 1, 1]^T$, $[\bar{\vec{e}}]_2 =  \frac{1}{\sqrt{6}} [-2, 1, 1]^T$ and $[\bar{\vec{e}}]_3 =  \frac{1}{\sqrt{2}} [0, 1, -1]^T$ to give
\begin{empheq}[box=\widefbox]{align}
\breve{\vec{\mathcal{C}}}_k^{Voigt} &=
\frac{C_{1111} + 2C_{1122}}{3} [\vec{q}]_1 \otimes [\vec{q}]_1
+ \frac{C_{1111} - C_{1122}}{6} [\vec{q}]_2  \otimes [\vec{q}]_2
+ \frac{C_{1111} - C_{1122}}{2} [\vec{q}]_3  \otimes [\vec{q}]_3 \nonumber \\
&+ C_{1212} [\vec{l}]_1  \otimes [\vec{l}]_1 + C_{1212} [\vec{l}]_2  \otimes [\vec{l}]_2 + C_{1212} [\vec{l}]_3  \otimes [\vec{l}]_3,
\end{empheq}
with $[\vec{q}]_1 = [1,1,1,0,0,0]^T$, $[\vec{q}]_2 = [-2,1,1,0,0,0]^T$ and $[\vec{q}]_2 = [0,1,-1,0,0,0]^T$. Similar but simpler expressions are found in 2-dimensions
\begin{align}
\breve{\vec{\mathcal{C}}}_k^{Voigt} =
\begin{bmatrix}
C_{1111} & C_{1122} & 0 \\
& C_{2222} & 0 \\
sym & & C_{1212}
\end{bmatrix} =
\sum_{i=1}^2\sum_{j=1}^2 \bar{\lambda}_i^{ \breve{\vec{\mathcal{C}}}_k^{Voigt} } [\bar{\vec{e}}]_i^j [\vec{d}]_j \otimes  [\bar{\vec{e}}]_i^j [\vec{d}]_j +  \bar{\lambda}_{3}^{ \breve{\vec{\mathcal{C}}}_k^{Voigt} } [\vec{l}]_1  \otimes [\vec{l}]_1
\end{align}
where the  \emph{scaling} and \emph{flip} eigenvectors are simply unit vectors
\begin{align}
[\vec{d}]_1 = [1, 0, 0]^T, \quad
[\vec{d}]_2 = [0, 1, 0]^T, \quad
[\vec{l}]_1 = [0, 0, 1]^T
\end{align}
and, the eigenvalues for neo-Hookean style energies are given by
\begin{align}
\bar{\lambda}_1^{ \breve{\vec{\mathcal{C}}}_k^{Voigt} }  = C_{1111} + C_{1122}, \quad
\bar{\lambda}_2^{ \breve{\vec{\mathcal{C}}}_k^{Voigt} }  = C_{1111} - C_{1122}, \quad
\bar{\lambda}_3^{ \breve{\vec{\mathcal{C}}}_k^{Voigt} }  = C_{1212}.
\end{align}
Similarly, in 2-dimensional Voigt format the eigenvectors of $\vec{H}_{\bar{W}}$ can always be chosen as $[\bar{\vec{e}}]_1 = \frac{1}{\sqrt{2}} [1, 1]^T$ and $[\bar{\vec{e}}]_2 =  \frac{1}{\sqrt{2}} [-1, 1]^T$ to give
\begin{empheq}[box=\widefbox]{align}
\breve{\vec{\mathcal{C}}}_k^{Voigt} =
\frac{C_{1111} + C_{1122}}{2} [\vec{q}]_1 \otimes [\vec{q}]_1 +
\frac{C_{1111} - C_{1122}}{2} [\vec{q}]_2  \otimes [\vec{q}]_2
+  C_{1212} [\vec{l}]_1  \otimes [\vec{l}]_1,
\end{empheq}
with $[\vec{q}]_1 = [1,1,0]^T$ and $[\vec{q}]_2 = [1,-1,0]^T$. Note that, the above eigen-decompositions hold true for both $\vec{C}$- and $\vec{b}$-based formulations.

\section{Symbolic code for computing tangent eigensystems}\label{app:b}
\begin{figure}
\centering
\includegraphics[scale=0.73]{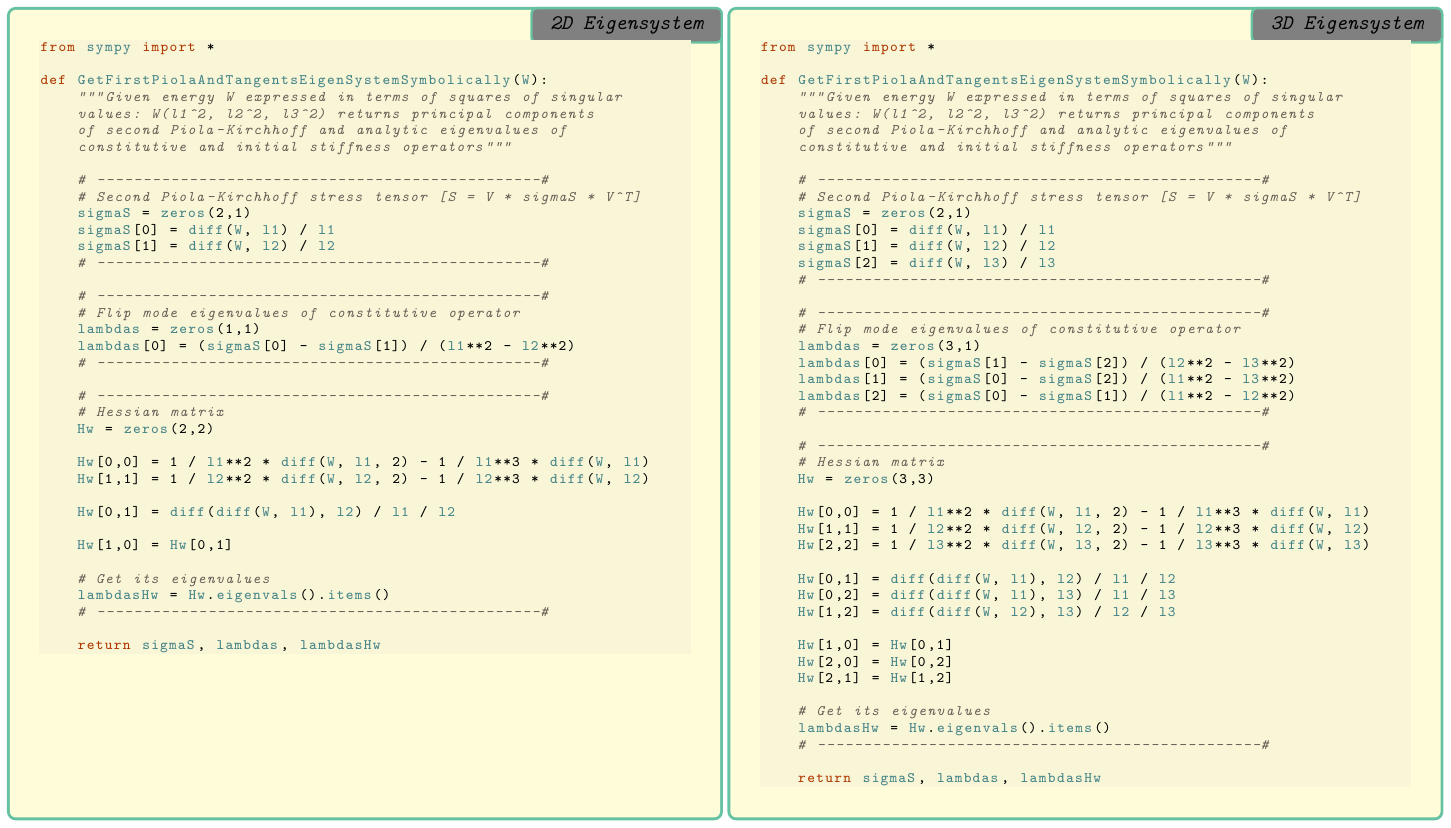}
\caption{Symbolic code for obtaining principal components of Second Piola-Kirchhoff (which are also eigenvalues of initial stiffness) and eigensystem of constitutive tangent operator in 2 and 3 dimensions.}
\label{fig:symcodeCbased}
\end{figure}

Symbolic code for the analytic eigensystems of $\vec{F}$-based formulation was given in \citet{Poya2023b}. In \autoref{fig:symcodeCbased} we provide the analytic eigensystems of $\vec{C}$-based formulation from which the analytic eigensystem of $\vec{b}$-based formulation can also be obtained using the push-forward operations defined in \eqref{eq:pushfd_Hw}, \eqref{eq:pushfd_flips} and \eqref{eq:pushfd_IS}.

\small{%
\bibliographystyle{model1-num-names}
\bibliography{bibliography}
}

\end{document}